\documentclass{aa}
\usepackage[varg]{txfonts}
\usepackage{epsfig}
\usepackage{natbib}
\bibpunct{(}{)}{;}{a}{}{,} 

\usepackage{graphicx}
\usepackage[]{hyperref}

\usepackage{siunitx}

\newcommand{\ms}{\:{\rm ms}}

\newcommand\kms{\:\rm{\,km\,s^{-1}}}

\newcommand\masy{\:\rm{\,mas\:yr^{-1}}}

\begin{document}

\title{A kinematic study of central compact objects \\
and their host supernova remnants}

\author{Martin G.~F.~Mayer\thanks{\email{mmayer@mpe.mpg.de}}\inst{1} \and Werner Becker\inst{1,2}}
\institute{Max-Planck Institut f\"ur extraterrestrische Physik, Giessenbachstrasse, 85741 Garching, Germany \and 
Max-Planck Institut f\"ur Radioastronomie, Auf dem H\"ugel 69, 53121 Bonn, Germany}

\date{Received 19 April 2021 /
Accepted 28 May 2021 }

\abstract
{Central compact objects (CCOs) are a peculiar class of neutron stars, primarily encountered close to the center of young supernova remnants (SNRs) and characterized by thermal X-ray emission. 
Measurements of their proper motion and the expansion of the parent SNR are powerful tools for constraining explosion kinematics and the age of the system.} 
{Our goal is to perform a systematic study of the proper motion of all known CCOs with appropriate data available. 
From this, we hope to obtain constraints on the violent kick acting on the neutron star during the supernova explosion, and infer the exact site of the explosion within the SNR. In addition, we aim to measure the expansion of three SNRs within our sample to obtain a direct handle on their kinematics and age.} 
{We analyze multiple archival \emph{Chandra} data sets, consisting of HRC and ACIS observations separated by temporal baselines between 8 and 15 years. 
We achieve accurate source positions by fitting the imaging data with ray-tracing models of the \emph{Chandra} point spread function. 
In order to correct for \emph{Chandra}'s systematic astrometric uncertainties, we establish a reference frame using X-ray detected sources in \emph{Gaia} DR2, to provide accurate proper motion estimates for our target CCOs.
Complementarily, we use our coaligned data sets to trace the expansion of three SNRs by directly measuring the spatial offset of various filaments and ejecta clumps between different epochs.}
{In total, we present new proper motion measurements for six CCOs. 
Within our sample, we do not find any indication of a hypervelocity object, and we determine comparatively tight upper limits ($<230\,\si{km.s^{-1}}$) on the transverse velocities of the CCOs in G330.2+1.0 and RX J1713.7$-$3946. 
We tentatively identify direct signatures of expansion for the SNRs G15.9+0.2 and Kes 79, at estimated significance of $2.5\sigma$ and $2\sigma$, respectively.   
Moreover, we confirm recent results by Borkowski et al., measuring the rapid expansion of G350.1$-$0.3 at almost $6000\,\si{km.s^{-1}}$, which places its maximal age at $600-700$ years, making this object one of the youngest Galactic core-collapse SNRs. 
The observed expansion, combined with the proper motion of its CCO which is much slower than previously predicted, implies the need for a very inhomogeneous circumstellar medium to explain the highly asymmetric appearance of the SNR.
Finally, for the SNR RX J1713.7$-$3946, we combine previously published expansion measurements with our measurement of the CCO's proper motion to obtain a constraining upper limit of 1700 years on the system's age.
}  
{}


\titlerunning{Kinematic study of CCOs \& SNRs}
\maketitle

\section{Introduction}

\begin{table*}[t!]
\renewcommand{\arraystretch}{1.15}
\caption{CCOs with basic properties of their host SNRs.}
\label{CCOTable}
\centering
\begin{tabular}{ccccc}
\hline\hline
SNR & CCO & Distance & Age & References\\
    &   & $(\si{kpc})$ & $(\si{yr})$ & \\
\hline
\object{G15.9+0.2} & \object{CXOU J181852.0$-$150213} & $7-16$ & $2900-5700$ & 1,2,3
\\
\object{Kes 79} & \object{CXOU J185238.6+004020} & $3.5-7.1$ & $4400-6700$ & 4,5,6,7 
\\ 
\object{Cas A} & \object{CXOU J232327.9+584842} & $3.4$ & $\sim350$ & 8,9 \\ 
\object{Puppis A} & \object{RX J0822$-$4300} & $1.3-2.2$ & $2000-5300$ & 10,11,12,13,14 
\\ 
\object{G266.1$-$1.2} (Vela Jr.) & \object{CXOU J085201.4$-$461753} & $0.5-1.0$ & $2400-5100$ & 15 
\\ 
\object{PKS 1209$-$51/52} (G296.5+10.0) & \object{1E 1207.4$-$5209} & $1.3-3.9$ & $\sim 7000$ & 16,17 
\\
\object{G330.2+1.0} & \object{CXOU J160103.1$-$513353} & $4.9-10.0$ & $\lesssim 1000$ & 18,19,20 
\\ 
\object{RX J1713.7$-$3946} (G347.3$-$0.5) & \object{1WGA J1713.4$-$3949} & $1.0-1.3$ & $1500-2300$ & 21,22,23,24 
\\ 
\object{G350.1$-$0.3} & \object{XMMU J172054.5$-$372652} & $4.5$ & $\sim 600$ & 25,26,27 
\\ 
\object{G353.6$-$0.7} & \object{XMMU J173203.3$-$344518} & $\sim 3.2$ & $\sim 27\,000 $& 28,29 
\\ \hline
\end{tabular}
\tablefoot{The two SNRs G349.7+0.2 \citep{G349CCO} and G296.8$-$0.3 \citep{G296CCO} have been claimed to potentially host CCOs. Since for neither of these, \emph{Chandra} data suitable for our purpose exists, these sources are not further considered in our work. 
}
\tablebib{
(1) \citet{G15Dist}, (2) \citet{Sasaki18}, (3) \citet{Maggi17}, (4) \citet{KesDist}, (5) \cite{KesCODist}, (6) \citet{DistKes3}, (7) \citet{KesAge1}, (8) \citet{Fesen06}, (9) \citet{Alarie14}, (10) \citet{Reynoso17}, (11) \citet{Reynoso03}, (12) \citet{Aschenbach15}, (13) \citet{Winkler88}, (14) \citet{Mayer20}, (15) \citet{Allen15}, (16) \citet{Giacani00}, (17) \citet{Roger88}, (18) \citet{MCClure}, (19) \citet{Williams18}, (20) \citet{Borkowski18}, (21) \citet{Fukui}, (22) \citet{Cassam04}, (23) \citet{Acero2017}, (24) \citet{Tsuji2016}, (25) \citet{Lovchinsky}, (26) \citet{Bitran97}, (27) \citet{Borkowski20}, (28) \citet{Maxted18}, (29) \citet{Tian08}.}
\end{table*}

It is generally accepted that collapsing massive stars, i.e.~core-collapse supernovae, leave behind a compact remnant, a neutron star (NS) or black hole. 
A natural consequence of asymmetries in core-collapse supernovae is a strong kick acting on the compact remnant \citep[e.g.][]{Wongwa13}, whose kinematics are thus directly connected to explosion properties. 
Proper motion studies of NSs are therefore a powerful tool which allows to set a lower limit on the kinetic energy and momentum transferred to the NS during the explosion of the progenitor. Beyond that, depending on its age, the proper motion of a NS may reveal its origin or even the exact supernova explosion site, which can be used to robustly infer the age of the NS and/or associated supernova remnant (SNR).

For radio pulsars, which make up the vast majority of known NSs, proper motion can be measured precisely for a comparatively large number of sources, using precise positions from pulsar timing or very-long-baseline interferometry. This allows for population studies of NS kinematics, as for example by \citet{hobbs05} and \citet{Verbunt17}. Such works establish the picture of NSs being very high-speed objects, with a mean three-dimensional velocity around $400 \kms$, and reliably measured projected velocities at least as high as $\sim800 \kms$.

In contrast to radio pulsars, for NSs exhibiting at best weak radio and optical emission, such as magnetars, X-ray-dim isolated neutron stars, or central compact objects in supernova remnants (CCOs), such measurements are much more challenging.  
While all three of these neutron star classes are characterized by primarily thermal X-ray emission, CCOs are particularly noteworthy for being exclusively young ($\lesssim 10^4\,\si{yr}$) and nonvariable X-ray sources, located close to the center of supernova remnants. Unlike radio pulsars, they show no signs of interaction with their surroundings, such as extended nebular or jet-like emission. Furthermore, their spectra can be described entirely using blackbody or neutron star atmosphere models with luminosities $\gtrsim10^{33}\,\si{erg.s^{-1}}$, without any unambiguous evidence for nonthermal magnetospheric emission \citep[see][]{Becker09,Gotthelf13,DeLuca17}. 
Despite intensive searches \citep[e.g.][]{Mignani08,Mignani19}, no point-like or diffuse counterparts to any CCO at radio, infrared or optical wavelengths have been found with present-day instrumentation. This places tight upper limits on the presence of potential binary companions and indicates no need for additional components to explain the observed spectral energy distribution.  

We give an overview\footnote{\label{DeLucaNote} For a more expansive review of CCOs and their characteristics, we refer to \citet{DeLuca17} and the overview table on CCOs at \url{http://www.iasf-milano.inaf.it/~deluca/cco/main.htm}.} of known CCOs in Table \ref{CCOTable}.  
Out of ten known CCOs, only for those in Kes 79, Puppis A, and PKS 1209$-$51/52, X-ray pulsations have been detected. These pulsations at $100-500\,\ms$ are likely attributable to the rotational modulation of the emission from hotspots on the neutron star surface \citep{Gotthelf10,Gotthelf05,Zavlin00}. Analysis of their spin-down points toward CCOs having a very weak magnetic dipole field. This, in combination with their steady emitting behavior, lead to their designation as ``anti-magnetars'' \citep{Gotthelf13,Halpern10}. 

An important issue is that, observationally, few orphaned CCOs, meaning older pulsars without apparent nearby SNR but in a similar location of $(P,\dot{P})$--space, are found \citep{Luo15}. It has therefore been suggested that the phenomenology of CCOs is caused by magnetic field burial due to fallback accretion of material shortly after the supernova explosion. The magnetic field is then believed to resurface and evolve into a macroscopic dipolar field on timescales $\sim 10^4\,\si{yr}$, placing the object in a different region of parameter space \citep[e.g.][]{Bogdan14,Luo15,Gourgou20}. 

Since they are exclusively detected in X-rays, the proper motion of CCOs can currently only be measured directly using data from the \emph{Chandra} X-ray Observatory, which presently is the only X-ray telescope providing sufficient spatial resolution for the task. Prior to this work, this has been attempted for five CCOs, with very diverse results: 
Most prominently, such measurements have been performed multiple times for RX J0822$-$4300, located in the SNR Puppis A \citep{HuiBeck06,Winkler07,Becker12,Gotthelf13,Mayer20}. Owing to the increasing temporal baseline, their precision has improved strongly over the years, the most recent value by \citet{Mayer20} being $ (80.4 \pm 7.7) \masy$, at $68\%$ confidence. This corresponds to a projected velocity of $(763 \pm 73)\kms$ for a distance of $2\,\si{kpc}$, which potentially places this CCO at the fast end of the NS velocity distribution.\footnote{The main source of uncertainty in the projected velocity is the distance to Puppis A. While, in the past, values around $2\,\si{kpc}$ were assumed, it has been most recently measured to be $1.3\,\si{kpc}$ \citep{Reynoso17}, implying a smaller velocity.}     
In contrast, for the CCO in PKS 1209$-$51/52, 
\citet{Halpern15} measured a much smaller proper motion of $(15 \pm 7) \masy$, which corresponds to a projected velocity below $180 \kms$ for a distance of $2\,\si{kpc}$. Only weak constraints could be placed on the proper motions of the CCOs in Cas A \citep{DeLaney13} and G266.2$-$1.2 \citep[Vela Jr.,][]{Mignani19}, owing to a paucity of X-ray bright field sources for astrometric frame registration. 
Finally, very recently, \citet{Borkowski20} showed that the CCO in G350.1$-$0.3 seems to exhibit only modest proper motion, contrary to expectations (see Sect. \ref{G350}).

Alternatively to a direct measurement, it can be feasible to infer the proper motion of a NS indirectly, if its origin (i.e.~the explosion site) and the age of the SNR are known sufficiently precisely \citep[e.g.][]{Lovchinsky,Fesen06}. However, if the estimate for the explosion site is solely based on the apparent present-day morphology of the SNR, this method is much more prone to systematic biases than the direct one. 
The main reason is that the apparent geometric center of a SNR does not necessarily coincide with its true explosion site. This was first shown by \citet{Dohm96}, who demonstrated numerically that SNRs expanding into interstellar medium with a density gradient can show significant offsets between their morphological center and their true explosion site, despite maintaining a relatively circular shape.

A previous study by \citet{NS_SNR_Connection} explored the relationship between the proper motions of 18 young neutron stars (four of which are CCOs), and the morphology of their host SNRs. Several of their proper motion measurements were obtained directly from X-ray imaging data, with others (including those for all four CCOs) inferred indirectly. 
They did not find any correlation between the magnitude of SNR asymmetry 
and the projected velocity of the neutron stars. However, they found that an anticorrelation exists between the direction of proper motion and the direction of brightest ejecta emission, a proxy to highest ejecta mass. A similar study based on six systems by \citet{Katsuda18} confirmed this observation. In addition, the latter work did find a positive correlation between the degree of SNR asymmetry and the NS kick velocity. These findings support a hydrodynamic kick mechanism, where the NS reaches its extreme velocity due to asymmetric ejection of matter during the explosion, rather than via a neutrino-induced channel.   

The most immediate handle on the age of a SNR is the direct measurement of its expansion.  
This approach yields a robust upper limit on the age, since any physically justifiable assumption about the expansion history of the SNR would result in an age estimate lower than that for free expansion.\footnote{A notable exception is the Crab nebula, whose expansion has been accelerated in the past, likely due to energy input from the central pulsar \citep{Trimble68,Nugent98}.}
Tracing the expansion of a SNR is in principle possible at all wavelengths which allow to resolve the relevant emission features at sufficiently high resolution, such as the radio, optical and X-ray regimes. 
Typically, the optical regime yields the most precise age constraints, since astrometric uncertainties are very small. Moreover, the optical emission usually originates from high density ejecta clumps, which are subject to little deceleration by the circumstellar medium. A popular example is Cas A, whose explosion date was quite precisely determined by \citet{Thorstensen01} and \citet{Fesen06} to around the year 1670. 

However, many young SNRs are not visible at optical wavelengths, often due to Galactic absorption. In this case, X-ray expansion measurements are a powerful tool to constrain age and internal kinematics, despite the lower spatial resolution and higher deceleration of the observed features. Examples for the viability of this method include the confirmation of the nature of G1.9+0.3 as the youngest Galactic SNR \citep{Borkowski14}, the analysis of the propagation of the north-western rim of the ``Vela Jr'' SNR \citep[G266.1$-$1.2,][]{Allen15}, or multiple recent efforts to trace the shock wave propagation of the TeV SNR RX J1713.7$-$3946 \citep{Acero2017,Tsuji2016,G347SWExp}. Measurements of precisely dated very young SNRs (such as Cas A, Tycho's, or Kepler's SNR) show that expansion as traced in X-rays typically appears somewhat decelerated compared to the expectation for free expansion \citep{CasXray3,Hughes00, Vink08}. This highlights that SNR expansion measurements generally provide only an upper limit on their true age.

CCOs together with their parent SNRs present an ideal target for combining kinematic information from both the neutron star and the supernova shock wave. Since such systems are exclusively young, one can expect to be able to measure large expansion velocities, corresponding to a significant fraction of their undecelerated value. In addition, a precisely measured proper motion of the CCO provides accurate constraints on the supernova explosion site, the exact origin of the SNR shock wave. Combining these constraints promises unbiased results on the SNR's kinematics and age, as both the starting point and the present-day location of the shock wave are known.  

The goal of this work is to establish a complete and internally consistent sample of proper motion measurements for CCOs and of expansion measurements for their host SNRs. 
We aim to provide direct constraints on the motion of all currently known CCOs (with appropriate data available) for which this task has not been previously performed. The ultimate target of this effort is to search for indications of violent supernova kicks, manifesting themselves in large velocities, and to constrain the location of the true center of targeted SNRs, with respect to their present-day morphology.
For the three SNRs for which an expansion measurement has not yet been carried out at the time of writing, we perform an independent expansion study using our combined data set.  

Our paper is organized as follows: We outline our data selection and reduction in Sect. \ref{Data} and describe our analysis strategy in Sect. \ref{Method}. We then give a detailed overview over the results for each individual CCO/SNR in Sect. \ref{CCO}. Finally, we discuss physical consequences of the compiled sample of proper motion and expansion measurements in Sect. \ref{Discussion}, and summarize our results in Sect. \ref{Summary}.
Unless stated otherwise, all errors given in this paper are computed at $68\%$, and all upper limits at $90\%$ confidence.

\section{Data selection and reduction\label{Data}}
Our initial list of potential targets for analysis consisted of the SNRs listed in Table \ref{CCOTable}. From this list, we excluded Puppis A, PKS 1209$-$51/52 (due to already precisely measured proper motion of the CCO) and Vela Jr.~\citep[due to the established lack of astrometric calibrators,][]{Mignani19}.  
Even though the proper motion of the CCO in Cas A has already been constrained directly \citep{DeLaney13}, we decided to include it in our target list, in order to apply our own methodology to the data. 

We searched the \emph{Chandra} Data Archive\footnote{\url{https://cda.harvard.edu/chaser/}} for existing imaging observations covering the remaining potential targets. We took into account data sets from both \emph{Chandra} imaging detectors, the High Resolution Camera \citep[HRC,][]{HRCRef} and the Advanced CCD Imaging Spectrometer \citep[ACIS,][]{ACISRef}. Whenever possible, we preferred all our observations to be taken with the same detector.
The archival data were required to span a minimum temporal baseline of five years in order to qualify for a proper motion measurement.
Suitable sets of observations which fulfill these criteria were found for six targets, the SNRs G15.9+0.2, Kes 79 (G33.6+0.1), Cas A, G330.2+1.0, RX J1713.7$-$3946 (G347.3$-$0.5) and G350.1$-$0.3. For G353.6$-$0.7, we found only a single suitable observation in imaging mode (taken in 2008), not permitting further analysis for our purpose. 
Out of the six target SNRs, the expansion of three (Cas A, G330.2+1.0, RX J1713.7$-$3946) has been measured directly in past studies  \citep{Fesen06,Borkowski18,Acero2017,Tsuji2016,G347SWExp}.\footnote{The expansion study of G350.1$-$0.3 by \citet{Borkowski20} was published during the preparation of this work.} This leaves us with the SNRs G15.9+0.2, Kes 79 and G350.1$-$0.3 as targets for an expansion measurement.  

A journal of all observations used in our analysis is given in the appendix in Table \ref{OBsTable}. 
For some of our targets, such as G330.2+1.0 and G350.1$-$0.3, there are multiple observations with very deep coverage, promising high-precision results. In contrast, for RX J1713.7$-$3946, we found only two shallow observations which contain the CCO, taken with different detectors. This makes a performing a meaningful measurement more challenging and prone to systematic effects. 
In all cases treated in our sample, the target is only covered at two epochs that are spaced far enough apart in time to perform a proper motion measurement, despite most having multiple closely spaced observations at one of the epochs.    

We reprocessed all archival data using the \emph{Chandra} Interactive Analysis of Observations \citep[CIAO,][]{CIAORef} task {\tt chandra\_repro} with standard settings, to create level 2 event lists calibrated according to current standards. For ACIS data, we enabled the option to use the Energy-Dependent Subpixel Event Repositioning ({\tt EDSER}) algorithm \citep{EDSER}, to be able to exploit on-axis data at optimal spatial resolution. 
We used CIAO version 4.12 and the calibration database (CALDB) version 4.9.0 throughout the work described in this paper. 
After the reprocessing step, we screened all data for soft proton flares, excluding the few affected time intervals with background levels $3\sigma$ above the respective quiescent average, yielding cleaned and calibrated event lists on which we based our subsequent analysis.   

\section{Methods \label{Method}}
Our main analysis strategy consisted of several steps: First, we searched for serendipitous field sources for astrometric calibration (Sect. \ref{CalSources}), which we then used to align the coordinate frames of the observations and simultaneously fit for the CCO's proper motion (Sect. \ref{PMMethod}). For the three target SNRs of our expansion study, we exploited the astrometrically aligned data set to quantify the radial motion of key features of the SNR (Sect. \ref{ExpMethod}). 

\subsection{Astrometric calibration sources \label{CalSources}}
\begin{figure*}[h!]
\centering
\includegraphics[width=18.4cm]{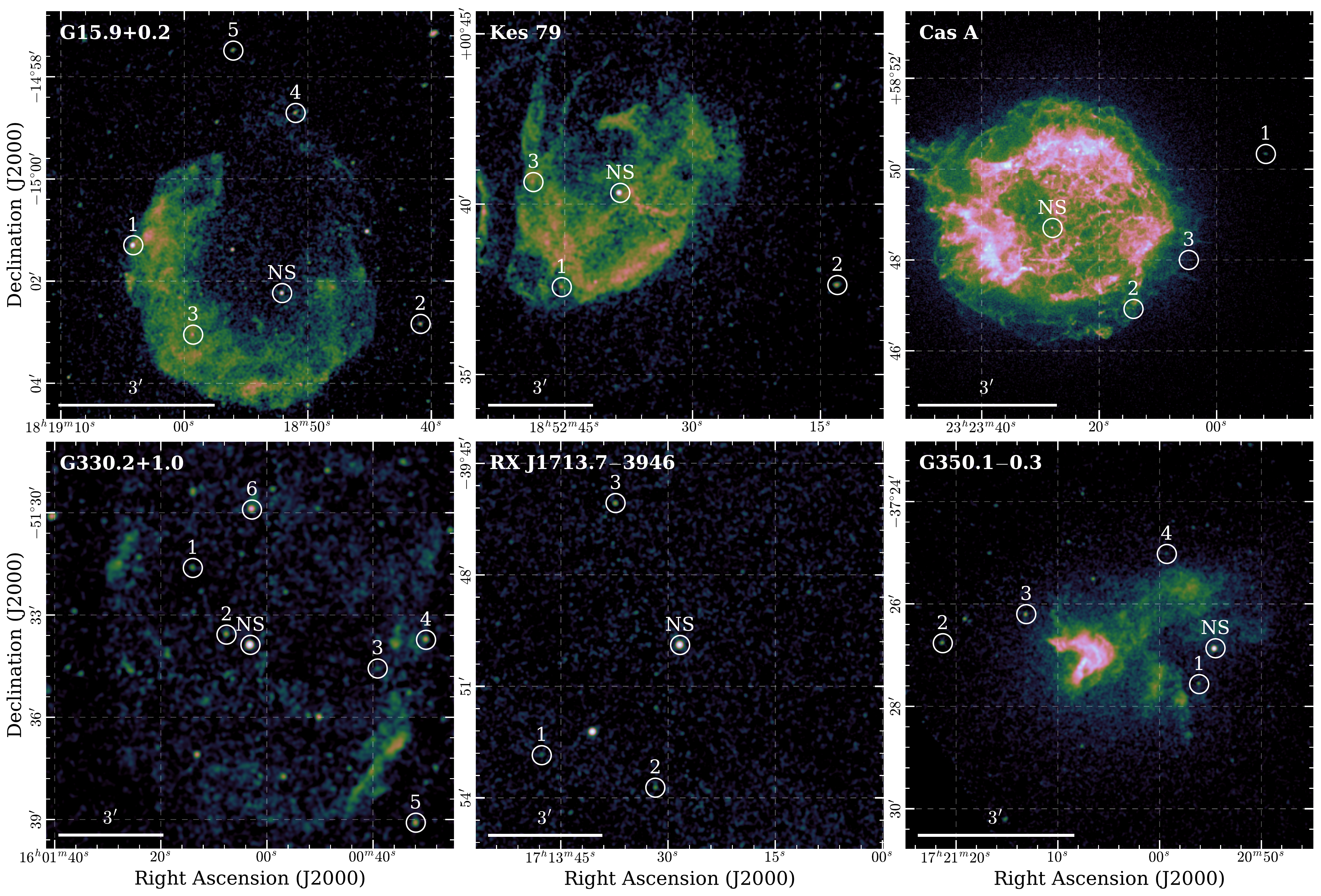}
\caption{Overview over all target SNRs, with positions of the neutron star (NS) and the calibrator sources (labelled as in Table \ref{CalibList}) indicated on exposure-corrected images of the respective SNR. We show the deepest available observation for all targets except for RX J1713.7$-$3946, where we used the ACIS observation (ID 5559) instead of the HRC observation, due to its lower intrinsic noise level.}
\label{StarOverview}
\end{figure*}

While the underlying principle of a proper motion analysis is rather straightforward, practically performing such a measurement in X-rays poses some challenges:
Owing to uncertainties in aspect reconstruction, absolute source positions in \emph{Chandra} images are only accurate to around $0.8\arcsec$ at $90\%$ confidence.\footnote{\url{https://cxc.harvard.edu/cal/ASPECT/celmon/}} 
In order to maximize the precision of the CCO's position and proper motion, it is thus necessary to align the coordinate system of our observations to an absolute reference, the International Celestial Reference System (ICRS). 
For this purpose, it is optimal to use serendipitous sources in the field of view with precisely known astrometric positions. Since most of our systems had not been studied in the context of a proper motion analysis previously, we searched for the presence of such calibrator sources systematically:

From the cleaned event lists, we extracted images over the entire detector area, which we binned at $1\times1$ and $2\times2$ detector pixels for ACIS and HRC, respectively. For ACIS, we included only events in the standard broad band, $0.5-7.0\, \si{keV}$. We then employed the CIAO task {\tt wavdetect} to perform wavelet-based source detection and rough astrometric localization at a detection threshold of $10^{-5}$. We chose this rather ``generous'' threshold since we required the presence of a source in all observations in order for it to qualify as a calibrator. This eliminates the need to strictly reject potentially spurious sources in the individual detection runs. 

For each of the observations, we matched the output source detection lists to the \emph{Gaia} DR2 catalog\footnote{\url{https://www.cosmos.esa.int/web/gaia/dr2}} \citep{GaiaMission,GaiaSummary}, which offers by far the best accuracy of all astrometric catalogs available at the time of writing. 
Those \emph{Gaia} sources with an X-ray counterpart within at most $1.5\arcsec$ in each observation made up our candidate list. We then visually inspected these candidates, excluding those sources which we deemed unreliable, for example because of large positional errors in X-rays or doubts in their correct optical identification. If many potential reference sources were found across the detector, we selected a subset of reliably identified and bright sources, spread evenly across the detector.

In further analysis steps, we occasionally decided to exclude objects from our list of frame registration sources. Reasons for this include the presence of a nearby overlapping source, the disagreement of positional offsets determined from different calibrator sources, or simply the failure to provide useful constraints compared to brighter sources.
We display the calibrator source positions as well as the location of the respective target CCOs within their SNRs in Fig.~\ref{StarOverview}. In addition, the list of astrometric reference sources used for our analysis is given in the appendix in Table \ref{CalibList}, where we also indicate which of the sources we decided to exclude in subsequent steps.

\subsection{Proper motion analysis of CCOs \label{PMMethod}}
Our method for measuring the proper motion of CCOs follows the approach used in \citet{Mayer20} and \citet{Becker12}.
The main difference to these previous works is the performance of a simultaneous fit for the object's proper motion and the astrometric calibration of individual observations, which provides a natural way to deal with observations of varying depth (Sect. \ref{TrafoPMFit}).  

\subsubsection{PSF modelling and fitting \label{PSFFit}} 
The first step is the precise measurement of the positions of all sources, that is of the CCO and calibrator objects, at each epoch.
We closely followed the procedure outlined and discussed in depth in \citet{Mayer20} to obtain models of the \emph{Chandra} point spread function (PSF) via ray-tracing simulations, using the ChaRT (\emph{Chandra} Ray Tracer) online tool\footnote{\url{http://cxc.harvard.edu/chart/index.html}} and the MARX software package\footnote{\url{http://space.mit.edu/CXC/MARX/}} \citep[][version 5.4.0]{Marx}. 
As in \citet{Mayer20}, we fitted the respective PSF models to source image cutouts, using the spectral and image fitting program Sherpa  \citep{Sherpa}.\footnote{\url{http://cxc.harvard.edu/sherpa/}} 
For each source in each observation, we obtained a profile of the logarithmic likelihood $\mathcal{L}_{ij}(x,y)$ of the source position on a finely spaced grid (typical spacing $\lesssim 20\,\si{mas}$) in the region around the best-fit. 
In our notation, the variables $x$ and $y$ describe relative positions as measured by \emph{Chandra} in a tangent plane coordinate system around the approximate location of the CCO, while we use the indices $i$ and $j$ to label the individual observations and objects (including $j=0$ for the CCO).

\subsubsection{Coordinate transformation and proper motion fit \label{TrafoPMFit}} 
In the next step, we corrected the astrometry of each observation for minor misalignment with the ICRS by applying a linear transformation to the \emph{Chandra} coordinate system. This transformation allows for a positional shift as well as a small stretch and rotation of the coordinate system. 
Prior to the analysis, two groups of ``interesting'' parameters are unknown:\footnote{Of course, other parameters such as source count rates and background levels are also unknown. However, for our purpose, the only globally interesting parameters are related to source positions.}
\begin{itemize}
    \item The astrometric solution of the CCO with the set of free parameters $\{\mu_{\alpha},\mu_{\delta},\alpha_0,\delta_0\}$,\footnote{We follow the convention that $\mu_{\alpha} = \dot{\alpha}\cos(\delta)$, so that the proper motion is directly proportional to the object's physical velocity. Positive $\mu_{\alpha}$ corresponds to motion from west to east.} 
    describing its proper motion and its position at a reference epoch $t_0$, in right ascension ($\alpha$) and declination ($\delta$), respectively. 
    \item A set of transformation parameters $\{\Delta x_{i},\Delta y_{i},r_{i},\theta_{i}\}$ per observation $i$, describing translation in the tangent plane, as well as minor scaling and rotation corrections to the coordinate frame. 
\end{itemize}
We make the realistic assumption that the errors on \emph{Gaia} positions of calibrator stars can be neglected when compared to the errors in measured X-ray source positions. Thus, the true\footnote{In the following, we refer to a position calibrated to the ICRS as ``true'' position, while a ``measured'' position describes the position determined in the coordinate system of a particular \emph{Chandra} observation.} 
calibrator source positions are effectively known prior to the observation. 
Therefore, for each of our targets, there is a set \mbox{$\mathbf{\Lambda}=\{\mu_{\alpha},\mu_{\delta},\alpha_0,\delta_0,\Delta x_{1},\Delta y_{1},...,r_{N},\theta_{N} \}$} of $4 + 4 \times N$ parameters (with $N$ the number of observations included), which are a priori unknown. 

In order to constrain their values, we calculate the logarithmic likelihood for a given set of parameters, $\mathcal{L}_{\rm{tot}}(\mathbf{\Lambda})$, in the following manner: For a given astrometric solution $\{\mu_{\alpha},\mu_{\delta},\alpha_0,\delta_0\}$, we compute the true position of the CCO $(x'_{i0},y'_{i0})$, at the epoch of observation $i$ according to: 
\begin{equation} \label{ProperMotionEq}
 \left( \begin{array}{c}
x'_{i0}\\
y'_{i0}\\
\end{array} \right) = 
\left( \begin{array}{c}
- \mu_{\alpha}\\
\mu_{\delta}\\
\end{array} \right) \cdot (t_{i}-t_{0})
+ 
\left( \begin{array}{c}
\alpha_0\\
\delta_0\\
\end{array} \right) ,
\end{equation}
where $t_{i}$ is the time of observation $i$. The fixed parameter $t_{0}$ describes the reference epoch, for which we always chose the time of the most recent observation in our respective data set. 
The true positions $(x'_{ij},y'_{ij})$ of the calibrator sources are computed analogously, using their known \emph{Gaia} astrometric solutions. 

These true positions are then converted to measured positions $(x_{ij},y_{ij})$, to account for a coordinate system shift modelled by the parameters $\{\Delta x_{i},\Delta y_{i},r_{i},\theta_{i}\}$:
\begin{equation} \label{CoordTrafo}
\small \left( \begin{array}{c}
x_{ij}\\
y_{ij}\\
\end{array} \right) =r_{i} \left( \begin{array}{cc}
\cos \theta_{i} & -\sin \theta_{i} \\
\sin \theta_{i} &  \cos \theta_{i} \\
\end{array} 
\right)  \left( \begin{array}{c}
x'_{ij}\\
y'_{ij}\\
\end{array} \right) +
 \left( \begin{array}{c}
\Delta x_{i}\\
\Delta y_{i}\\
\end{array} \right) .
\end{equation}
The likelihood of a given measured position is evaluated by interpolating the logarithmic likelihood grid $\mathcal{L}_{ij}(x,y)$, which we compute as described in the previous section.
By summing over the contributions of all $N$ observations and $S$ sources, the total likelihood for a set of parameters $\mathbf{\Lambda}$ can then be straightforwardly calculated:
\begin{equation}\label{FinalLike}
    \mathcal{L}_{\rm{tot}}(\mathbf{\Lambda}) = 
    \sum_{i=1}^{N} \sum_{j=0}^{S} \mathcal{L}_{ij} \left ( x\!=\!x_{ij}(\mathbf{\Lambda}),\, y\!=\!y_{ij}(\mathbf{\Lambda}) \right ). 
\end{equation}
In order to extract the most likely proper motion solution, it is necessary to derive the posterior probability distribution for $\mathbf{\Lambda}$ which is implied by our likelihood function in its $(4 + 4 \times N)$--dimensional parameter space.
To achieve this in practice,
we used the UltraNest\footnote{\url{https://johannesbuchner.github.io/UltraNest/}} software \citep{Buchner21}, which employs the nested sampling Monte Carlo algorithm MLFriends \citep{Buchner16,Buchner19}, to robustly tackle this multi-dimensional problem. We used uninformative (flat, truncated) priors on all variables, since we wanted our study to be as independent as possible from external assumptions. 

This method is equivalent to performing the PSF fits to all sources at all epochs simultaneously, under the constraints of NS motion at constant velocity and known positions of all calibrators. The advantage compared to a truly simultaneous implementation is that our method allows for very efficient evaluation of the likelihood function via interpolation on a precomputed grid. 
With our approach, sources with large measured positional uncertainties only add little weight to the fit, and thus do not ``wash out'' the final result. Furthermore, covariances between parameters as well as non-Gaussian errors are properly taken into account.

\begin{figure}
\centering
\includegraphics[width=1.0\columnwidth]{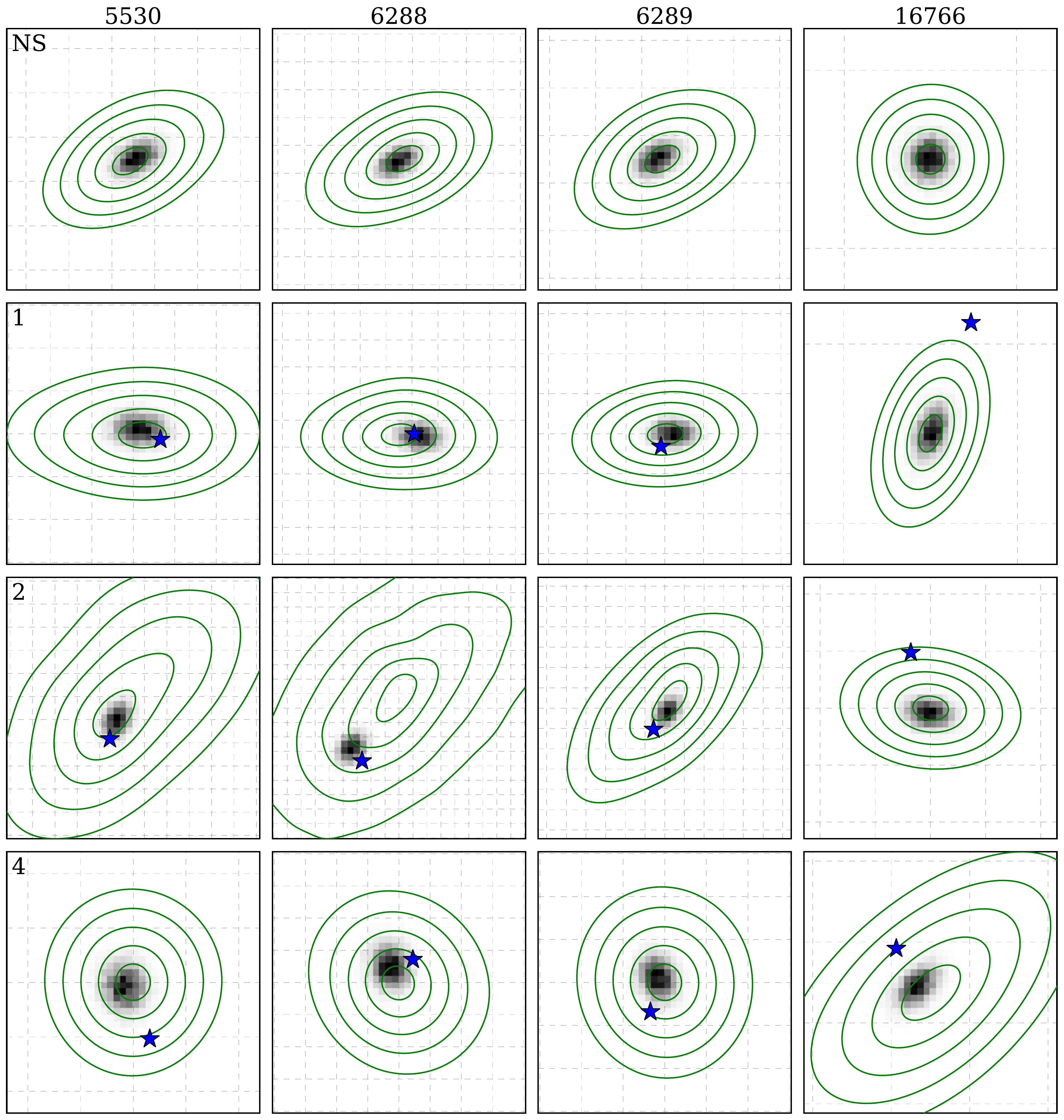}
\caption{Illustration of PSF-fit likelihood for the source position and the prediction by our simultaneous proper motion fit for the CCO 
of G15.9+0.2. 
Each column shows data from the individual epochs (with the observation ID displayed in the title), while each row represents one particular object. 
The one to five sigma contours of the PSF-fit likelihood $\mathcal{L}_{ij}(x,y)$ are shown in green, and the predicted positions obtained from our simultaneous fit (see posterior distribution in Fig. \ref{G15Corner}) are indicated using a grayscale 2D histogram.
The grid in each plot is spaced by $0.3\arcsec$ in right ascension and declination.
We display the true source position for our calibrator stars (known from \emph{Gaia}) with a blue star. 
The uncertainty contours for observation 16766 are by far the smallest, since the corresponding exposure was much deeper than at the other epochs (at $\sim90\,\si{ks}$).
}
\label{ExamplePostPred}
\end{figure}

However, the main caveat of this approach is that it neglects the possible presence of any systematic offsets beyond the linear coordinate transformation allowed by our fit. Such could be caused by subtle differences between different detectors and roll angles or by the possible misidentification of a frame registration source with the wrong astrometric counterpart. 
To identify such potential systematic effects, we visualized our model's predictions in the following way: For each point in the posterior sample of our fit, 
we computed the expected location of each source in the \emph{Chandra} coordinate system $(x_{ij},y_{ij})$. The distribution of these locations was then compared to the measured uncertainty contours, or equivalently the contours of $\mathcal{L}_{ij}(x,y)$. 
This allowed us to check whether the results given by our simultaneous fit are sensible for all objects, and to recognize potentially misidentified astrometric calibrators. 
An example of such a comparison, illustrating the strongly varying scales of our astrometric uncertainties, is shown in Fig. \ref{ExamplePostPred}. 

\subsubsection{Correction for effects of Solar motion and Galactic rotation \label{GalRot}} 
Ultimately, we are attempting to extract physically meaningful quantities, such as neutron star velocity, with reference to the ``rest frame'' of the SNR. Therefore, it is necessary to take into account the expected motion of the NS from Galactic rotation alone \citep[see e.g.][]{hobbs05, Halpern15}. 
If this effect were not considered, our estimate of the physical CCO velocity would be contaminated by the motion of the sun and the rotation of the Galactic disk.

To determine the local standard of rest (LSR) of a system, we assumed the linear Galactic rotation curve and solar velocity given in \citep{GalRot}. 
For each system, the expected velocity for co-rotation with the disk was expressed as two components of proper motion in equatorial coordinates, $(\mu^{\rm{LSR}}_{\alpha},\mu^{\rm{LSR}}_{\delta})$.
From its distance $d$, proper motion, and LSR, it is then possible to determine the peculiar (projected) velocity of an object, $v^{*}_{\rm{proj}}$ as
\begin{equation}
    v^{*}_{\rm{proj}} = \mu^{*}_{\rm{tot}} \times d = \sqrt{\left ( \mu_{\alpha}-\mu^{\rm{LSR}}_{\alpha} \right )^2+ \left (\mu_{\delta}-\mu^{\rm{LSR}}_{\delta}\right )^2\,} \times d. 
\end{equation}

One complication here is that, for some of our targets, the SNR distance $d$ is not constrained precisely.
For these cases, we assumed a flat probability distribution over the interval between its lower and upper limit (see Table \ref{CCOTable} for available constraints). The resulting distribution of $(\mu^{\rm{LSR}}_{\alpha},\mu^{\rm{LSR}}_{\delta})$ was then appropriately convolved with our distribution of the proper motion of the CCO. 
In practice, we found that the uncertainty on the LSR is significantly smaller than the uncertainty on the proper motion of all our targets. 

A further technical issue is that, due to our choice of flat priors on $(\mu_{\alpha},\mu_{\delta})$ in Sect.~\ref{TrafoPMFit}, the resulting probability distribution for $\mu^{*}_{\rm tot}$ is biased toward large values. This is because, in the transformation to $\mu^{*}_{\rm tot}$, one essentially integrates over concentric circles of increasing radius, introducing an effective prior $\propto \mu^{*}_{\rm tot}$, and always leading to a peak of the distribution at non-zero $\mu^{*}_{\rm tot}$.
Therefore, in order to not overestimate the projected velocity of our CCOs, we reweighted the posterior distributions by a factor $\propto 1/\mu^{*}_{\rm tot}$. 
The effect of this correction decreases with increasing significance of non-zero proper motion.

\subsection{Expansion measurement of SNRs \label{ExpMethod}}
When studying the dynamics of a young core-collapse SNR, a complementary tool to the analysis of the neutron star's proper motion is the measurement of the SNR's expansion. It provides the most immediate constraint on the shock velocity in its shell, and, by backward extrapolation, can be used to provide a very reliable constraint on its maximum age. Therefore, we took advantage of our results from the previous section, and measured the expansion of three CCO-hosting SNRs (G15.9+0.2, Kes 79, G350.1$-$0.3) in the following way:    

First, we used the optimal transformation parameters $\{\Delta x_{i},\Delta y_{i},r_{i},\theta_{i}\}$ for each observation $i$ from Sect. \ref{TrafoPMFit}, to perform the astrometric calibration of our data set.  
Since we fitted for the motion of the CCO and the frame calibration simultaneously, this has the advantage that, for observations taken close to each other in time, the position of the CCO (which is often the brightest source) is forced to be identical. This practically provides an additional reliable source for coordinate frame calibration. 
We aligned the data from all observations to a common coordinate system, using the CIAO task {\tt wcs\_update} with a {\tt transformfile} customized to contain the optimal transformation parameters derived in Sect.~\ref{TrafoPMFit}. All data from observations taken at the same epoch were then merged and reprojected to a common tangent plane using {\tt reproject\_obs}.
The systematic error in the coordinate frame alignment can be conservatively estimated from the combined statistical uncertainties of $\{\Delta x_{i},\Delta y_{i}\}$, yielding typical values between $0.1^{\prime\prime}$ and $0.3^{\prime\prime}$. 

In order to define the features of interest for our measurement, we extracted exposure-corrected images of each SNR in the standard ACIS broad band ($0.5-7.0 \,\si{keV}$). 
We then used SAOImage ds9 \citep{ds9} to interactively define rectangular regions containing emission features which we found suitable to trace the expansion of the SNR, for example sharp filaments or individual clumps. 
Since we are mostly interested in measuring radial motion, the orientation of these boxes was adapted to what we deemed the probable direction of motion given the shape of the feature and its location within the SNR.\footnote{Note that even for potential misalignment with the true direction of motion as large as $20^{\circ}$, the measured expansion speed would only deviate by $1-\cos(20^{\circ}) = 6\%$ from the true value.} 
In order to avoid any possible confirmation bias toward the detection of expansion, we refrained from excluding regions in retrospect, even if their motion turned out to be almost unconstrained by the available data.

\begin{figure}
\centering
\includegraphics[width=0.8\linewidth]{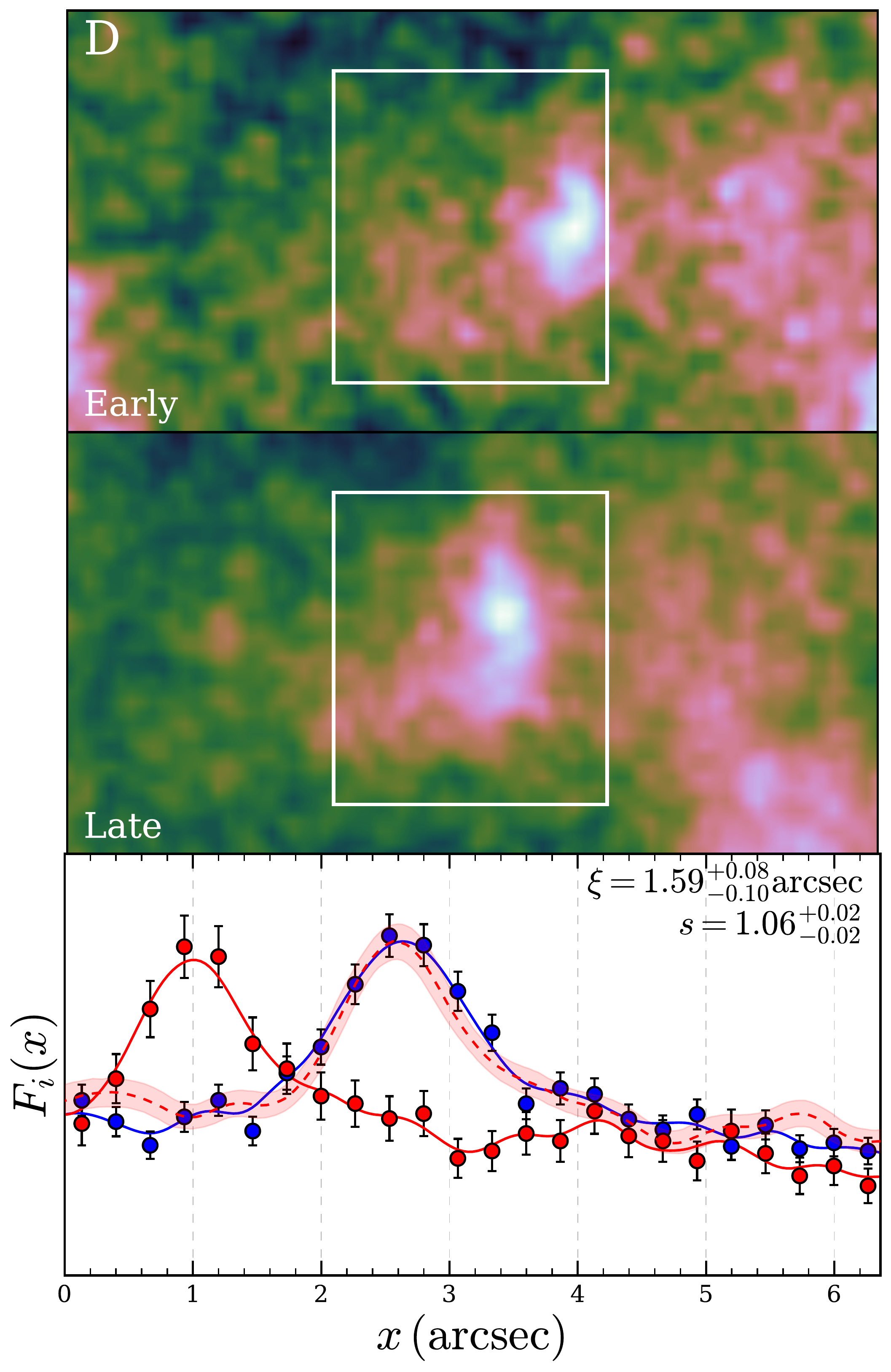}
\caption{Example expansion measurement in region D of G350.1$-$0.3 (see Sect.~\ref{G350}).
The top and middle panels depict the exposure corrected images at the early and late epoch, and the white rectangle marks the measurement region. 
The bottom panel displays the one-dimensional flux profiles $F_{i}(x)$ at the early (red) and late (blue) epoch. The markers display the binned distributions with errors, and the solid lines reflect the flux profiles smoothed by a Gaussian kernel. 
The dashed red line corresponds to the profile of the early epoch, shifted and rescaled by the indicated parameters (see upper right corner), with the shaded area displaying the associated uncertainty from bootstrapping. 
Note that we define the $x$ coordinate such that it increases in radial direction, meaning from west to east in this case.  
}
\label{ExpExample}
\end{figure}

At this point, it would be possible to determine the motion of an individual feature and its errors, for instance by fitting a model for the flux profile to the data of both epochs \citep[e.g.][]{Xi19}. 
However, since some clumps and filaments targeted in this work show a rather complex profile, it seems infeasible to empirically model their shape, because the required number of free parameters would be  high, or one would need to hand-tune the model for each feature.    
Alternatively, one could directly compare the observed emission profiles via a custom likelihood function (often some variant of $\chi^2$) quantifying their deviation, in one \citep[e.g.][]{G347SWExp, Tsuji2016},
or two dimensions \citep[e.g.][]{Borkowski20}.
However, after some initial testing, we found that such a likelihood function behaves quite erratically in many cases, meaning that small sub-pixel variations in the shift of the profiles lead to large ``jumps'' in the likelihood. This makes a direct determination of errors prone to systematics and irreproducible for varying bin sizes. 

Thus, in order to avoid biased results or underestimated errors, we used a bootstrap resampling technique \citep{bootstrap} to estimate uncertainties for the motion of all our features: 
We extracted the set of X-ray event coordinates $\mathbf{x}_{i}$ (with $i=1,2$ labeling the early/late observational epoch) along the suspected direction of expansion, as well as the corresponding exposure profile, within the rectangular region of each feature.  
For both epochs, we created $N=1000$ randomized realizations of the observed data set $\{\mathbf{x}^{*}_{1,i},...,\mathbf{x}^{*}_{N,i}\}$, via sampling with replacement from the list of observed events in the region. 
Thereby, the statistical uncertainty of the emission profile is approximated by the intrinsic scatter of our resampled data sets. 

From each resampled data set $\mathbf{x}^{*}_{n,i}$, we extracted one-dimensional flux profiles $F_{i}(x)$ describing the emission along the (expected) direction of motion. For each bin $x$, we estimated a simple Gaussian flux error as $\sigma_{i}(x) = F_{i}(x)/\sqrt{C_{i}(x)}$, where $C_{i}(x)$ describes the number of events in the bin. 
In order to estimate the optimal shift between the epochs, we defined the function $f$ which quantifies the deviation between the emission profiles at early and late epochs, in dependence of a positional shift $\xi$ and an amplitude scale factor $s$:
\begin{eqnarray}\label{ExpLike}
f(\xi, s) &=& \sum_{x} \frac{\left(s F^{}_1(x \!-\! \xi) \,-\, F_2(x)\right)^2}{ s^2\sigma_1^2(x \!-\! \xi) \,+\, \sigma_2^2(x) }, 
\end{eqnarray}
where $s$ was introduced to account for possible flux changes of the feature. 

By minimizing $f$ for each of the $N$ realizations, we obtained a sample approximately distributed according to the likelihood of $\xi$.
From this distribution, we extracted the angular speed $\mu_{\rm exp}$ of the emission feature given by $\mu_{\rm exp} = \xi / \Delta t$, with $\Delta t$ being the difference between the average event times of the early and late epochs. This is related to the projected expansion velocity via $v_{\rm exp} = \mu_{\rm exp} d $, where $d$ corresponds to the (assumed) distance to the SNR.

The last step of estimating the global minimum of $f$ is however complicated by the noisy nature of our function on small spatial scales. The reason for this is that the binning of events required to obtain an image is only an approximation to the underlying continuous flux distribution of a feature, which introduces a ``granularity'' in $f$.
We resolved this issue by applying a Gaussian kernel to smooth the 
flux histograms $F_{i}(x)$ before computing $f$. 
The width of the kernel was chosen such as to not over-smooth the prominent features in each region, but at the same time eliminate most of the purely statistical bin-to-bin fluctuations. 
Other approaches, like smoothing only one of the profiles or applying Poisson statistics to directly compare the two unsmoothed profiles, were found to typically yield similar or larger intrinsic errors.
The basic principle of our approach is visualized for an example region in G350.1$-$0.3 in Fig.~\ref{ExpExample}.

The main advantage of our method is that we only extracted the optimal value of $\xi$ for each realization. We therefore did not need to perform any ``stepping'' through parameter space to obtain the errors from a given likelihood threshold, as we would consider this likely to yield underestimated statistical errors. Furthermore, our method allowed us to implicitly account for statistical fluctuations in both measurement epochs, which would be neglected when treating the deeper observation as a definitive ``model'' without intrinsic errors.

As a simple practical test of our approach, we used two late-time data sets of G350.1$-$0.3 (ObsIDs 21118 and 21119) and applied our method, as if trying to measure expansion. Naturally, the true shift of features between these two observations is essentially nonexistent due to the small time difference of around one day.
Therefore, one expects the results of our experiment to cluster around zero, tracing not the motion of features but the intrinsic noise of our method. 
We found that the overall deviation from zero of our measurements can be satisfactorily accounted for with statistical and systematic errors. More quantitatively, the sum of normalized squared deviations from zero, $\chi^2 = \sum_{k=1}^{K} \xi_{k}^2/\Delta \xi_{k}^2$, was found to be $\chi^2 = 19.1$ for $K=18$ regions (i.e. ``degrees of freedom''), when accounting only for statistical uncertainties. Adding in quadrature the estimated systematic error of $0.1^{\prime\prime}$, we obtained $\chi^2 = 14.3$. 
The expected $68\%$ central interval of a $\chi^2$ distribution with 18 degrees of freedom is $(12.1,23.9)$.  Therefore, our test demonstrates that we obtain reasonable estimates for our total uncertainty, and that our overall error estimates are likely rather conservative.

\section{Results \label{CCO}}

\subsection{G15.9+0.2 \label{G15}}
G15.9+0.2 is a small-diameter Galactic SNR, first detected in the radio by \citet{Clark73}. In X-rays, its morphology resembles that of an incomplete, but very symmetric, circular shell. It shows a north-western ``blowout'' region with only weak apparent emission, likely due to reduced density of the circumstellar medium \citep{Sasaki18}. Its distance is only constrained to a relatively uncertain range of values \citep[$7-16$ kpc,][]{G15Dist}, and accordingly, the age shows correlated uncertainties of factor two \citep[$2900-5700$ years,][]{Sasaki18}. 

The bright CCO CXOU J181852.0$-$150213 is quite clearly offset by around $35''$ from the apparent geometrical center of the SNR toward west or south-west. As first noted by \citet{Reynolds2006}, this seems to imply a quite large transverse velocity, or a proper motion of around $10 \masy$ for an assumed age of 3500 years.

Unfortunately, the available data at the early epoch are split into three non-consecutive observations of $5.0$, $9.2$, and $15.1\,\si{ks}$ length, with the detector aimpoint being located around $3\arcmin$ north of the CCO, which reduces the usability of this data set for our purpose. In contrast, the late observation is very deep (with an exposure around $90 \,\si{ks}$) and aimed directly at the CCO, resulting in very small statistical errors on its late-time position.

\begin{figure}
\centering
\includegraphics[width=1.0\columnwidth]{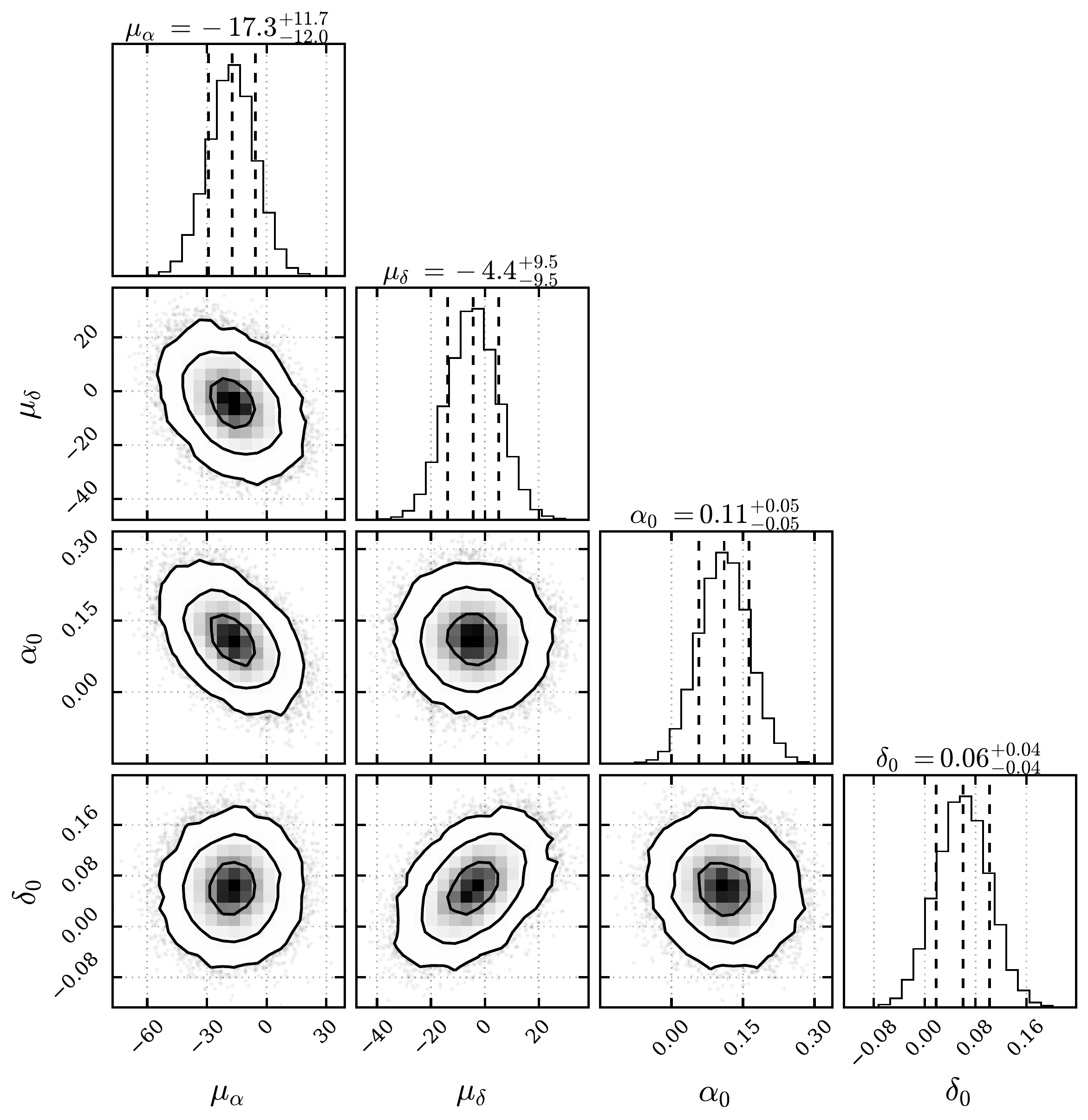}
\caption{Corner plot displaying the joint and marginal posterior distributions for the parameters of the astrometric solution for the CCO in G15.9+0.2. The astrometric calibration parameters $\{\Delta x_{i},\Delta y_{i},r_{i},\theta_{i}\}$ have been marginalized over, and are not shown here in order to keep the figure uncluttered (see Table \ref{AllPMTable} for constraints on all parameters). In the 2D correlation plots, we indicate the $1\sigma$, $2\sigma$, and $3\sigma$ contours, meaning the smallest regions containing $39.3\%$, $86.5\%$, and $98.9\%$ of the total probability mass, with solid lines. In the marginalized 1D histograms, we indicate the median and $68\%$ central interval of the posterior probability distribution, with dashed lines. The units of proper motion and the astrometric zero-point are $\si{mas.yr^{-1}}$ and $\si{arcsec}$, respectively.
This plot and analogous figures have been created using the {\tt corner.py} package \citep{Corner}.}
\label{G15Corner}
\end{figure}

The resulting posterior distribution from our astrometric fit can be seen in Fig. \ref{G15Corner}.
The $68\%$ central credible intervals for the proper motion are given by  $(-17 \pm 12, -4 \pm 10) \masy$, in right ascension and declination, respectively.
This seems to agree well with the expected proper motion direction given the CCO's present-day location in the SNR. 

While this cannot be counted as an unambiguous detection of significant proper motion for the CCO, a proper motion of the order of $15 \masy$ toward west seems realistic given the SNR geometry. It would imply a rather large projected velocity of the NS around $700 \kms$ when scaling to a distance of $10\,\si{kpc}$. The formal $90\%$ upper limit of $25 \masy$ (corresponding to $1200 \kms$ for that distance) on the total proper motion with respect to the LSR of G15.9+0.2 is physically unconstraining.

\begin{figure}
\centering
\includegraphics[width=0.9\columnwidth]{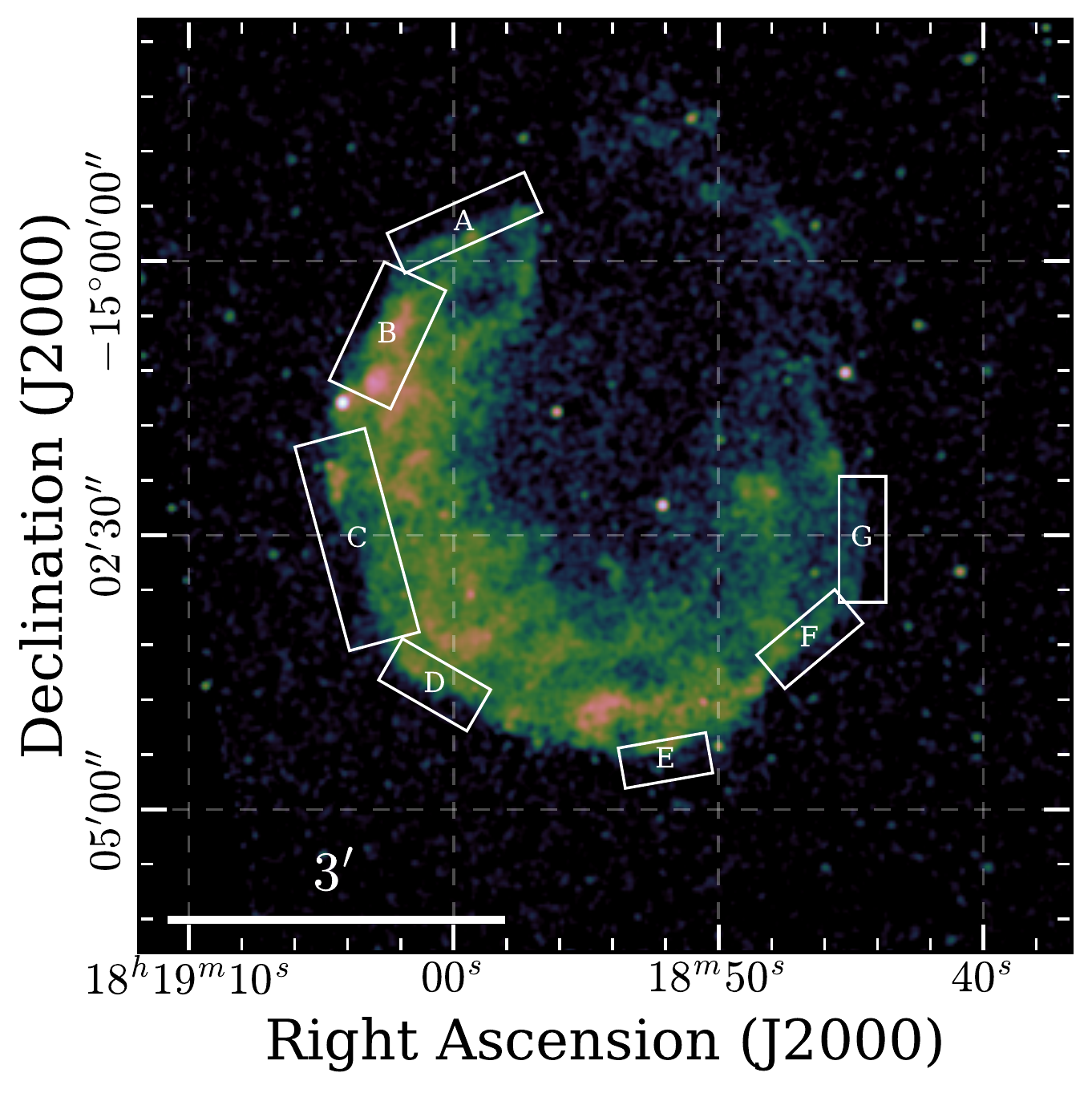}
\caption{SNR G15.9+0.2 with regions used for the expansion measurement indicated. We show a smoothed, exposure-corrected image of the SNR with logarithmic scaling of the color map. We overlay the boxes from which we extracted the one-dimensional emission profiles for our expansion measurement, and label them alphabetically for easier identification and to avoid confusion with numeric labels for the astrometric calibrator stars.}
\label{G15Image}
\end{figure}

\begin{figure}
\centering
\includegraphics[width=1.0\columnwidth]{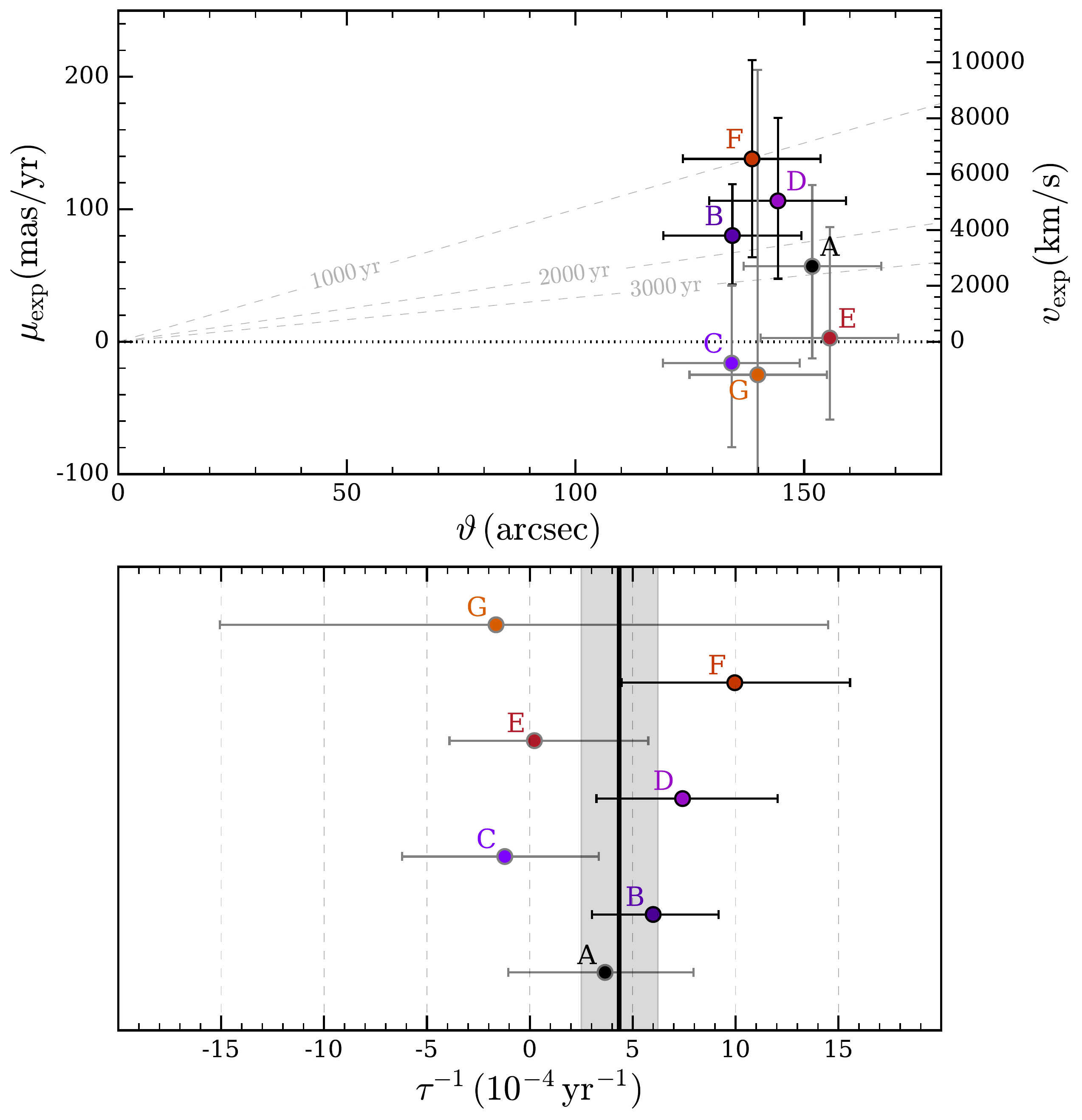}
\caption{Constraints on the expansion of G15.9+0.2. 
Top: For each feature (labels as in Fig. \ref{G15Image}), we plot the measured angular expansion speed $\mu_{\rm exp}$, and the corresponding projected velocity $v_{\rm exp}$ for a distance of $10 \,\si{kpc}$, against its angular distance to the SNR center $\vartheta$. 
To assess their respective significance, we have used black error bars for those measurements which enclose $> 95\%$ of the total probability mass on either side of $0$, and gray for the others.
In addition, we indicate the expected free expansion speed for ages of 1000, 2000, and 3000 years with dashed lines.
Bottom: Expansion rate $\tau^{-1}$ as measured in each individual region. The thick black line with the gray-shaded area represents the combination of the individual measurements (or weighted average), assuming uniform expansion. All error bars are at $68\%$ confidence.
The data underlying this figure is given in Table \ref{ExpTable}.
}
\label{G15Exp}
\end{figure}

The almost perfectly circular shape of G15.9+0.2 allowed us to define quite intuitive regions along its outer rim for an expansion measurement. They are illustrated in Fig. \ref{G15Image}. As can be seen, all regions trace the motion of the edge of the SNR shell, and due to the observed morphology, we would expect to observe quite similar expansion speeds along all directions.  
Minor problems faced in their definition were the bright star ``1'' superimposed on the eastern shell, and a detector chip gap crossing the southernmost part of the remnant, which we both avoided with our choice of regions.

In order to define a rough reference point for the expansion of the SNR, we used SAOImage ds9 \citep{ds9} to approximate the shell of the SNR with a circle. Its center at $(\alpha, \delta) = (18^{h}18^{m}54.\!\!^{s}2, -15^{\circ}01^{\prime}55^{\prime\prime})$ corresponds to the probable location of the explosion, with an assumed $1\sigma$--error of $15^{\prime\prime}$, corresponding to around $10\%$ of the SNR's radius.
This is an intuitive way to estimate the origin of the SNR's expansion since the CCO is clearly offset with respect to the apparent center, and its proper motion is not constrained very precisely at this point. 
This reference point allowed us to obtain a rough estimate of the angular distance $\vartheta$ travelled by the shock wave at the location of each region. 

The results of our expansion analysis are displayed in Fig. \ref{G15Exp} (for a full illustration of the data at both epochs, see Fig.~\ref{G15FullExp}).  Due to the comparatively low exposure time available in the early epoch, the statistical error bars on the proper motion of the SNR shell are quite large.   
We estimate that the systematic error introduced by astrometric misalignment of the observations is at most $0.15^{\prime\prime}$ (see Table \ref{AllPMTable}), corresponding to a proper motion error around $15\masy$. Therefore, the overall effect of this systematic error is rather small, also since there does not seem to be any bias toward detecting motion along a certain direction in our results. 

We attempt to give a quantitative estimate of significance and rate of expansion now: For each region, we combined the measured expansion speed $\mu_{\rm exp}$ with its distance from the SNR center $\vartheta$ to obtain an estimate of the expansion rate $\tau^{-1} = \mu_{\rm exp}/\vartheta$. In order to combine the constraints from individual boxes, we made the simplest possible assumption, which is that the SNR is expanding uniformly in all directions. The large size of the measured error bars presently does not warrant more complex (albeit more realistic) models, such as an azimuthal variation of the expansion velocity.
Our assumption allowed us to directly multiply the inferred likelihoods from all regions to estimate an error-weighted average expansion rate of G15.9+0.2. 
In this process, we marginalized over systematic uncertainties introduced by astrometric misalignment of observations and the unknown location of the SNR center. This means we recalculated $\mu_{\rm exp}$ and $\vartheta$ many times for random samples drawn from possible center locations and systematic  shifts, and summed up the combined distributions of $\tau^{-1}$ afterwards. Thereby, we removed the need to artificially convolve distributions. 

As can be seen in the lower panel of Fig. \ref{G15Exp}, our combined best estimate for the current average rate of expansion is $\tau^{-1} = (4.3\pm1.9)\,\times10^{-4}\,\si{yr^{-1}}$.  
This finding is moderately significant at around $2.5\sigma$, as the probability for the true value to be less than or equal to zero is found to be $0.6\%$. 
Given the almost perfectly circular shape of the SNR, and the measured error bar sizes, presently there appears to be no obvious deviation from uniform expansion.
Thus, while it is infeasible to constrain the kinematics of individual portions of the shell at this point, the statistical distribution of the results from our seven regions supports non-zero expansion. 
During the refinement of our analysis strategy, we attempted several techniques to measure the motion of the rim, and also attempted a simpler form of astrometric alignment of the two epochs (applying only a constant 2D shift of the coordinate system). For all these cases, we found that the finding of overall expansion still holds.

It may therefore be tempting to state that, assuming perfectly free expansion, our average expansion rate would correspond to an age $\tau \sim 2500 \,\si{yr}$. 
Of course however, given the available data and the size of statistical errors, this value constitutes only a zeroth-order estimate for the SNR's age and is therefore to be taken with caution until the expansion of G15.9+0.2 has been independently measured. 

\subsection{Kes 79 \label{Kes}}
Kes 79 (G33.6+0.1) is a bright mixed-morphology SNR, whose X-ray appearance is dominated by multiple filaments and shells, as well as the bright central X-ray source CXOU J185238.6+004020 \citep{KesMorph}. In the radio and mid-infrared regimes, its morphology appears similarly complex and filamentary \citep{Giacani09,KesAge1}. 
The distance of Kes 79 is not very well constrained. One of the most recent estimates was inferred from \ion{H}{I} absorption \citep{KesDist}, yielding a  distance of 3.5 kpc, whereas CO observations \citep{KesCODist} suggest a distance of 5.5 kpc. The age of Kes 79 has been estimated via detailed X-ray spectroscopy to lie between 4.4 and 6.7 kyr \citep{KesAge1}.

\begin{figure}
\centering
\includegraphics[width=1.0\columnwidth]{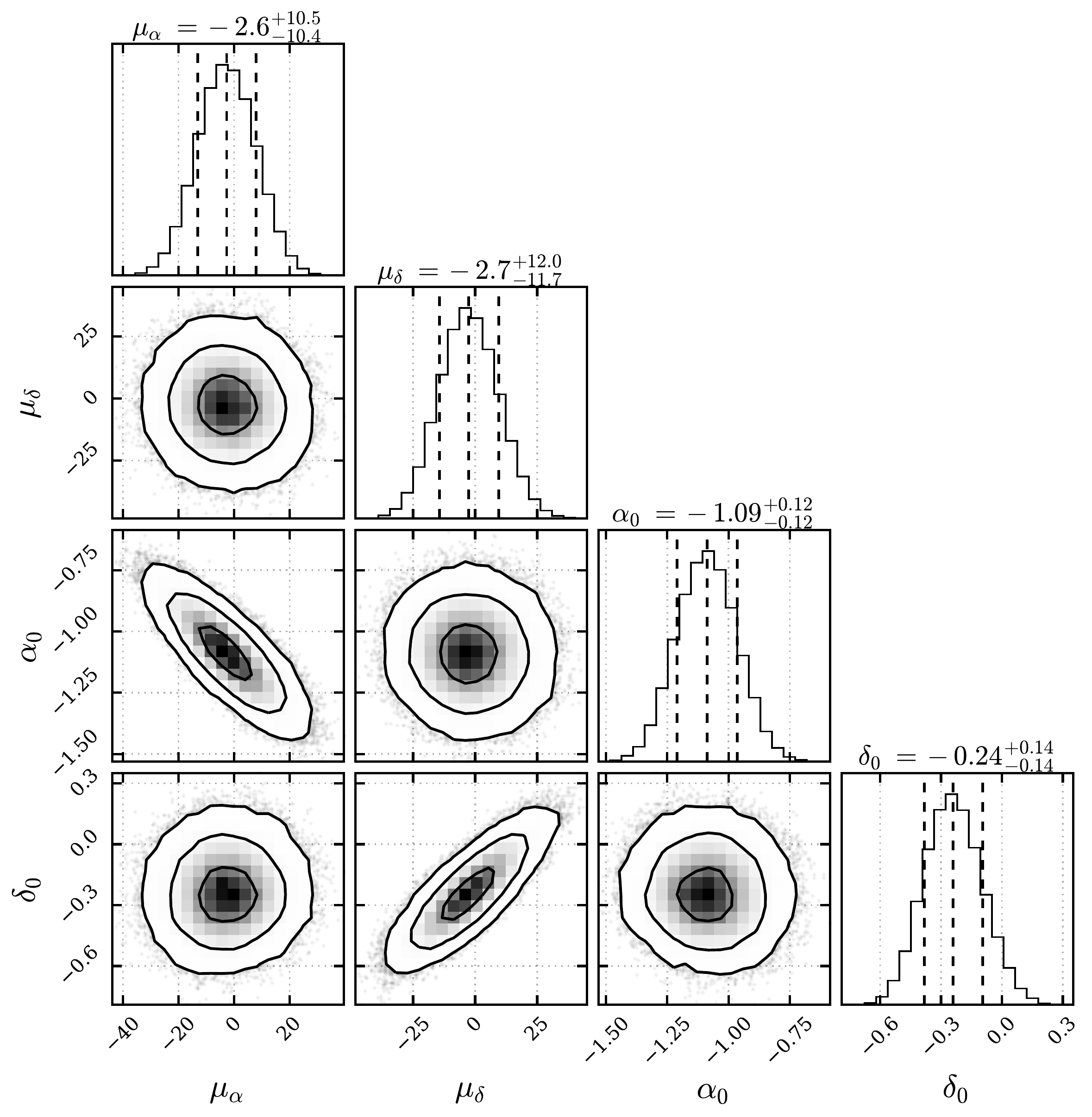}
\caption{Corner plot (as in Fig. \ref{G15Corner}) showing the posterior distribution of the astrometric solution for the CCO in Kes 79. See Table \ref{AllPMTable} for constraints on all parameters.}
\label{KesCorner}
\end{figure}

Our measurement of the CCO's proper motion (Fig. \ref{KesCorner}) is perfectly consistent with zero, the $68\%$ central credible intervals being $(-3_{-10}^{+11},-3_{-11}^{+12})\masy$. This can be converted to a $90\%$ upper limit on the CCO's peculiar proper motion of around $19 \masy$. Equivalently, its tangential velocity is constrained at $<450\kms$, when scaled to a distance of $5 \,\si{kpc}$. 

While the limit on the CCO's transverse velocity is not very constraining by itself, our measurement has a noteworthy implication on the neutron star's timing properties, particularly its period derivative: The magnitude of the Shklovskii effect, an entirely kinematic contribution to the measured period derivative \citep{Shklovskii}, can now be constrained to $< 4 \times 10^{-19}\,\si{s.s^{-1}}$ at $90\%$ confidence (for a distance of $5 \,\si{kpc}$). This corresponds to $ <5\%$ of the measured period derivative \citep{Halpern10}, which justifies to neglect this contribution to first order. 

\begin{figure}
\centering
\includegraphics[width=0.9\columnwidth]{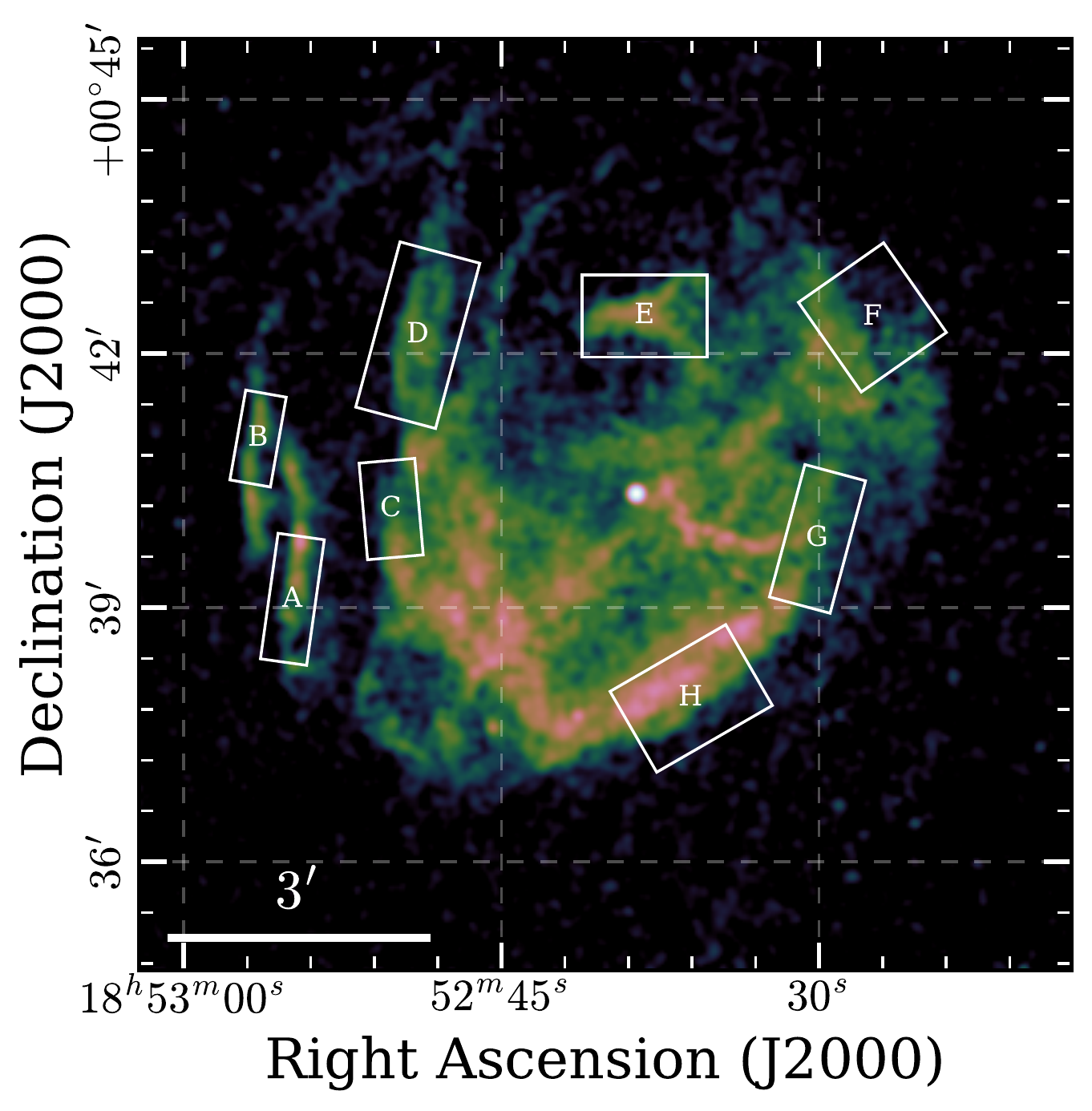}
\caption{SNR Kes 79 with regions used for the expansion measurement indicated (as in Fig.~\ref{G15Image}).}
\label{KesImage}
\end{figure}

\begin{figure}
\centering
\includegraphics[width=1.0\columnwidth]{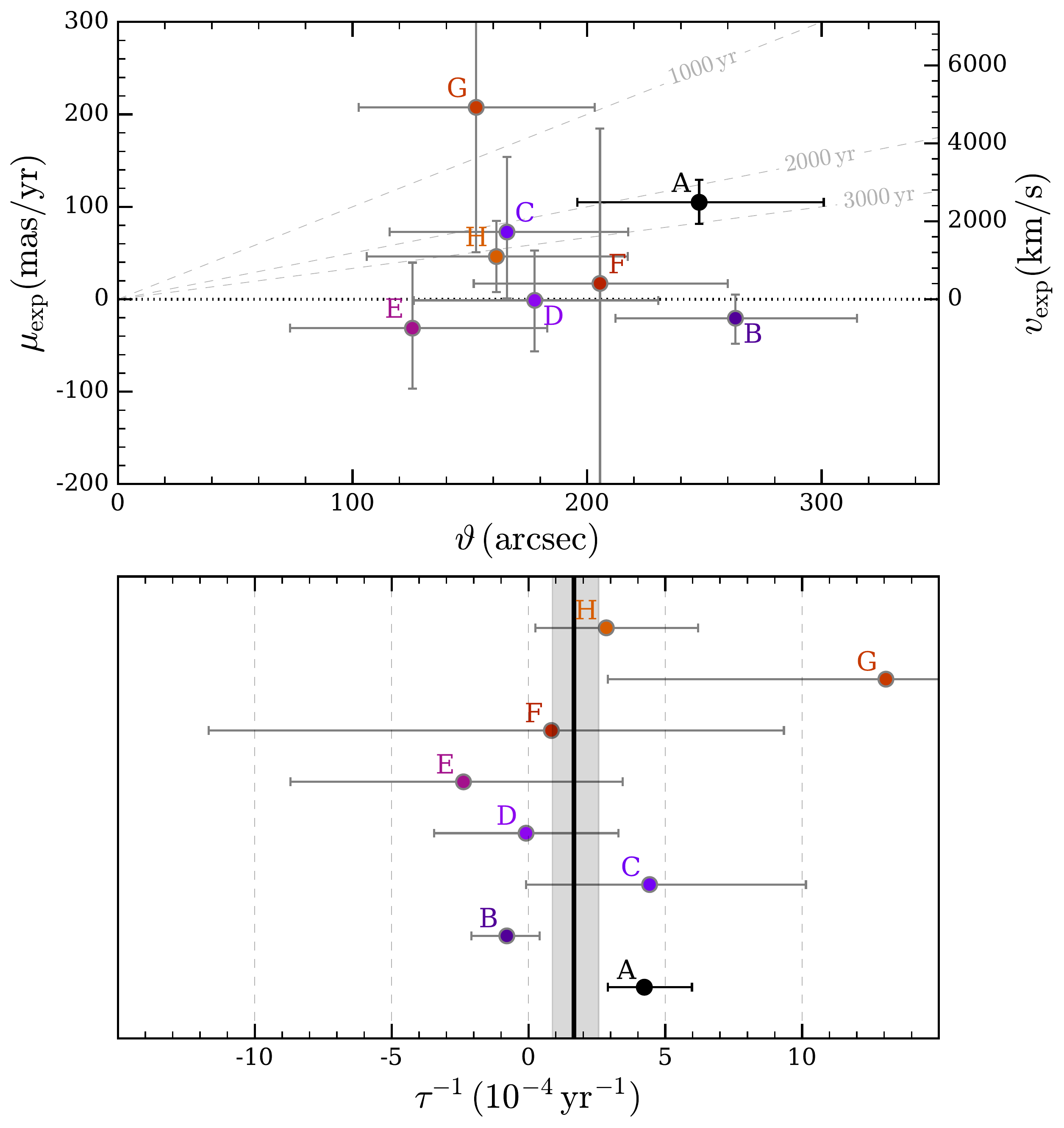}
\caption{Constraints on the expansion of Kes 79 (as in Fig. \ref{G15Exp}). The assumed distance to Kes 79 for the conversion to $v_{\rm exp}$ is $5.0 \, \si{kpc}$. The data underlying this figure is given in Table \ref{ExpTable}.}
\label{KesExp}
\end{figure}

As can be seen in Fig. \ref{KesImage}, Kes 79 exhibits several sharp filaments and shell fragments, which are in principle ideally suited for tracing the propagation of the supernova shock wave through the interstellar medium. However, our data set has several caveats: First, the exposure times of all observations are comparatively low (at $29.5$, $9.8$, and $10.0\, \si{ks}$), such that statistical fluctuations in the measured flux profiles are quite high. Second, for all three observations, the telescope pointing directions and roll angles were different. Since we attempted to avoid all the chip gaps of the ACIS-I detector, we had to exclude several emission features from our region definitions, which would otherwise increase the measurable signal quite significantly (e.g.~close to regions A, B, C, H). In order to quantify the distance between our emission features and the SNR center, we made use of our results in Sect. \ref{DiscExp}, establishing our estimated explosion site as (relatively uncertain) reference point. 

The measured proper motion $\mu_{\rm exp}$ of the shock wave in the individual regions is displayed in the upper panel of Fig. \ref{KesExp}. Again, we observe that the overall picture is visually dominated by large statistical uncertainties, with most points scattered around 0, and several regions appearing almost unconstrained. 
However, the important exception is filament A as it, on its own, shows evidence for non-zero proper motion at $\sim 4 \sigma$ statistical significance. In contrast, no regions show stronger deviations from zero than $\sim 1 \sigma$ in the negative direction, which would be expected if this result was simply due to underestimated errors. 
We conservatively estimate the systematic astrometric error from coordinate system alignment to around $0.25^{\prime\prime}$ (see Table \ref{AllPMTable}), corresponding to an error on the angular motion around $17 \masy$. 
While the proximity of region A to a chip gap seems concerning for our overall interpretation, even an inspection by eye of the shock profiles in region A (see Fig. \ref{KesFullExp}) shows an obvious shift in the exposure-corrected flux profile, strengthening our confidence in the observed result. 

Analogously to Sect. \ref{G15}, we combined the measurements of individual regions to constrain the uniform expansion rate of Kes 79 at $\tau^{-1} = 1.7^{+0.9}_{-0.8}\times10^{-4}\,\si{yr^{-1}}$. The probability for the true value to be less than or equal to zero is estimated to $1.6\%$.
Thus, the nominal statistical significance of non-zero expansion is still low (at around $2\sigma$). However, it should be noted that it is possible that certain features, traced for instance by the filament in region A, propagate more freely than others which could be more strongly decelerated. Therefore, the assumption of uniform expansion is likely unrealistic given the complex X-ray and radio morphology of Kes 79. Indirect signs of non-uniform expansion have actually been found in spatially resolved X-ray spectroscopy of Kes 79 by \citet{KesAge1}. If the expansion is indeed nonuniform, a combination of likelihoods is technically not valid, since one does not truly measure the same underlying quantity in different regions.

\subsection{Cas A \label{Cas}}
Cas A (G111.7-2.1) is the youngest known Galactic core-collapse SNR at an age of around 350 years \citep[e.g.][]{Fesen06, Alarie14}. 
Its X-ray morphology can be clearly separated into an inner, thermally emitting ejecta-dominated shell, and an outer forward shock region, which exhibits mainly synchrotron emission and expands at around $5000\kms$ \citep{CasXray1,CasXray2,CasXray3}. 
The distance to Cas A has been precisely constrained to $(3.33\pm0.10)\,\si{kpc}$ via an analysis of the expansion of filaments in the optical \citep{Alarie14}. 

The CCO CXOU J232327.9+584842 was discovered in the \emph{Chandra} first-light image of Cas A \citep{Tananbaum99}. 
There exists a past proper motion analysis for the CCO performed by \citet{DeLaney13}, which yielded a final measurement of the transverse velocity of $(390\pm400)\kms$. This is equivalent to a total proper motion of $(24 \pm 25) \masy$ toward south-west, for their assumed distance of $3.4\,\si{kpc}$. Nevertheless, since they used different statistical methods than we do, and did not employ any PSF modelling, we decided to reanalyze the data to produce a comparable measurement to the other objects in this paper. 

Our initial search for detectable calibrator sources yielded only one source common to all five observations  (labelled as ``1'' here), which is identical to the one found in \citet{DeLaney13}. However, instead of using quasi-stationary flocculi of Cas A for image registration, we decided to stack the late-time observations to increase our chance to detect faint sources. To do this, we aligned the observations using point-like sources found by {\tt wavdetect} which are highly significant in all observations ($>10\sigma$). These correspond mostly to bright clumps of emission of the SNR which can be regarded as effectively stationary on the relevant timescales ($<10$ days).  
We applied the CIAO task {\tt wcs\_match} to find the optimal transformation between two respective epochs, and then reprojected and merged all late-time observations using the {\tt wcs\_update} and {\tt reproject\_obs} tasks.

Using this combined late-time data set, we detected two more point sources common to early and late observations, and with close positional match to \emph{Gaia} stars (see Fig.~\ref{StarOverview} and Table \ref{CalibList}). Due to their location in the outer regions of Cas A (see Figure \ref{StarOverview}), we made some further considerations to ensure their correct identification: We verified that their spectral nature is softer than that of the surrounding emission, using an archival $1\,\si{Ms}$ ACIS observation (PI U.~Huang, ObsIDs 4634$-$4639). 
Furthermore, we emphasize that if these sources were actually associated to ejecta clumps of the SNR rather than being foreground sources, they would be expected to show very large proper motion due to their location at the edge of the remnant shell. Such behavior would be clearly recognizable with respect to the secure source 1. 
Since we did not observe this, we concluded that all three astrometric calibrators are probably correctly identified.

\begin{figure}
\centering
\includegraphics[width=1.0\columnwidth]{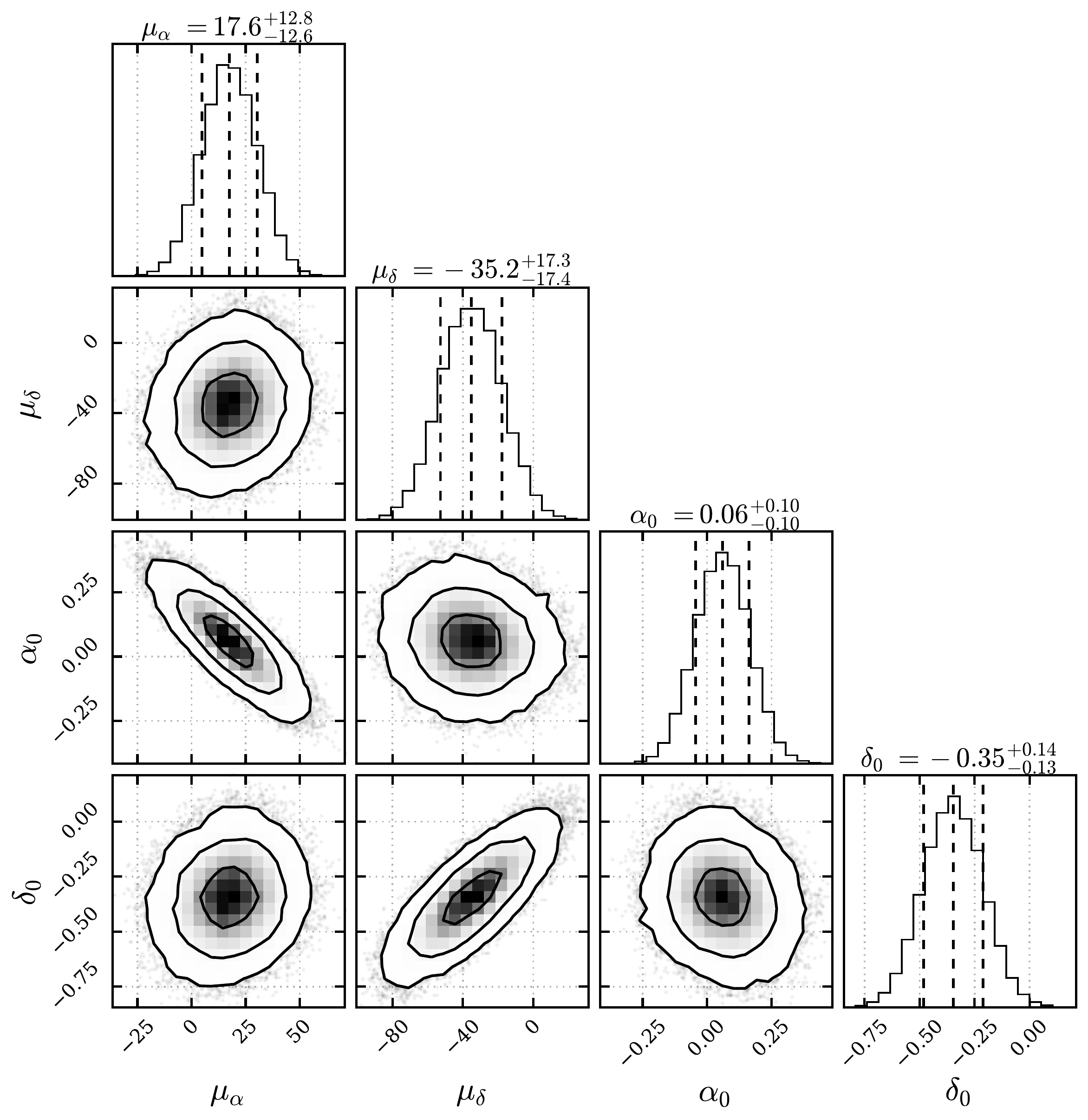}
\caption{Corner plot (as in Fig. \ref{G15Corner}) showing the posterior distribution of the astrometric solution for the CCO in Cas A. See Table \ref{AllPMTable} for constraints on all parameters.}
\label{CasCorner}
\end{figure}

The results from our proper motion measurement are displayed in Fig. \ref{CasCorner}. We find evidence for non-zero proper motion directed toward the southeast, with a best-fit value of $(18^{+12}_{-13},-35^{+17}_{-18})\masy$. An analogous fit to the data from the late epochs taken individually yields a similar result, showing that the impact of our merging technique is rather small.

Converting our 2D probability distribution for $(\mu_{\alpha},\mu_{\delta})$ to a distribution for the absolute value of proper motion, we obtain a $68\%$ central credible interval $\mu^{*}_{\rm{tot}} = 35^{+16}_{-15} \masy$. This corresponds to a transverse physical velocity of $(570\pm260)\kms$ at the distance to Cas A of $3.4\,\si{kpc}$. Here, including the effect of Galactic rotation does not considerably alter the result.
\citet{Fesen06} performed an analysis of the expansion of high-velocity optically emitting ejecta knots, yielding a precise estimate for the explosion date of Cas A (around the year 1670). Connecting the explosion site inferred in \citet{Thorstensen01} with the present-day location of the CCO, they constrained its velocity to around $350\kms$ at a position angle $169^{\circ}\pm8^{\circ}$. Our measurement is marginally consistent with both that indirect estimate and the analysis of \citet{DeLaney13}.

\subsection{G330.2+1.0 \label{G330}}
The SNR G330.2+1.0 is located at a distance of at least $4.9 \,\si{kpc}$ \citep{MCClure}, and can be seen as a faint but complete shell in X-rays. Its emission is almost entirely nonthermal in nature, except for a region in the east showing thermal emission \citep{Williams18}. 
Analysis of shock-velocity and filament expansion points toward a rather young age of the remnant, likely below 1000 years \citep{Williams18,Borkowski18}.
As part of their expansion analysis, \citet{Borkowski18} noted that no evidence for ``discernible motion'' of the CCO CXOU J160103.1$-$513353 was found, which we attempt to further quantify here.

\begin{figure}
\centering
\includegraphics[width=1.0\columnwidth]{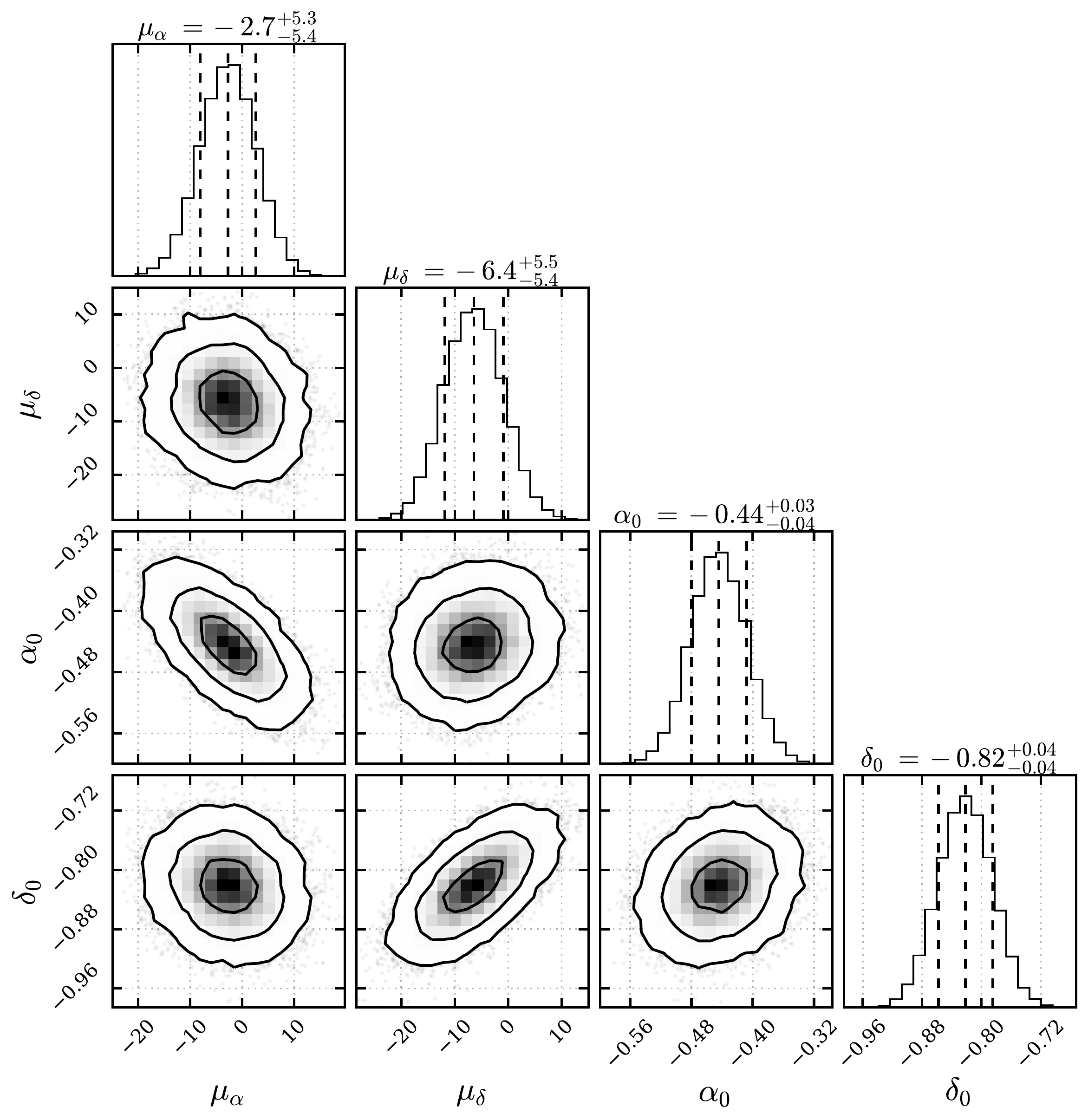}
\caption{Corner plot (as in Fig. \ref{G15Corner}) showing the posterior distribution of the astrometric solution for the CCO in G330.2+1.0. See Table \ref{AllPMTable} for constraints on all parameters.}
\label{G330Corner}
\end{figure}

We can see the posterior distribution for the astrometric solution of the CCO in Fig. \ref{G330Corner}. We find a very small proper motion toward south of $(-2.7_{-5.4}^{+5.3},-6.4^{+5.5}_{-5.4})\masy$. There are no obvious systematic discrepancies between the posterior predictions from our fit and the PSF fit contours, and we find no need to artificially add a systematic error component to the likelihood contours. 
Thus, due to the very long exposure times of the observations at early and late epochs, and the large number of clean astrometric calibrators, we find the small statistical error bars of our result plausible. 

The expected proper motion of an object co-rotating with the Galactic disk at the location of G330.2+1.0 at a distance between 4.9 and 10.0 kpc is $(-4.9\pm0.8,-4.5\pm0.6)\masy$.
Thus, the measured proper motion of the CCO is perfectly consistent with zero when effects of Galactic rotation are taken into account.
We quote the corresponding $90\%$ upper limit on the CCO's peculiar proper motion, which lies at $9.9\masy$. This corresponds to a transverse kick velocity below $230 \kms$, assuming a distance of 5 kpc. 

This case demonstrates that, even given the ``poor'' spatial resolution of X-ray data when compared to the optical, effects of Galactic rotation cannot generally be neglected for proper motion measurements.
Furthermore, our very tight constraints on its motion imply that the maximum angular distance travelled by the CCO during the ``lifetime'' of the G330.2+1.0 ($\lesssim1000$ years) is only around $10\arcsec$. This is close to negligible compared to the SNR's present-day radius around $5 \arcmin$, making CXOU J160103.1$-$513353 quite literally a central compact object.    

\subsection{RX J1713.7$-$3946 \label{G347}}
RX J1713.7$-$3946 (G347.3$-$0.5) is one of the brightest known emitters of ultra-high-energy gamma rays \citep[e.g.][]{HessG347}. 
It appears quite luminous in X-rays and comparatively dim in the radio regime, with its X-ray spectrum being absolutely dominated by nonthermal synchrotron emission \citep{Okuno18}. 
Morphologically, the SNR can be described as elliptically shaped, with the bright western shell apparent to consist of multiple ring-like structures which consist of thin filaments on smaller scales \citep{Cassam04}.
The distance to the SNR is estimated to be around $1.0-1.3$ kpc, based on CO and \ion{H}{I} observations of clouds interacting with the shock wave \citep{Fukui,Cassam04}. 
The expansion of the outermost SNR filaments has been directly measured in three quadrants \citep{Acero2017,Tsuji2016,G347SWExp}, yielding projected shock velocities up to $3900 \kms$ and implying an age for RX J1713.7$-$3946 between 1500 and 2300 years.

The position of the CCO 1WGA J1713.4$-$3949 is slightly offset from the apparent center of the SNR \citep{AschPfeff} by around $4.5^{\prime}$ toward south-west. Since the system is located quite nearby and the SNR age is likely rather low, it therefore seems reasonable to expect significant proper motion. For instance, assuming the geometric center to correspond to the explosion site, we would expect motion of around $130 \masy$ for an age of 2000 years. 

There are three archival \emph{Chandra} observations of the CCO,\footnote{Due to the large extent of the SNR, the CCO is outside the field of view of the numerous observations of the SNR shell.} one from 2005 using ACIS-I, one from 2013 using HRC-I, and one from 2015 (ID 15967) using ACIS-I. The latter observation was carried out using only a subarray of the detector, and with an offset aimpoint. Therefore, it is not suitable for our analysis, leaving us with only two moderately deep observations.

Due to the intrinsically high noise level in the HRC observation, it proved very difficult to find adequate reference sources which can be convincingly cross-matched to a \emph{Gaia} counterpart. 
In order to avoid having to rely on potentially erroneous optical associations, we decided to modify our approach, measuring relative source offsets between our two observations directly instead of measuring absolute positions. To achieve this, for each source, we convolved the PSF-fit likelihood contours of the late epoch with the ``mirrored'' contours of the early epoch (meaning we set $x \rightarrow -x$, $y \rightarrow -y$). This effectively subtracts the two positions from each other, resulting in a probability distribution for the astrometric offset between the two observations for a given object. To account for the unknown possible proper motion of our calibrators, we convolved their offset distributions with a Gaussian of width $\sigma = 10 \masy \times \Delta t$, where $\Delta t = 7.9 \,\si{yr}$ corresponds to the time difference between the two epochs. We then performed our astrometric fit with only $\{\mu_{\alpha},\mu_{\delta}\}$ and a single set of frame registration parameters $\{\Delta x,\Delta y,r,\theta\}$ as free parameters. The advantage of this method is that it does not require precisely knowing the true source positions, and ``only'' assumes their motions to not be much larger than $10 \masy$. 

\begin{figure}
\centering
\includegraphics[width=0.7\columnwidth]{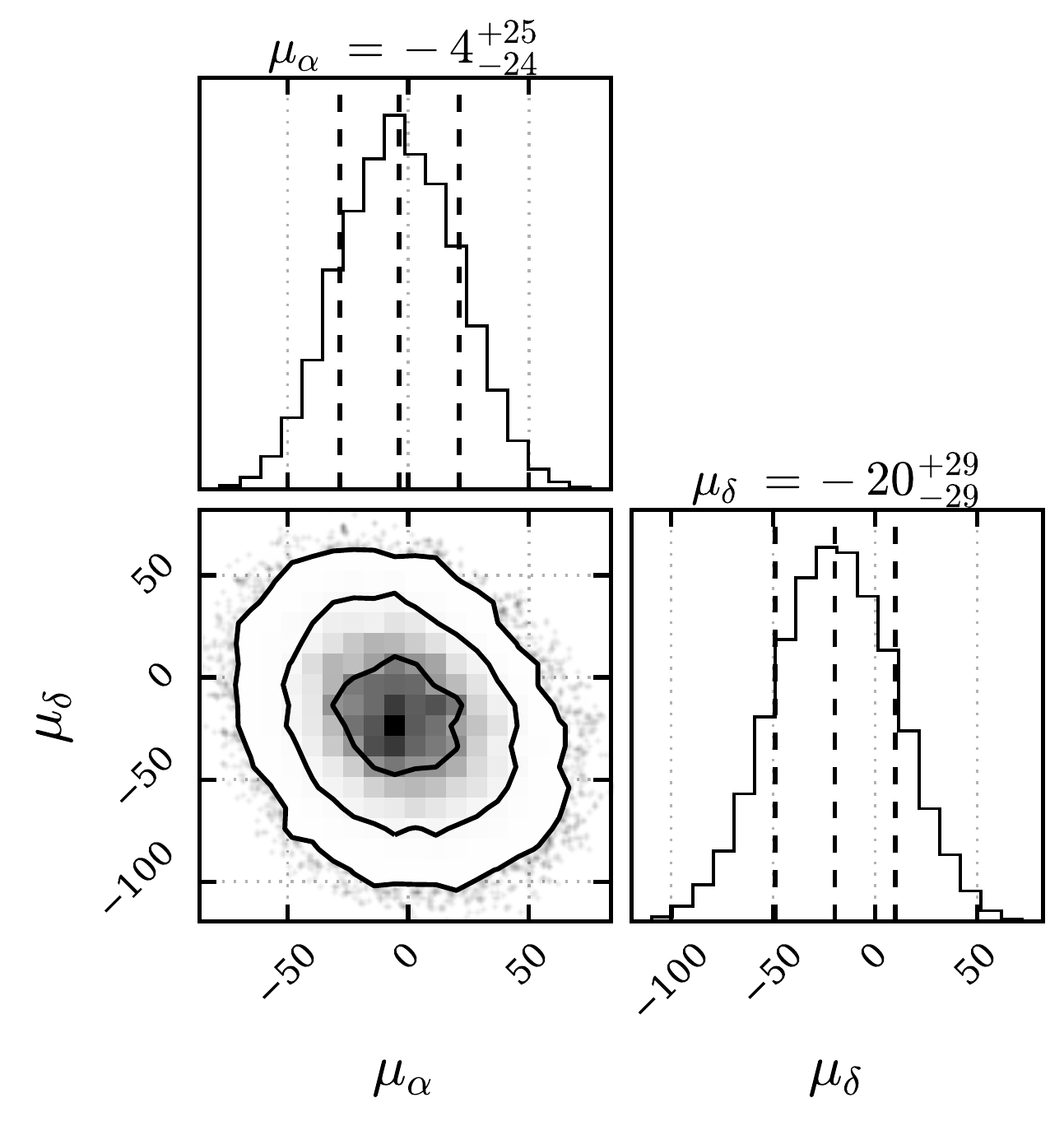}
\caption{Corner plot (as in Fig. \ref{G15Corner}) showing the posterior distribution of the astrometric solution for the CCO in G347.3-0.5. For this target, a relative astrometric frame registration method was applied as described in the text. We therefore measured no absolute positions, but only relative source displacements, which correspond to the components of proper motion. See Table \ref{AllPMTable} for constraints on all parameters.}
\label{G347CornerRel}
\end{figure}

The resulting posterior distribution can be seen in Fig. \ref{G347CornerRel}. 
We measured a proper motion of $(-4^{+25}_{-24}, -20\pm29) \masy$, which is consistent with zero within its quite large statistical uncertainties. We verified this result by checking that for both of our calibrators taken as the sole astrometric reference (applying only a simple translation of the coordinate system), we also obtained proper motion consistent with zero. 
This demonstrates that our result is not overly biased by our method or choice of calibration sources. We estimate the $90\%$ credible upper limit on the peculiar proper motion to be around $48\masy$, corresponding to a limit on the 2D space velocity of around $230 \kms$ for a distance of $1\,\si{kpc}$. 
A future observation would certainly allow to greatly reduce these large error bars on the CCO's proper motion. Given the issues we faced, it would probably be advantageous to use the ACIS detector for such a task, as it is generally better suited for the detection of faint sources than the HRC.

\subsection{G350.1$-$0.3 \label{G350}}
G350.1$-$0.3 exhibits a very unusual morphology at multiple wavelengths: In the radio, it shows a distorted and elongated shape \citep{Gaensler08}, and in X-rays, a very bright irregularly shaped clump in the east dominates the overall morphology with only weak emission otherwise. 
The most likely explanation for the bright clumpy X-ray emission is the interaction of hot, metal-rich ejecta with a molecular cloud, based on which a distance of $4.5 \,\si{kpc}$ seems likely \citep{Bitran97, Gaensler08}. \citet{Lovchinsky} performed spectral analysis of a deep ACIS-S observation of the SNR and from that estimated an age of $600-1200$ years for the SNR. 

The bright X-ray source XMMU J172054.5$-$372652 is located $3^{\prime}$ west of the dominant emission region, quite offset from the apparent center of the remnant. 
Assuming the origin of the CCO to be located at the apparent center of emission of the SNR, \citet{Lovchinsky} predicted a projected velocity of $1400-2600\kms$ for the NS. This would require an extremely strong kick to have acted on the NS during the supernova explosion, and the implied proper motion of $65-130 \masy$ should be easily detectable for us.

During the preparation of this work, \citet{Borkowski20} published a study on the expansion and age of G350.1$-$0.3.\footnote{Just prior to the submission of this work, a similar paper by \citet{Tsuchioka21} became available as preprint, which however did not influence our work substantially, anymore.}
They provided a detailed analysis of the motion of many individual emission clumps as well as a brief measurement of the CCO's proper motion. Their main results include the measurement of rapid expansion of the SNR, constraining its age to be at most around $600 \,\si{yr}$. Moreover, they detected comparatively slow motion of the CCO toward north, at significant uncertainty. Using our independent methods, we aim to confirm their measurement of the SNR's expansion and obtain quantitative constraints on the proper motion of the CCO.

The total exposure of the data set at both early and late epochs is exquisite, at around $100$ and $200\,\si{ks}$, respectively. The data are therefore ideal for obtaining precise constraints on the motion of both the CCO and the SNR ejecta in the nine-year time span between the two epochs.

\begin{figure}
\centering
\includegraphics[width=1.0\columnwidth]{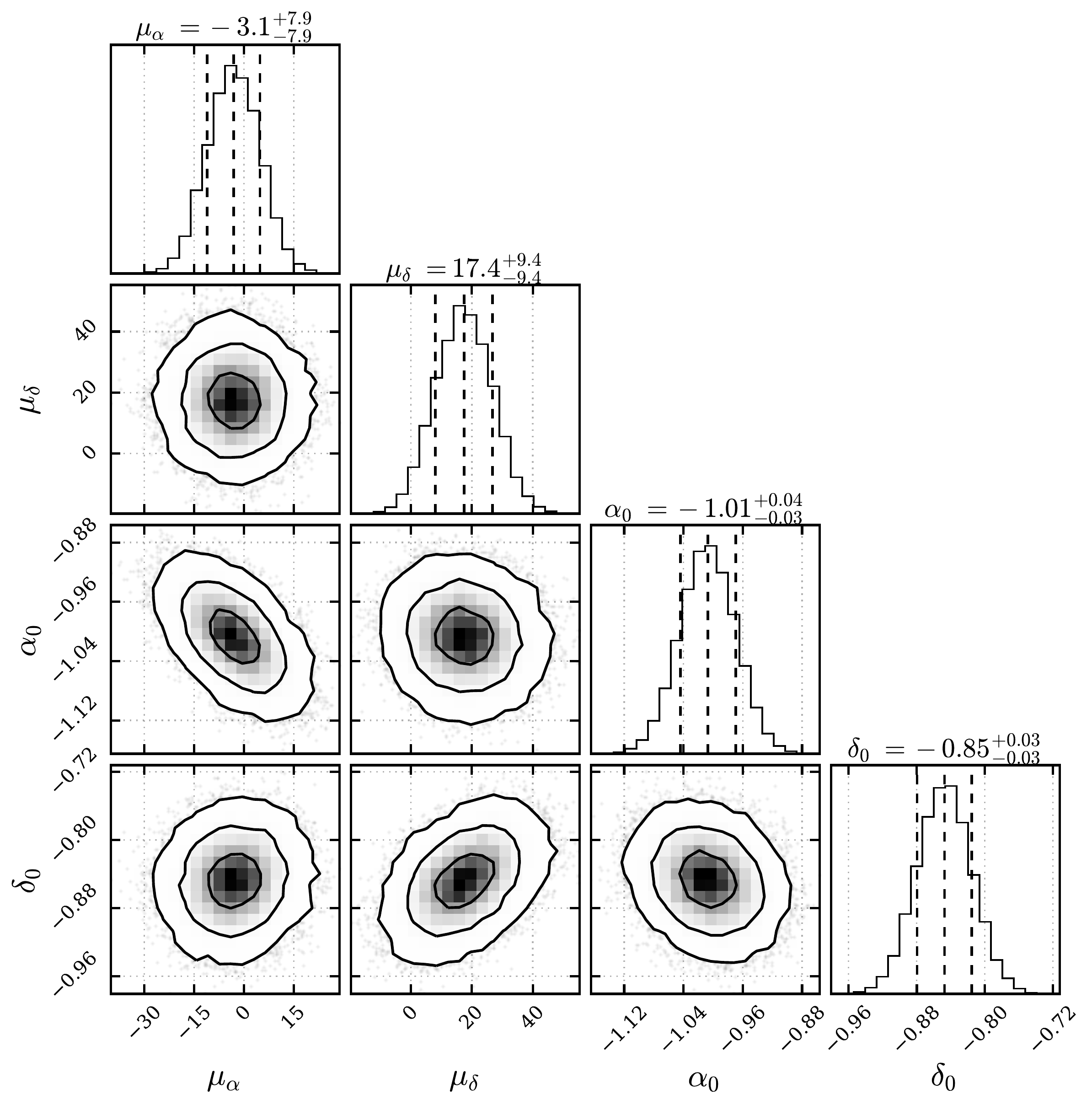}
\caption{Corner plot (as in Fig. \ref{G15Corner}) showing the posterior distribution of the astrometric solution of the CCO in G350.1-0.3. See Table \ref{AllPMTable} for constraints on all parameters.}
\label{G350Corner}
\end{figure}

From our proper-motion fit, where we treated all five late-time observations independently, we obtained the posterior distribution shown in Fig. \ref{G350Corner}. We find that the CCO appears to be moving northward (with large uncertainty), at $(-3\pm 8, 17^{+10}_{-9})\masy$. 
From this, we obtain a $68\%$ central credible interval for the peculiar proper motion of $\mu^{*}_{\rm{tot}} = 15^{+10}_{-9} \masy$, corresponding to a transverse velocity component of $320^{+210}_{-190}\kms$ at a distance of 4.5 kpc. 
\citet{Borkowski20}, using positions from Gaussian fits to the CCO and field sources, found a proper motion of $(-5,14)\masy$, which is consistent with our finding of slow motion approximately due north.  
Both measured values are certainly much lower in magnitude (and quite likely different in direction) than predicted from the SNR geometry in \citet{Lovchinsky}. The implications of this will be discussed in Sect. \ref{G350Disc}.

\begin{figure}
\centering
\includegraphics[width=0.9\columnwidth]{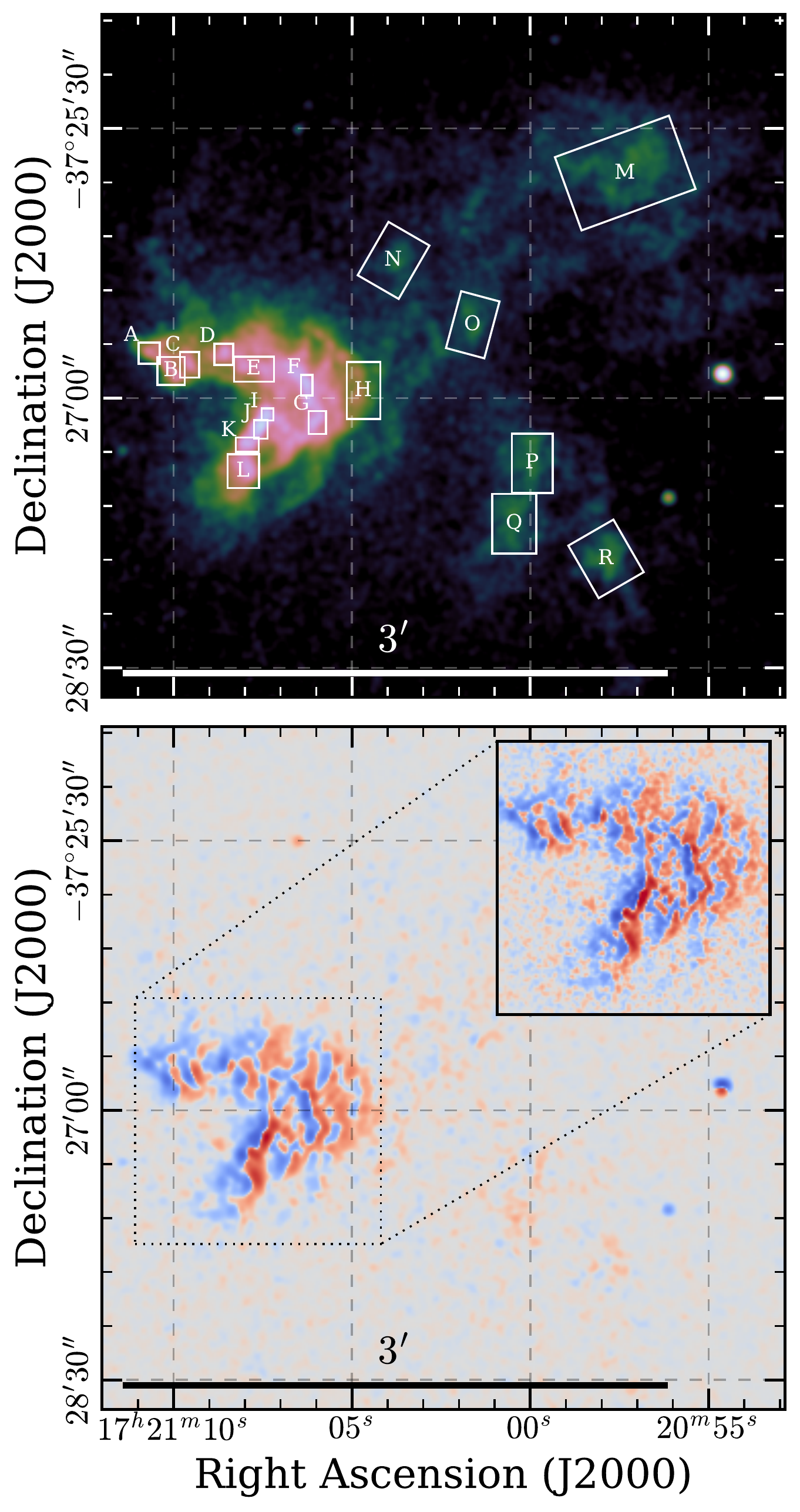}
\caption{SNR G350.1$-$0.3 with regions used for the expansion measurement indicated. 
The bottom panel shows a difference image between the early and late epochs. We use a symmetric logarithmic color scale to highlight features on small scales, with blue corresponding to the late-epoch emission being brighter, and red being the opposite. For the main image, we smoothed the data with a Gaussian kernel of $1^{\prime\prime}$, while in the inset we show the region of the bright clump smoothed with only $0.5^{\prime\prime}$}
\label{G350Image}
\end{figure}

\begin{figure}
\centering
\includegraphics[width=1.0\columnwidth]{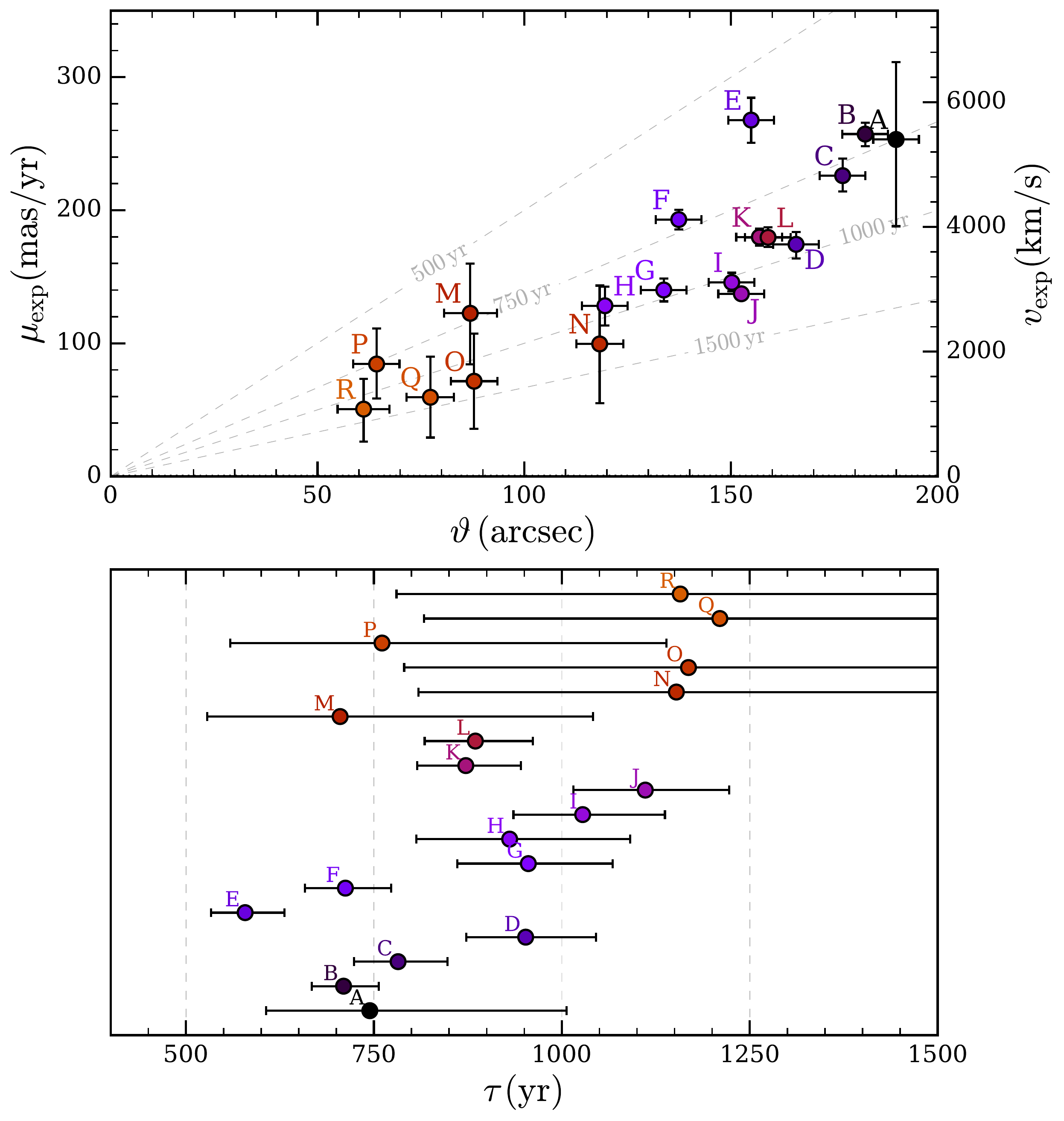}
\caption{Expansion and age estimate for G350.1$-$0.3. Top panel as in Fig. \ref{G15Exp}, with an assumed distance of $4.5\,\si{kpc}$.
Bottom: Constraints on the expansion age $\tau$ from each individual region, with error bars at $68\%$ confidence. 
The data underlying this figure are given in Table \ref{ExpTable}.}
\label{G350Exp}
\end{figure}

As a first test for the expansion of G350.1$-$0.3, we performed a simple image subtraction of the early-time from the late-time image, to check for obvious changes between the two epochs. 
The lower panel of Fig. \ref{G350Image} displays the differential between the merged, exposure-corrected and smoothed images.  
There is very obvious evidence for substantial eastward motion of the bright ejecta clump, which appears more significant toward its outer edge. In addition, the image nicely confirms the measured proper motion of the CCO, with its position showing a small but visible offset between the two epochs. 
This image is a very useful qualitative confirmation of the SNR's rapid expansion, independent of the exact method used in quantitative analysis.

As can be seen in the upper panel of Fig. \ref{G350Image}, the complex morphology and bright emission of the eastern clump allowed us to define numerous small-scale features of emission. Their propagation can be measured in order to infer the internal kinematics and global expansion behavior of the ejecta. 
In contrast, for the much fainter emission in the north and south of the SNR, we defined regions of diffuse emission on much larger scales, which we hoped would allow us to trace the interaction of the shock wave with the interstellar medium. 

For each feature, we used the indirectly inferred location of the explosion site from Sect. \ref{DiscExp} to estimate the angular distance $\vartheta$ which has been traversed since the supernova.
The measured expansion speeds $\mu_{\rm exp}$ for all 18 regions can be seen in the upper panel of Fig. \ref{G350Exp} (for a full illustration of the data at both epochs, see Fig.~\ref{G350FullExp_1}). We estimate systematic uncertainties due to astrometric frame registration to be smaller than $0.1^{\prime\prime}$, or equivalently $11\masy$. 
The proper motion of most features is much better constrained than for those in Sects.~\ref{G15} and \ref{Kes}. Therefore, we combined our probability distributions for $\vartheta$ and $\mu_{\rm exp}$ directly to obtain the free expansion age of the SNR, defined as $\tau = \vartheta / \mu_{\rm exp}$. The resulting constraints on $\tau$ are displayed in the lower panel of Fig. \ref{G350Exp}.

As we can see, there is spectacular evidence for significant expansion in almost all regions investigated. The maximal observed expansion speed is around $250\masy$, corresponding to a projected shock wave velocity close to $6000\kms$ for a distance of $4.5\,\si{kpc}$. 
In combination with its angular distance from the SNR center, we find the lowest expansion age for region E, at $\tau=(580 \pm 50) \,\si{yr}$, including systematic errors. These findings are in great agreement with those from \citet{Borkowski20}, who measured the fastest expansion at a very similar location (labelled ``B3'' by them, finding $\tau = 590\,\si{yr}$) despite a different region shape. 
Among the ``runners-up'' in terms of likely expansion rate are the regions A, B, F, M, P, all of which show values consistent with $\tau \sim700\,\si{yr}$. However, none of these regions independently confirms that $\tau < 700\,\si{yr}$.
In principle, the highest undecelerated expansion age $\tau$ can be considered a hard upper limit on the true age.\footnote{Mathematically, the expansion at this evolutionary stage can ideally be described by the relation between radius and age $R \propto t^{\beta}$, with $\beta$, the deceleration parameter, expected to range between $0.4$ and $1$ for the Sedov-Taylor stage \citep{Sedov,Taylor} and free expansion, respectively. The true age $t$ is therefore related to the expansion age $\tau = R/\dot{R}$ via $t = \beta \tau \leq \tau$.}
However, it seems important to ask why region E would indicate expansion at a higher rate than regions B or C. Its location does not appear to be ``special'' in the sense that it is not located close to the edge of the bright interacting ejecta clump. Naively, one would expect the least decelerated expansion there, if the bent shape of the clump is caused by the direct interaction of the shock wave with an obstacle. 
We thus choose to place a more conservative upper limit of $700\, \si{yr}$ on the true age of G350.1$-$0.3, which nonetheless confirms it as one of the three youngest known Galactic core-collapse SNRs \citep[see][]{Borkowski20}. 

The internal kinematics of the bright eastern clump are found to vary quite significantly with location: In particular, its southern elongated feature shows a significant velocity gradient, with the regions I and J being decelerated more strongly than the southern regions K and L. Similarly, the easternmost regions B and C display quite rapid motion when compared with the rest of the clump, especially to the nearby region D.  
Furthermore, upon closer inspection of flux profiles at early and late epochs, it becomes evident that some features in the clump, for example regions J and K, have evolved in shape and decreased in brightness (see Fig.~\ref{G350FullExp_2}) over the time span of nine years. The general trend seems to be that the X-ray emission in these regions becomes less ``clumpy'' and more diffuse as the SNR evolves. 
All these findings trace the interaction of the ejecta with an X-ray dark obstacle \citep{Gaensler08} which appears to cause the unique outward bent shape of the clump. Clearly, the interaction is the strongest close to the center of the clump, leading to the lowest measured expansion rates there, and higher velocities for its eastern and southern parts. In this scenario, if parts of the shock wave encounter a reduced ambient density after having passed by the obstacle, morphological changes of associated emission features could be a natural consequence. 

Finally, we have clearly detected the motion of the larger, more diffuse emission features in the north and south of the SNR (labelled M to R). The expansion rates we found here are a lot less certain, but on average appear comparable to or slightly smaller, than those in the bright emission clump. 
Here, there are some subtle differences between our findings and those of \citet{Borkowski20}, for instance for the faint regions M and O (labelled ``NNE'' and ``K'' by them). In these regions, they inferred rapid, almost undecelerated expansion, corresponding to expansion ages $<700 \,\si{yr}$, with quite small statistical errors. While our measurement for region M is formally consistent with theirs, we find much larger statistical errors in both regions M and O, making our constraints on the local degree of deceleration of the shock wave rather weak.  
These differences in uncertainty could be caused by different shapes or orientations of the measurement regions. However, we think the most likely origin is the difference in analysis methods, with our one-dimensional resampling technique yielding more conservative errors than a direct comparison of two-dimensional flux images. 

Overall, we did not find significant evidence for spatially varying degrees of deceleration within the faint diffuse emission regions.
The fact that every single region of the SNR shows (more or less) significant signatures of expansion away from a common center clearly proves that the unusual morphology of the SNR is not caused by the superposition of two unrelated objects, as was conjectured by \citet{Gaensler08}. Instead, G350.1$-$0.3 is indeed a single and highly peculiar SNR.

\section{Discussion \label{Discussion}}
\begin{table*}[t!]
\renewcommand{\arraystretch}{1.5}
\caption{Overview over direct proper motion measurements and exact positions for all CCOs.}
\label{PMTable}
\centering
\resizebox{\textwidth}{!}{
\begin{tabular}{ccccccccccc}
\hline\hline
SNR & CCO & 
$\alpha_0$ (J2000.) & $\delta_0$ (J2000.) & $t_0$ &
$\mu_{\alpha}$ & $\mu_{\delta}$ & $\mu_{\rm tot}$ & $d^{\prime}$ & $v_{\rm{proj}}$ & Reference \\
    &   & 
(h:m:s) & (d:m:s) & (MJD) &
$(\si{mas.yr^{-1}})$  & $(\si{mas.yr^{-1}})$ & $(\si{mas.yr^{-1}})$ & $(\si{kpc})$ & $(\si{km.s^{-1}})$ & \\
\hline
G15.9+0.2 & CXOU J181852.0$-$150213 & 
18:18:52.072$^{+0.004} _{-0.004}$ &$-$15:02:14.05$^{+0.04} _{-0.04}$ & 57\,233 &
$-17\pm12$ & $-4\pm10$ & $<25$ & $10$ & $<1200$ & This work \\ 
Kes 79 & CXOU J185238.6+004020 &
18:52:38.561$^{+0.008} _{-0.008}$ & +00:40:19.60$^{+0.15} _{-0.14}$ & 57\,441 &
$-3_{-10}^{+11}$ & $-3_{-11}^{+12}$ & $<19$ & $5.0$ & $<450$ & This work \\ 
Cas A & CXOU J232327.9+584842 &
23:23:27.932$^{+0.013} _{-0.013}$ & +58:48:42.05$^{+0.13} _{-0.13}$ & 55\,179 &
$18^{+12}_{-13}$ & $-35^{+17}_{-18}$ & $35^{+16}_{-15}$ & $3.4$ & $570\pm260$ & This work \\ 
Puppis A & RX J0822$-$4300 & 
08:21:57.274$^{+0.009}_{-0.010}$ & $-$43:00:17.33$^{+0.08}_{-0.08}$ & 58\,517 &
$-74.2^{+7.4}_{-7.7}$ & $-30.3\pm 6.2$ & $80.4\pm7.7$ & $2.0$ & $763 \pm 73$\tablefootmark{b} & 1 \\ 
G266.1$-$1.2 (Vela Jr.) & CXOU J085201.4$-$461753 & 
08:52:01.37\tablefootmark{a} & $-$46:17:53.5\tablefootmark{a} & 51\,843 &
...\tablefootmark{c} & ...\tablefootmark{c} & $<300$ & $1.0$ & $<1400$ & 2,3 \\ 
PKS 1209$-$51/52 & 1E 1207.4$-$5209 &
12:10:00.913$^{+0.003} _{-0.003}$ & $-$52:26:28.30 $^{+0.04} _{-0.04}$ & 54\,823 &
...\tablefootmark{c} & ...\tablefootmark{c} & $15\pm7$ & $2.0$ & $<180$ & 4 \\
G330.2+1.0 & CXOU J160103.1$-$513353 & 
16:01:03.148$^{+0.004} _{-0.004}$ & $-$51:33:53.82$^{+0.04} _{-0.04}$ & 57\,878 &
$-2.7_{-5.4}^{+5.3}$ & $-6.4^{+5.5}_{-5.4}$ & $<9.9$ & $5.0$ & $<230$ & This work \\ 
RX J1713.7$-$3946& 1WGA J1713.4$-$3949 &
17:13:28.30\tablefootmark{a} & $-$39:49:53.1\tablefootmark{a} & 56\,360 & 
$-4^{+25}_{-24}$ &  $-20\pm29$ & $<48$ & $1.0$ & $<230$ & This work \\ 
G350.1$-$0.3 & XMMU J172054.5$-$372652 & 
17:20:54.585$^{+0.003} _{-0.003}$ & $-$37:26:52.85$^{+0.03} _{-0.03}$& 58\,308 &
$-3\pm8$ & $17^{+10}_{-9}$ & $15^{+10}_{-9}$ & $4.5$ & $320^{+210}_{-190}$ & This work \\ 
G353.6$-$0.7 & XMMU J173203.3$-$344518 & 
17:32:03.41\tablefootmark{a} & $-$34:45:16.6\tablefootmark{a} & 54\,584 &
... & ... & ...\tablefootmark{d} & $3.2$ & ...\tablefootmark{d} & 5,6 
\\ \hline
\end{tabular}
}
\tablefoot{To provide a complete four-parameter astrometric solution, we display the best-fit positions of the CCO $(\alpha_0, \delta_0)$ at a given epoch $t_0$, corresponding to the latest available observation of the respective target. 
The values for the projected physical velocity $v_{\rm{proj}}$ are scaled to an assumed distance $d^{\prime}$, without the inclusion of any additional errors to account for uncertainties in $d^{\prime}$. 
The measurements of total proper motion $\mu_{\rm tot}$ and projected velocity  $v_{\rm{proj}}$ from this work have been corrected for the effect of Galactic rotation.
All our measurements and errors correspond to the median and $68\%$ central interval of the underlying probability distribution. All upper limits derived in this work are at $90\%$ confidence. \\
\tablefoottext{a}{The positions without listed uncertainties have not been corrected for \emph{Chandra}'s absolute astrometric inaccuracy, and therefore have estimated $1\sigma$--errors on the order of $0.4\arcsec$, corresponding to an $0.8\arcsec$ radius of the 2D $90\%$ confidence region.
}
\tablefoottext{b}{The most recent distance measurement of around $1.3 \,\si{kpc}$ \citep{Reynoso17} to Puppis A would imply a transverse velocity of $v_{\rm{proj}} = (496\pm47)\kms$ for the CCO.}
\tablefoottext{c}{No explicit values for $\mu_{\alpha}, \mu_{\delta}$ were given.}
\tablefoottext{d}{Due to the lack of suitable data, no proper-motion measurement currently exists for the CCO of G353.6$-$0.7.}
}
\tablebib{(1) \citet{Mayer20}, (2) \citet{Mignani07}, (3) \citet{Mignani19}, (4) \citet{Halpern15}, (5) \citet{Halpern10b}, (6) \citet{Maxted18}.}
\end{table*}

\subsection{Proper motion and neutron star kinematics \label{DiscPM}}

This work constitutes the first step toward the ambitious goal of building a sample of consistently measured transverse velocities for all central compact objects. At the present time, only four members of the class (the CCOs in Puppis A, Cas A, G350.1$-$0.3 and PKS 1209$-$51/52) have a measured non-zero proper motion. However, we hope that future observations will aid in reducing error bars and obtaining more precise constraints for those CCOs for which only upper limits on the proper motion could be established here. 

We display an overview of the astrometric solutions of all known CCOs, derived in this work and previous studies, in Table \ref{PMTable}. Our work almost doubles the number of CCOs with quantitative proper-motion measurements from five to nine. 
For several systems in this work, the results of our proper motion measurements were markedly different from reasonable expectations. For instance, for both G350.1$-$0.3 and RX J1713.7$-$3946, the present-day location of the CCO within the SNR suggests measurable non-zero proper motion. For RX J1713.7$-$3946, we were unable to obtain any significant signature of proper motion despite the small distance to the system. For G350.1$-$0.3, we did measure a mildly significant non-zero value, which is however considerably smaller than what the SNR morphology would suggest. This is similar to the case of PKS 1209$-$51/52, where \citet{Halpern15} measured very small proper motion, despite a striking offset of the CCO from the apparent SNR center. These examples illustrate that estimates of the neutron star kick purely based on SNR geometry have to be interpreted with care, and in most cases cannot replace a direct measurement. 
As demonstrated by \citet{Dohm96}, an inhomogeneous circumstellar medium can easily result in an offset between the SNR's explosion site and its apparent center, as the shock wave experiences different degrees of deceleration in different directions, leading to a distortion of the remnant's morphology. This effect was investigated in detail for Tycho's SNR by \citet{Williams13}, who showed that the observed density and shock velocity variations along its outer rim imply a significant offset between the apparent and true SNR center, by up to $20\%$ of its radius.
Therefore, even if a central neutron star appears clearly offset from its host's morphological center, such an offset need not be due to its proper motion, but may also be caused -- solely or to some fraction -- by the SNR’s distorted shape.    

The two most important factors limiting the precision of the measurements in this work are the availability of reliable astrometric calibration sources and their photon statistics.
For instance, the uncertainties on CCO proper motion in G15.9+0.2 and Kes 79 are primarily due to the (comparatively) short exposures at the former and latter of the two epochs, respectively, which lead to a small number of photons available for precise astrometric localization. 
In contrast, for RX J1713.7$-$3946, we were hindered by the lack of observed X-ray sources with reliable astrometric counterparts in the HRC observation. 
We argue that the main reason for this is the smaller effective area and higher intrinsic background rate of the \emph{Chandra} HRC when compared to the ACIS instrument. In principle, the design of the HRC as a microchannel plate instrument \citep{HRCRef} is ideal for astrometric measurements. However, such can only be performed in an absolute manner in the presence of frame calibration sources, for which the ACIS usually possesses better detection prospects, unless their X-ray emission is very soft.

By combining the measured total proper motion $\mu_{\rm{tot}}$ with an estimate for the distance $d$ to the SNR, one can constrain the projected velocity of the CCO in the plane of the sky $v_{\rm{proj}} = \mu_{\rm{tot}}d$. This can be viewed as a lower limit to the physical velocity of the neutron star, and thus serves as a proxy to the violent kick experienced by the proto-neutron star at its birth. 
Therefore, by comparing the fastest measured neutron star velocities to theoretical considerations and numerical simulations of core-collapse supernovae, one can attempt to constrain the kick mechanism and the degree of explosion asymmetry. 

Currently, the CCO with the largest securely measured velocity is RX J0822$-$4300 in Puppis A, with a value of $v_{\rm proj} = (763 \pm 73)\kms$, scaled to a distance of $2 \,\si{kpc}$.
As is shown in Table \ref{PMTable}, none of the other known CCOs (except possibly those in G15.9+0.2 and Vela Jr.~for which sufficient precision has not yet been achieved in a direct measurement) are likely to show a projected velocity in excess of $\sim 800\kms$. 
A likely mechanism by which large neutron star kick velocities are achieved is the ``gravitational tug-boat'', in which an asymmetric ejecta distribution accelerates the proto-neutron star by exerting a gravitational pull in the direction of slow and massive ejecta clumps \citep{Wongwa13}. As it has been shown that kick velocities above $1000 \kms$ can be achieved in realistic explosion scenarios \citep{Janka17}, none of the presently measured CCO velocities are in serious conflict with theoretical expectations.

At the present time, the observed distribution of CCO velocities shows no obvious departure from that for radio pulsars. For instance, assuming a Maxwellian velocity distribution\footnote{The projection of a 3D Maxwellian into the observed two dimensions yields a Rayleigh distribution \mbox{$p(v_{\rm{proj}}) \propto v_{\rm{proj}}\times \exp \left( - v^2_{\rm{proj}}/2\sigma^2 \right)$.}} with one-dimensional $\sigma = 265\kms$ as given by \citet{hobbs05}, the probability of measuring $v_{\rm{proj}} \leq 250 \kms$ is around $35\%$. Therefore, observing three such cases in our sample is unsurprising. However, the current sample of measured CCO velocities is neither large nor precise enough for definite conclusions on the velocity distribution of CCOs to be drawn.   

Most likely, the task of performing a statistically meaningful comparison between kick velocities of radio pulsars and CCOs will require a lot of time and observational effort to complete (including the detection of new CCOs).
However, it may ultimately help in shedding light on the question of what is the fundamental difference driving the phenomenological diversity of young neutron stars. At the present time, it is unclear how the difference in magnetic field strengths between CCOs, rotation-powered pulsars, and magnetars is related to the conditions before, during, or after the supernova explosion. 

\subsection{SNR expansion and explosion sites \label{DiscExp}}

\begin{figure*}
\centering
\includegraphics[width=18.4cm]{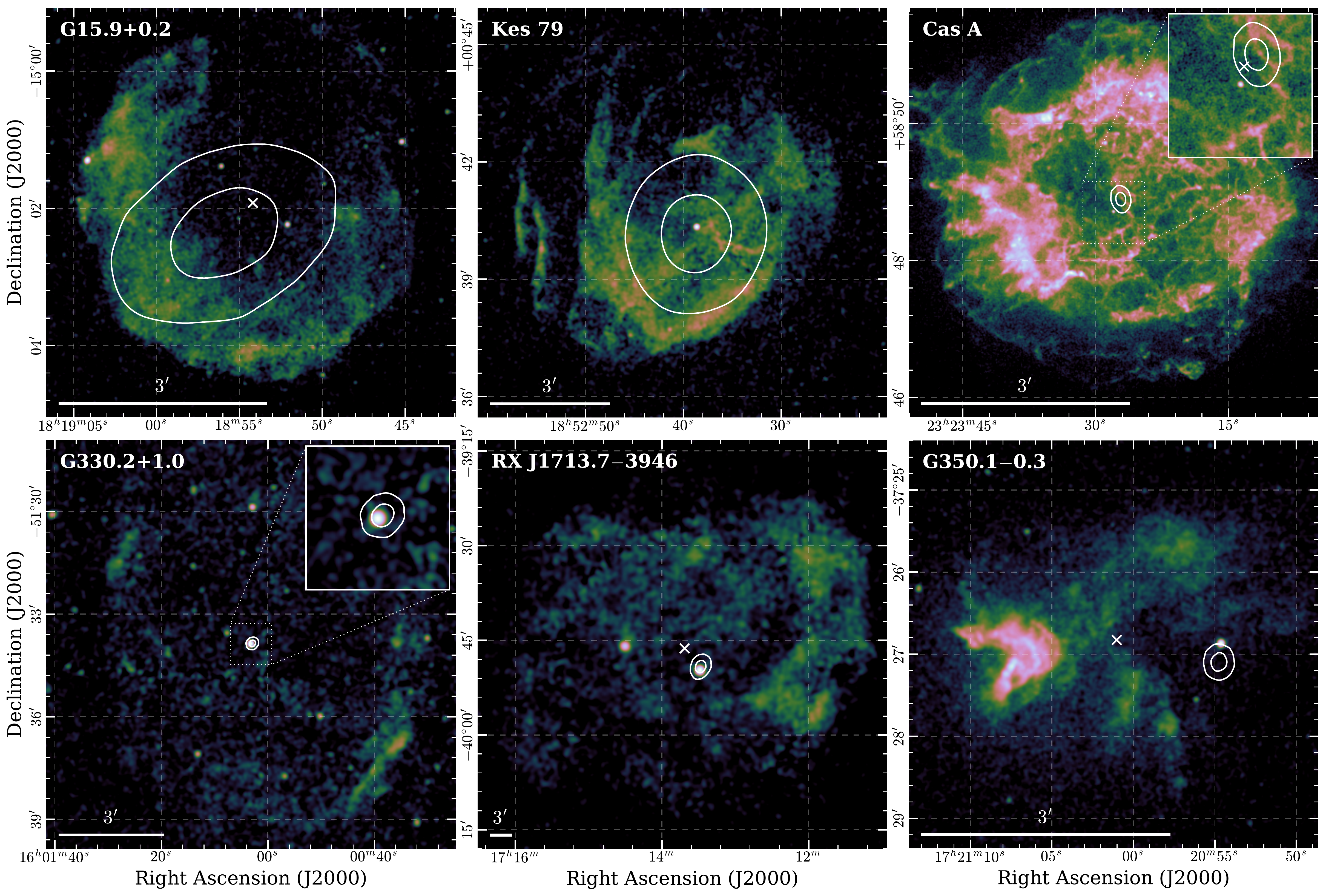}
\caption{Constraints on the explosion sites of the six SNRs targeted in this paper.
We show exposure-corrected images of the SNRs (with logarithmic intensity scaling), overlaid with $1\sigma$ and $2\sigma$ contours for the most likely origin of the CCO. 
The apparent geometrical centers of G15.9+0.2 (estimated in this work), RX J1713.7$-$3946 \citep{AschPfeff} and G350.1$-$0.3
\citep{Lovchinsky} are indicated with white crosses. For Cas A, we plot the precisely constrained expansion center from \citet{Thorstensen01} in an inset to make it more easily distinguishable.
All images were created from the \emph{Chandra} data used in this work, except for RX J1713.7$-$3946, where we used an archival \emph{ROSAT} observation.}
\label{CenterOverview}
\end{figure*}

Our proper motion measurements provide a constraint on the CCO's position over time, back to the explosion date of the supernova. This provides additional input to the physical interpretation of the present-day expansion and morphology of the SNR, as the location of its true center can be estimated.  
With this in mind, we inferred explosion sites of our six SNRs by extrapolating our posterior distribution of the CCO's proper motion backwards from its present-day location, to the assumed supernova explosion time. 
We then obtained the smallest two-dimensional regions containing $39.3\%$ and $86.5\%$ of the total probability mass, corresponding to $1\sigma$ and $2\sigma$ constraints on the explosion location, which can be regarded as the true center of the SNR. 
For this purpose, we assumed the following input ages, along the lines of available constraints (Table \ref{CCOTable}): $4000\,\si{yr}$ (G15.9+0.2), $5000\,\si{yr}$ (Kes 79), $340\,\si{yr}$ (Cas A), $1000\,\si{yr}$ (G330.2+1.0), $2000\,\si{yr}$ (RX J1713.7$-$3946), and $700\,\si{yr}$ (G350.1$-$0.3).

We compare the explosion sites to the present-day morphology of our six SNRs in Fig.~\ref{CenterOverview}. 
The extent of RX J1713.7$-$3946 is much larger than the \emph{Chandra} field of view, which is why we used an archival \emph{ROSAT} PSPC observation \citep{AschPfeff} to illustrate the characteristic structure of this SNR.

\subsubsection{G15.9+0.2 and Kes 79: Direct evidence for expansion?}
It is immediately obvious that, for some SNRs, our proper motion measurements are physically much less constraining than for others. 
For instance, the extent of the uncertainty contours of the explosion sites of G15.9+0.2 and Kes 79 is almost comparable to the extent of the two SNRs themselves. 
The reason for this is the comparatively shallow exposure of part of the used data sets, in combination with the small intrinsic angular size and large age of the SNRs.

Nevertheless, we have measured possible direct signatures of expansion for both remnants, which at this point appear slightly more significant for G15.9+0.2. If the true age of G15.9+0.2 were indeed lower than the $4000 \,\si{yr}$ which we assumed for estimating the explosion site, its uncertainty contours would ``shrink'' closer toward the CCO. The resulting constraints would likely still be consistent with what we estimate as the geometric center of the SNR.

For both objects, we have demonstrated the feasibility of an exploratory proper motion and expansion study, whose constraints could be significantly improved with future \emph{Chandra} follow-up observations. For instance, a dedicated observation of Kes 79 around the year 2025 with a depth of $\sim 30 \,\si{ks}$ would extend the measurement baseline to $\sim 25 \,\si{yr}$ in combination with the 2001 observation. This would be expected to reduce the errors on proper motion and expansion by a factor $\sim 2$.
For G15.9+0.2, given that the very deep late-time observation was carried out in 2015, a followup observation appears to be feasible only from around 2025. For instance, a $\sim 60 \,\si{ks}$ observation would effect an error reduction by a factor $\sim 1.9$.
In combination, reduced error bars on the motion of both the CCO and the SNR shell would provide insights into the kinematics and temporal evolution of filaments at the shock front as well as the SNR age. Moreover, a more accurate measurement of the proper motion of the CCO in G15.9+0.2 would further constrain its kick velocity. This would allow verifying if its offset location within the SNR is indeed caused by a rather violent kick, or simply due to asymmetric expansion of the SNR.

\subsubsection{G330.2+1.0: Rapid shockwave expansion away from a stationary CCO}
In contrast to G15.9+0.2 and Kes 79, we have provided stringent upper limits on the proper motion of the CCO of G330.2+1.0. Thus, its present-day location can be regarded as an almost perfect approximation to the SNR expansion center. 
The most obvious application of this is a comparison with the present-day expansion of the SNR. \citet{Borkowski18} measured rapid motion away from the CCO, at up to $\sim 9000\kms$ (for a distance of $5 \,\si{kpc}$) for most parts of the SNR shell. In contrast, they detected a far smaller rate of expansion for parts of the south-west rim, as well as significant temporal decline in its brightness. This, together with the relatively symmetric present-day morphology of the SNR, points toward a quite recent collision of the blast wave with a dense cloud in the south-western region causing significant deceleration, after almost free expansion for the majority of the SNR's lifetime 
\citep{Borkowski18}. 

\subsubsection{G350.1$-$0.3: A very young and highly peculiar SNR \label{G350Disc}}

\begin{figure}
\centering
\includegraphics[width=0.9\columnwidth]{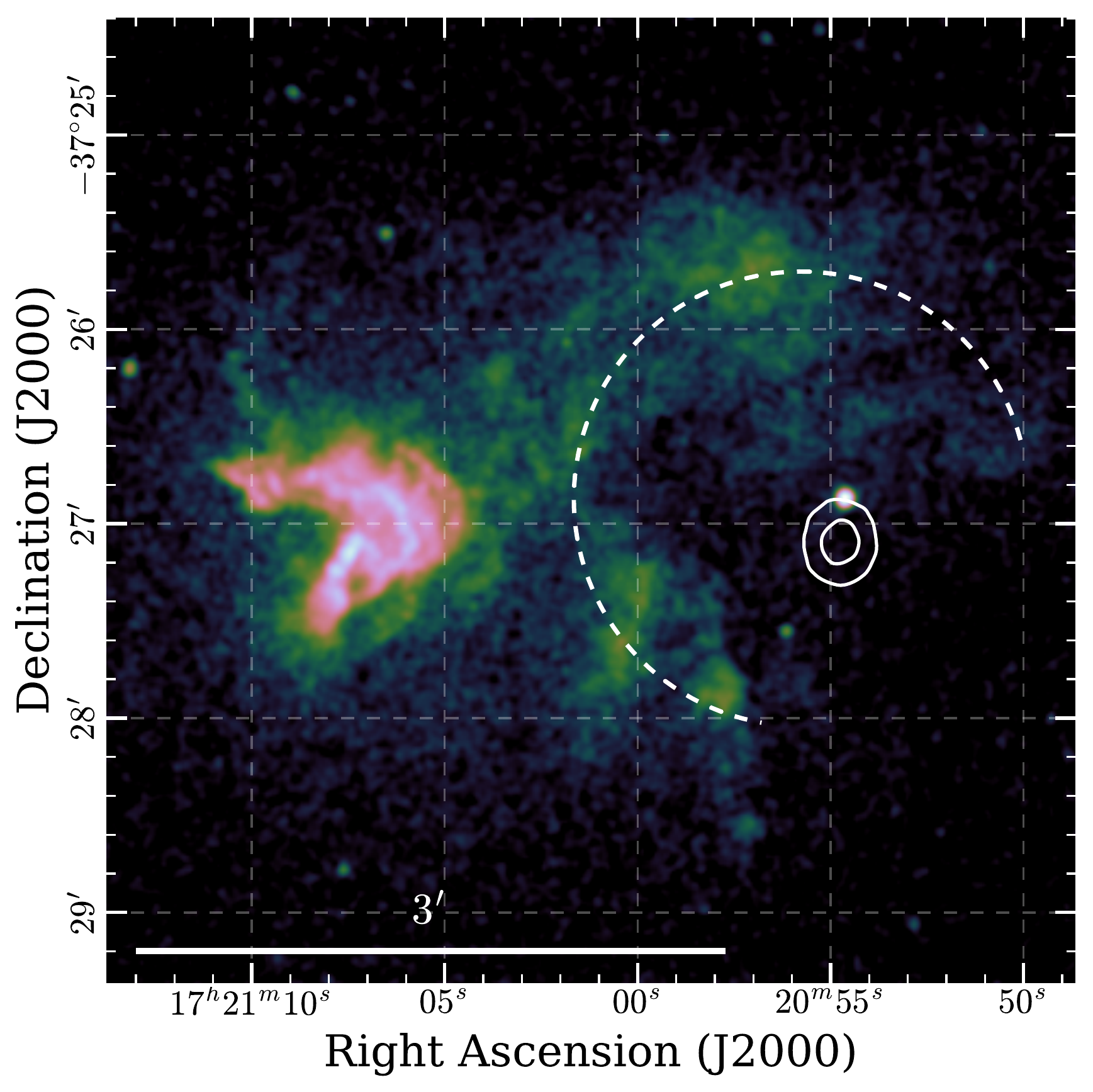}
\caption{Potential diffuse shell of G350.1$-$0.3.
We show the exposure-corrected image of G350.1$-$0.3 and its explosion site as in Fig.~\ref{CenterOverview}, overlaid with the incomplete circular shell tracing the SNR's diffuse emission (dashed line).
}
\label{G350Sketch}
\end{figure}

Contrary to all other SNRs which host a CCO, the explosion site of G350.1$-$0.3 which was deduced in this work is found to be strongly offset from the remnant's emission center, but matches the position of the CCO closely (see Fig. \ref{CenterOverview}). Consequently, the striking asymmetry of the SNR is further enhanced by our inferred explosion site: Only the rapidly propagating eastern ``half'' of the remnant appears to be visible at all, while the supposed half west of its true center, if present, remains almost entirely undetected.
In any case, conservation of momentum requires the presence of a large amount of mass (either in the form of a compact object or ejecta) moving toward the west. 

It is interesting to observe that, when ignoring the presence of the bright eastern clump, the remaining part of the SNR appears to form an almost circular incomplete shell, centered approximately on our inferred explosion site (indicated in Fig.~\ref{G350Sketch}). In this context, the bright eastern clump is required to be connected to fast ejecta as it is located far outside this potential shell, and shows the highest physical expansion velocities.
As significant expansion is detected in all investigated regions, including the more diffuse northern and southern parts of the shell, it seems conceivable that the initial explosion imprinted a quadrupolar asymmetry on the ejecta distribution. 

We consider a possible explanation for the missing western SNR shell to be a large density gradient in the surrounding interstellar medium \citep{Gaensler08}. The alternative, a dipolar asymmetry in the amount of ejected material, would necessarily have imprinted a very strong westward kick on the CCO. Given our measurement of its modest northward proper motion, this seems rather unlikely. The detection of rapid eastward expansion of the shockwave at up to $5000-6000\kms$ here and in \citet{Borkowski20} only strengthens this conclusion, since it increases the momentum attributed to the bright eastern clump and thus the implied recoil. Furthermore, the expansion measurement leads to a precisely constrained explosion site, due to the now certain very low age of the SNR.
Therefore, it can be stated with high confidence that the circumstellar material around G350.1$-$0.3 is highly inhomogeneous: It likely exhibits a density gradient from east to west and a significant localized density enhancement in the east. This eastern clump causes, as we have shown, a local velocity gradient in the expanding ejecta, which manifests itself in the fascinating outward-bent shape of the interacting shock wave. 

Given our observations, we can infer that the true extent of the remnant is necessarily greater than apparent in X-rays, with a radius of at least $3.5\arcmin$ (the approximate distance between the bright eastern ejecta clumps and the current location of the CCO), larger than the $2.5\arcmin$ estimated in \citet{Lovchinsky}. Since the remnant is larger, but also expanding much faster, than predicted, their age estimate of $600-1200$ years is still consistent with the $600-700$ years found by \citet{Borkowski20} and deduced in our analysis. 

The overall multi-wavelength morphology of G350.1$-$0.3 is quite rich, and can aid in understanding its peculiar appearance in X-rays. 
As shown by \citet{Lovchinsky}, the bright eastern clump has an obvious counterpart in the mid-infrared ($24\,\si{\mu m}$), likely due to dust shocked by the interaction of the supernova ejecta with a molecular cloud.
In its fainter regions, the $24\,\si{\mu m}$ image of the SNR shows features corresponding to the southern part of the diffuse X-ray shell, but little apparent emission in the north.
In contrast, as displayed by \citet{Gaensler08}, the SNR as viewed in the radio domain (at $4.8\,\si{GHz}$) exhibits a horizontal ``bar'', coincident with the northern part of the shell seen in X-rays.  
However, there appears to be no radio emission originating from the southern region. 
Keeping these points in mind, it is interesting to observe that the X-ray emission in these two regions also shows spectral differences, with the north emitting harder and the south softer radiation \citep[see Fig. 1 in][]{Borkowski20}. 
We interpret these findings as a further signature of inhomogeneous circumstellar medium, where denser circumstellar material in the south causes stronger interaction with the supernova blast wave, and thus leads to infrared emission from shocked or heated dust. Less dense material in the north interacts weakly with the shock wave, allowing it to propagate more freely, consistent with the observed radio and harder X-ray emission. \citet{Borkowski20} came to a related conclusion, demonstrating that the northern part of the SNR shows little evidence for the presence of ejecta, and its emission is thus likely attributable to the propagation of the supernova blast wave through a comparatively thin medium.

Given the youth of G350.1$-$0.3, we checked the catalog of ``guest stars'' by \citet{HistSNe} for any potential historical counterpart. There was no obvious match to our SNR, except possibly one reported guest star in AD 1437, which would imply an age $\sim 580 \, \si{yr}$. The specified position of this historic event is $(\alpha, \delta) = (16^{h}55^{m}, -38^{\circ})$ which places it about five degrees from the position of G350.1$-$0.3 \citep{HistSNe}. This spatial discrepancy and the possible nova explanation preferred by the authors argue against an association of this ``guest star'' with G350.1$-$0.3. 
In fact, the observed hydrogen column density $N_{H} \sim 4\times10^{22}\,\si{cm^{-2}}$ \citep{Lovchinsky} implies a visual extinction by around $A_{V} \sim 22\,\si{mag}$, according to the relation of \citet{Predehl95}. Combining this with an assumed peak absolute magnitude $M_{V} \sim -18$ for a core-collapse supernova at a distance of $4.5\,\si{kpc}$ results in an observed apparent magnitude $m_{V} \sim 18$. It therefore seems very unlikely that the supernova associated to G350.1$-$0.3 would have been observable with the naked eye at all. 

\subsubsection{RX J1713.7$-$3946: Evidence for non-uniform expansion history?}
For RX J1713.7$-$3946, we have determined that the supernova explosion site is likely located within $\sim 2 \arcmin$ of the present-day location of the CCO, which is somewhat at odds with the apparent geometrical center \citep{AschPfeff}. 
The expansion of the SNR has been independently measured along three sectors of its shell, in the south-east, south-west and north-west, in three individual studies \citep{Acero2017,Tsuji2016,G347SWExp}. Interestingly, these works all showed that the maximal expansion speed of the shockwave is remarkably uniform, on the order of $4000 \kms$ (or $800\masy$), in all directions.\footnote{All three studies measured the motion of several clumps or filaments. However, here, we focus on the highest respective velocities, since these are expected to be closest to the undecelerated speed of the shock wave in that particular region.}

\begin{figure}
\centering
\includegraphics[width=1.0\columnwidth]{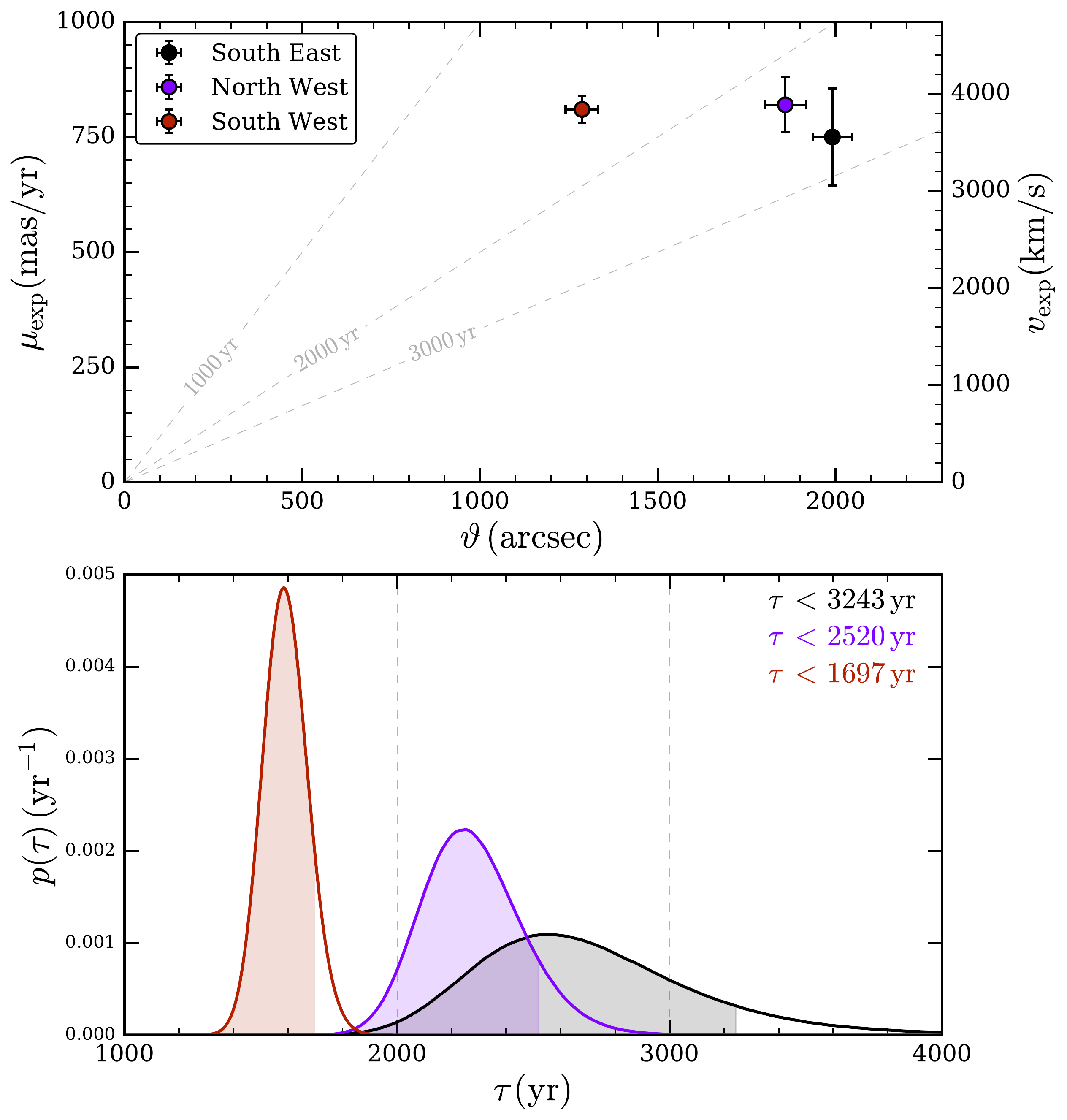}
\caption{Expansion and age estimate for RX J1713.7$-$3946. The top panel shows the measured maximal proper motion $\mu_{\rm{exp}}$ of filaments in the expanding shell versus their angular distance $\vartheta$ from the explosion site. 
We indicate values from the literature for the south-east \citep{Acero2017}, north-west \citep{Tsuji2016}, and south-west \citep{G347SWExp} regions, respectively. For the proper motion, we display the error bars as given in the respective paper, scaled to $68\%$ confidence and including systematic errors for \citet{Acero2017}.
The bottom panel shows the derived likelihood and corresponding $90\%$ confidence upper limits for the undecelerated expansion age $\tau$, inferred for the respective filaments. }
\label{G347Age}
\end{figure}

With our determination of the expansion center of the SNR, we can now provide updated constraints on the age of RX J1713.7$-$3946, by combining the location of its true center with the  proper motion of the fastest ``blobs'' in the three regions, in analogy to Sect. \ref{G350}:
Comparing our inferred explosion site with the estimated position of ``blob A'' in the south-west \citep{G347SWExp}, we find the angular distance covered by the shock wave up to today to be around $\vartheta = 1290\arcsec \pm 50\arcsec$. Together with its measured proper motion of $\mu_{\rm{exp}} = (810\pm30)\masy$, this yields an undecelerated expansion age of $\tau = (1590 \pm 80)\,\si{yr}$. Since it is likely that the shock wave has been somewhat decelerated during its expansion history, we quote only the corresponding $90\%$ upper limit on the expansion age of the SNR at $\tau < 1700\, \si{yr}$. 
An analogous investigation using the reported values and errors for the north-west \citep{Tsuji2016} and the south-east \citep{Acero2017} rims yields higher formal limits of $\tau < 2520\, \si{yr}$ and $\tau < 3240\, \si{yr}$, respectively, as illustrated in Fig.~\ref{G347Age}. 
Note here that assuming a different age as input to the computation of the SNR's explosion site would slightly alter the size of the uncertainties on its expansion age. However, it would have little effect on the overall result, since only significant non-zero motion of the CCO would bias our computation in any direction. 

Therefore, if we assume all input expansion velocities and errors to be reliable, we can infer that the shock wave must have been decelerated non-uniformly, leading to by far the largest present-day expansion rate inferred for the south-west filament. Generally, it can be considered unlikely that any physical mechanism could have substantially accelerated the blast wave in the past, while it is very probable for the shock wave to have experienced some degree of deceleration due to past interaction with the surrounding medium.
Therefore, we choose to quote the smallest out of the three measurements of $\tau$, corresponding to a true age of $<1700\,\si{yr}$.
This is a quite constraining result, as recent age estimates have been in the range $1500-2300\,\si{yr}$ \citep{Acero2017,Tsuji2016}, depending on the exact model assumed for ejecta and ISM density profiles. In particular, our value is still consistent with the tentative association of RX J1713.7$-$3946 with the historical ``guest star'', a possible supernova, observed by Chinese astronomers in the year 393 \citep{SN393}, if we assume parts of the south-west shell to be expanding almost freely.

However, it is somewhat puzzling why the shock wave would show the least decelerated expansion along the south-western direction, in particular for ``blob A''.
The observed X-ray emission from ``blob A'' is much softer than its surroundings with a power law spectral index $\Gamma = 2.74\pm 0.07$ \citep{G347SWExp}. Using the simplest assumptions, this implies a shock velocity of only $\sim2800\kms$\citep{Okuno18}, around $1000\kms$ smaller than the measured value. Additionally, X-ray and molecular gas observations imply that the shock wave is expanding in a cavity blown by the progenitor in the south-east, whereas it has likely relatively recently struck the dense cavity wall in the west \citep{Cassam04,Fukui12}. Thus, one would naively expect to measure the least decelerated expansion of the shock wave in the south-east of RX J1713.7$-$3946.      
We therefore emphasize that our age estimate is largely based on the published proper motion of only a single ``blob'' in the south-west of the SNR by \citet{G347SWExp}, and is therefore to be taken with caution. In order to confirm these constraints, future follow-up observations and/or independent reanalyses are needed to verify both the proper motion of the CCO and that of ``blob A''.  

In this context, we consider again the double ring-like morphology of RX J1713.7$-$3946 in X-rays, which is most striking for the western shell (see Fig. \ref{CenterOverview}). As pointed out by \citet{Cassam04}, both the inner and outer apparent ring exhibit a deformed elliptical structure. This structure appears nearly continuous around the entire SNR, and for the inner part, is matched well by the SNR's radio morphology. It is interesting to observe that the southern and eastern portions of this inner ring appear to be located quite close to our derived explosion site, as could be expected from the reverse shock. However, due to the overall weakness of thermal X-ray emission \citep{Katsuda15} and the radio detection of the inner ring, such an interpretation seems quite unlikely for this feature.
Alternatively, one could also imagine that the complicated morphology stems from the superposition of two unrelated SNRs. This however can be considered unlikely due to the observed dominance of non-thermal X-ray emission across the entire remnant \citep[e.g.][]{Okuno18} and the observed expansion in multiple regions. 
The most likely scenario is that the SNR's peculiar morphology is caused by the projection of different parts of the shock wave, expanding into an anisotropic wind-blown cavity from the progenitor star. Locally varying collision times with the cavity wall are also a reasonable explanation for the observed differences in the relative expansion rate of the SNR.

\subsubsection{Cas A: Adding independent constraints to optical expansion measurements}
Finally, for Cas A, the situation is quite unique: While X-ray measurements of the propagation of the supernova shock wave have been performed \citep[e.g.][]{Vink98,CasXray3}, their general target was to trace the exact location and propagation of the forward shock along the SNR shell. These studies were able to demonstrate that the forward shock of Cas A travels at a speed around $(4200 - 5200)\kms$, implying significant deceleration of the shockwave compared to free expansion. 
Direct expansion and age measurements for this infant SNR can be performed with much higher precision by tracing the motion of clumpy filaments in optical line emission. Prominent examples are the works by \citet{Fesen06} and \citet{Thorstensen01}, who determined the explosion date of Cas A to around the year 1670.

Our constraint on the explosion site via neutron star proper motion is presently far less precise than the constraints from optical filament expansion. 
However, the accuracy of the latter result depends on the presence of systematic effects, for example from non-symmetric deceleration of ejecta knots.
If we imagine being able to use a potential future observation to extend our temporal baseline from around 10 to more than 20 years, we can envisage greatly reduced error bars on our X-ray measurement of the CCO's proper motion.
For instance, we estimate that a further HRC observation of similar depth as the 1999 data set ($\sim 50\,\si{ks}$) would reduce the error bars on the proper motion by a factor $\sim2.5$. 
The trajectory of the neutron star, which is almost perpendicular to the motion of the fastest optical knots \citep{Fesen06}, could then serve as a complementary and completely independent ingredient to measuring the age and location of the expansion center of Cas A. 
The main advantage is that, due to its enormous density, the neutron star can be regarded as a ``bullet'', moving through its surroundings at effectively constant velocity. This makes its proper motion a useful tool to search for systematics in the motion of the fastest optical ejecta knots, possibly improving the accuracy of the age estimate for Cas A.

As pointed out by \citet{Fesen06}, it is interesting to observe that the neutron star in Cas A appears to be moving in a direction unrelated to that of the prominent ``jets'' of high-velocity ejecta. Instead, the CCO moves into the direction roughly opposite the brightest ejecta emission \citep{NS_SNR_Connection}. 
This is likely a manifestation of the neutron star kick mechanism, in which the proto-neutron star experiences the ``gravitational tug'' of slow and massive ejecta during the first seconds of the explosion \citep{Wongwa13}.

\section{Summary\label{Summary}} 
In this work we have analyzed several sets of archival \emph{Chandra} observations with the aim of deducing proper motion measurements consistently for all known central compact objects with suitable data.
Important elements of our analysis were the systematic search for serendipitous sources for astrometric frame calibration, the fitting of models of the \emph{Chandra} PSF to the data to obtain unbiased source positions, and the alignment of measured positions with a calibrated reference frame. 
The final results were obtained by a simultaneous fit to the data from all observational epochs, to provide accurate estimates of the proper motion of the target. In this way, we attempted to exclude biases by PSF morphology, aspect reconstruction uncertainties, or the effects of Galactic rotation.  

In total, we have presented six new measurements of neutron star proper motion, four of which had not been directly targeted prior to this work. Thus, we have approximately doubled the existing sample of CCOs with quantitatively constrained kinematics. 
We have found moderately significant non-zero proper motion only for two objects in our sample, the neutron stars in Cas A and G350.1$-$0.3, with no evidence for any hypervelocity CCOs. In contrast, we have set comparatively tight upper limits on the motion of several objects in our sample. For instance, the transverse physical velocities of the CCOs in G330.2+1.0 and RX J1713.7$-$3946 are constrained at $<230 \kms$ at distances of $5 \,\si{kpc}$ and $1 \,\si{kpc}$, respectively.

Complementarily, we have used our astrometrically coaligned data sets to directly measure the expansion of three SNRs, by comparing the one-dimensional emission profiles of characteristic features at different epochs.
For the SNRs G15.9+0.2 and Kes 79, we have applied our methods to comparatively shallow data sets to systematically search for signatures of expansion. 
For both objects, we found tentative evidence for the expansion of fragments of the SNR shell away from its center, at $2-2.5\sigma$ formal statistical significance. 
However, at the present time, the suboptimal exposure times and pointing configurations of the archival data sets prohibit a more detailed analysis of internal kinematics or an exact age estimate.

For the much younger SNR G350.1$-$0.3, we have qualitatively confirmed, using our independent methods, recently published results by \citet{Borkowski20}, most importantly the SNR's rapid expansion at close to $6000\kms$. Conservative interpretation of our measurement implies an upper limit on the SNR's age of $700$ years, which places G350.1$-$0.3 among the three youngest Galactic core-collapse SNRs. 
Additionally, we have directly confirmed that kinematics within the SNR vary on small scales within its bright eastern clump, likely due to the strong interaction of ejecta with a molecular cloud. Furthermore, expansion is visible also for much fainter emission features, proving that the peculiar observed morphology of G350.1$-$0.3 is in fact not due to chance superposition of two SNRs. 

We have used our CCO proper motion measurements to constrain the exact explosion locations of the six SNRs in our sample. For instance, for the young SNR G330.2+1-0, we have shown that its supernova occurred within only $10^{\prime\prime}$ of the present-day location of the CCO.
A very interesting finding is the confirmation of a striking asymmetry in the SNR G350.1$-$0.3, whose explosion site is located far west of its apparent geometric center. 
This finding has severe implications on the properties of the SNR, requiring an ultra-asymmetric explosion or strong density inhomogeneities in the circumstellar environment of one of the youngest and most bizarre Galactic SNRs. 

Finally, we have combined several published expansion measurements of RX J1713.7$-$3946 with our estimate of the supernova explosion site. We found that the current expansion rate significantly differs between individual regions. In particular, we have shown that part of the south-western shock wave has likely experienced much weaker past deceleration than other regions, since otherwise its present-day velocity could not be reconciled with its origin.
Using the most stringent resulting constraint, we have determined an upper limit of $1700$ years on the age of RX J1713.7$-$3946. 

Some of our findings, as any multi-epoch astrometric analysis, would certainly profit from future \emph{Chandra} follow-up observations, ideally similar to the 19-year campaign to measure the proper motion of the CCO in Puppis A \citep[see][]{Mayer20}. 
For instance, a future \emph{Chandra} observation to be taken a few years from today would allow us to constrain the age and neutron star kick velocity of G15.9+0.2 and Kes 79, via an analysis of the kinematics of the CCO and the SNR shell. 
Additionally, a more precise trajectory of the neutron star could be used to indirectly constrain deceleration of optical ejecta filaments in Cas A, identifying potential systematic errors on its age estimate. Similarly, combining X-ray measurements of expansion for RX J1713.7$-$3946 with a more precise measurement of the proper motion of its CCO may help in supporting or rejecting its association with the historical SN 393. 
We hope that some of these exciting questions can be addressed during the lifetime of \emph{Chandra}, as it is the only instrument allowing for the required high precision in X-ray astrometry -- now, as in the foreseeable future.

\begin{acknowledgements}
We would like to thank Johannes Buchner for helpful discussions on statistical problems. 
We are grateful to the anonymous referee for the provided constructive criticism and useful suggestions.
We appreciate the reliable support by the \emph{Chandra} helpdesk at several stages of our analysis. 
MGFM acknowledges support by the International Max-Planck Research School on Astrophysics at the Ludwig-Maximilians University (IMPRS).  
\\
We acknowledge the use of the \emph{Chandra} data archive and software provided by the \emph{Chandra} X-ray Center (CXC) in the application packages CIAO and Sherpa.
This work has made use of data from the European Space Agency (ESA) mission
{\it Gaia} (\url{https://www.cosmos.esa.int/gaia}), processed by the {\it Gaia}
Data Processing and Analysis Consortium (DPAC,
\url{https://www.cosmos.esa.int/web/gaia/dpac/consortium}). Funding for the DPAC
has been provided by national institutions, in particular the institutions
participating in the {\it Gaia} Multilateral Agreement.
\\
This research made use of Astropy,\footnote{\url{http://www.astropy.org}} a community-developed core Python package for Astronomy \citep{astropy:2013, astropy:2018}. Further, we acknowledge the use of the Python packages Matplotlib \citep{Hunter:2007}, SciPy \citep{SciPy}, and NumPy \citep{NumPy}.  
\end{acknowledgements}

\bibliographystyle{aa} 
\bibliography{Citations}

\onecolumn
\begin{appendix}

\section{Observation log and astrometric calibrators}
\begin{table*}[h!]
\renewcommand{\arraystretch}{1.15}
\caption{Journal of \emph{Chandra} observations used for our analysis} \label{OBsTable}
\centering
\begin{tabular}{cccccccc}
\hline\hline
SNR & Detector &   Obs. ID & Date & Exposure (s) & R.A. & Dec. & Roll\\
\hline
G15.9+0.2 & ACIS-S &  5530  &  2005 May 23 & 9062
&274.7150 & $-$14.9841 & 105.8 \\ 
            &    ACIS-S    &    6288  &  2005 May 25  &  4896  
            &274.7149& $-$14.9841 & 105.8 \\ 
            &    ACIS-S    &    6289  &  2005 May 28  & 14\,934  
            &274.7150 & $-$14.9841 & 105.8   \\ 
            &    ACIS-S    &    16766  &  2015 Jul 30/31  &   92\,021   
            & 274.7197 & $-$15.0420 & 259.1 \\ \hline  
Kes 79 & ACIS-I &   1982  &   2001 Jul 31/Aug 01  & 29\,571   
            &283.1585 &+0.6642 & 234.3 \\  
                                & ACIS-I & 17453 &  2015 Jul 20  & 9647   
            &283.1545 &+0.6019 &215.0 \\
                                & ACIS-I & 17659&  2016 Feb 23  & 9926   
            & 283.1594 & +0.6173&77.2 \\ \hline
Cas A & HRC-I & 1505 & 1999 Dec 19 & 48\,720   
            & 350.8566 & +58.8101& 287.1\\
                                & HRC-I &  12057 & 2009 Dec 13 & 10\,465   
            & 350.8570 & +58.8105 &287.0 \\
                                & HRC-I &  12059 & 2009 Dec 15 & 12\,294   
            & 350.8568 & +58.8106 & 287.0 \\
                                & HRC-I &  12058 & 2009 Dec 16 & 8865   
            & 350.8571 & +58.8105 & 287.0 \\
                                & HRC-I &  11240 & 2009 Dec 20 & 12\,400   
            & 350.8568 & +58.8106 & 287.0 \\ \hline
G330.2+1.0  &    ACIS-I    & 6687 &  2006 May 21/22 & 49\,977   
            & 240.2458 & $-$51.5714 & 3.3 \\
                                    &    ACIS-I    & 19163 &  2017 May 02/03 & 74\,143   
            & 240.2304 &$-$51.5508 & 30.2 \\
                                    &    ACIS-I    &  20068 &  2017 May 05/06 & 74\,140   
            & 240.2315 & $-$51.5497 & 30.2 \\ \hline
RX J1713.7$-$3946 &    ACIS-I    &  5559 &  2005 Apr 19 & 9645 
            & 258.3739 & $-$39.8275 & 72.5 \\
                                        &    HRC-I    &   13284 &  2013 Mar 08/09 & 37\,018   
            & 258.3724 & $-$39.8271 & 85.8 \\ \hline
G350.1$-$0.3 &    ACIS-S    & 10102 &  2009 May 21/22 & 82\,976 
            &260.2653 & $-$37.4388 & 57.9 \\
                                        &   ACIS-S    &   20312 &  2018 Jul 02 & 40\,579   
            & 260.2711 &$-$37.4497 & 296.2 \\
                                        &   ACIS-S    &   20313 &  2018 Jul 04 & 19\,840   
            & 260.2702 & $-$37.4491 & 296.2 \\
                                        &   ACIS-S    &    21120  & 2018 Jul 05 & 37\,619   
            & 260.2707 & $-$37.4494 & 296.2 \\
                                        &   ACIS-S    &    21119  &  2018 Jul 07/08 & 48\,289   
            &260.2709 & $-$37.4496 & 296.2 \\
                                        &   ACIS-S    &    21118  &  2018 Jul 08/09 & 42\,907   
            &260.2703 & $-$37.4492 & 296.2 \\
\hline
\end{tabular}
\tablefoot{
The column ``Exposure'' lists the effective unvignetted exposure time of each observation, referring to the sum of all good time intervals, corrected for the dead time fraction of the specific detector.
The columns ``R.A.'' and ``Dec.'' specify the right ascension and declination of the telescope pointing, given in decimal degrees, respectively. The column ``Roll'' describes the telescope roll angle west of north, given in degrees. }
\end{table*}
                
\begin{table*}[h!]
\renewcommand{\arraystretch}{1.15}
\caption{Overview over astrometric calibrator objects for all targets.}
\label{CalibList}
\centering
\resizebox{\textwidth}{!}{
\begin{tabular}{cccccccc}
\hline\hline
SNR & \multicolumn{2}{c}{Designation} &\multicolumn{2}{c}{Position (Epoch 2015.5)} & \multicolumn{2}{c}{Proper Motion} & Used\tablefootmark{a} \\  
 & Number & \emph{Gaia} DR2 Source ID        & R.A. (ICRS)          & Dec. (ICRS)         & $\mu_{\alpha}$  & $\mu_{\delta}$\\
    &     &               & (h:m:s)       & (d:m:s)     & (mas yr$^{-1}$) & (mas yr$^{-1}$) \\
\hline
G15.9+0.2 & 1 & 4146177809117217024 & 18:19:04.1362(0) & $-$15:01:17.914(0) & $-2.99 \pm 0.05$ & $-11.13 \pm 0.05$ & Y\\ 
 & 2 & 4146176679563359744 & 18:18:40.8608(0) & $-$15:02:50.272(0) & $-4.74 \pm 0.12$ & $-11.63 \pm 0.10$ & Y\\ 
 & 3 & 4146177396800326016 & 18:18:59.2871(0) & $-$15:03:02.877(0) & $0.58 \pm 0.36$ & $-0.19 \pm 0.34$ & N\\ 
 & 4 & 4146178977348360192 & 18:18:50.9880(0) & $-$14:58:42.607(0) & $0.01 \pm 0.09$ & $-5.59 \pm 0.08$ & Y\\ 
 & 5 & 4146179114787324800 & 18:18:56.0305(0) & $-$14:57:29.414(0) & $2.16 \pm 0.16$ & $0.31 \pm 0.15$ & N\\ \hline 
Kes 79 & 1 & 4266507644409347712 & 18:52:45.3536(0) & +00:37:33.883(0) & $2.71 \pm 0.34$ & $-5.48 \pm 0.37$ & Y\\ 
 & 2 & 4266506952913023488 & 18:52:13.0122(0) & +00:37:37.579(0) & $-2.48 \pm 0.58$ & $-4.97 \pm 0.50$ & Y\\ 
 & 3 & 4266508464741698816 & 18:52:48.6778(0) & +00:40:38.852(0) & $-1.11 \pm 0.45$ & $-2.45 \pm 0.41$ & Y\\ \hline 
Cas A & 1 & 2010478284367990016 & 23:22:51.6664(0) & +58:50:19.792(0) & $15.64 \pm 0.09$ & $0.30 \pm 0.08$ & Y\\ 
 & 2 & 2010477356655102592 & 23:23:14.1850(0) & +58:46:55.429(0) & $13.78 \pm 0.04$ & $-0.16 \pm 0.04$ & Y\\ 
 & 3 & 2010477253575869184 & 23:23:04.7862(0) & +58:48:00.022(0) & $-1.22 \pm 0.03$ & $-1.53 \pm 0.03$ & N\\ \hline
G330.2+1.0 & 1 & 5981660915203664384 & 16:01:13.9674(0) & $-$51:31:37.490(0) & $-5.01 \pm 0.23$ & $-5.03 \pm 0.15$ & Y\\ 
 & 2 & 5981660743404965120 & 16:01:07.6243(0) & $-$51:33:34.908(0) & $10.35 \pm 0.18$ & $9.56 \pm 0.13$ & Y\\ 
 & 3 & 5981637791098864128 & 16:00:39.0792(0) & $-$51:34:34.539(0) & $-9.87 \pm 0.07$ & $-10.89 \pm 0.05$ & N\\ 
 & 4 & 5981640814756329088 & 16:00:29.9561(0) & $-$51:33:43.894(0) & $-6.15 \pm 0.23$ & $-7.64 \pm 0.17$ & Y\\ 
 & 5 & 5981636932105333632 & 16:00:31.8387(0) & $-$51:39:05.986(0) & $0.49 \pm 0.08$ & $-1.31 \pm 0.06$ & Y\\ 
 & 6 & 5981661774197144448 & 16:01:02.8106(0) & $-$51:29:54.717(0) & $-18.74 \pm 0.30$ & $-19.66 \pm 0.23$ & Y \\ \hline 
RX J1713.7$-$3946 & 1 & 5972217652195355392 & 17:13:47.7129(0) & $-$39:52:51.150(0) & $-2.82 \pm 0.07$ & $-8.38 \pm 0.05$ & N\\ 
 & 2 & 5972124679069981440 & 17:13:31.7884(0) & $-$39:53:43.687(0) & $0.37 \pm 0.09$ & $-1.99 \pm 0.06$ & Y\tablefootmark{c} \\
 & 3 & 5972266138099405696 & 17:13:37.3502(15) & $-$39:46:04.196(20) & ...\tablefootmark{b} & ...\tablefootmark{b} & Y\tablefootmark{c} \\ \hline 
G350.1$-$0.3 & 1 & 5972890935584476288 & 17:20:56.1029(0) & $-$37:27:34.016(0) & $-3.07 \pm 0.18$ & $-4.57 \pm 0.14$ & Y\\ 
 & 2 & 5972880112266871936 & 17:21:21.3165(0) & $-$37:26:46.025(0) & $-2.48 \pm 0.17$ & $-1.02 \pm 0.12$ & Y\\ 
 & 3 & 5972880558943480448 & 17:21:13.1090(0) & $-$37:26:11.996(0) & $1.06 \pm 0.10$ & $1.50 \pm 0.07$ & N\\ 
 & 4 & 5972892516132452608 & 17:20:59.2872(0) & $-$37:25:01.514(0) & $2.41 \pm 0.37$ & $2.25 \pm 0.25$ & N\\ 
\hline
\end{tabular}
}
\tablefoot{We display the astrometric solutions for all our initially selected reference sources as listed in the \emph{Gaia} DR2 catalog \citep{GaiaSummary}. We state the rounded $1\sigma$ uncertainties of Right Ascension and Declination at the reference epoch in parentheses to illustrate their negligible character for almost all objects. The proper motion components along the Right Ascension and Declination axes are labelled as $\mu_{\alpha}$ and $\mu_{\delta}$, respectively. \\
\tablefoottext{a}{The column ``Used'' indicates whether an individual source was included in the final proper motion analysis or excluded for a specific reason (see subsections on the individual objects in Sect. \ref{CCO}).}
\tablefoottext{b}{No proper motion information in \emph{Gaia} DR2.}
\tablefoottext{c}{For these two sources, the absolute \emph{Gaia} positions were not used in our analysis since we considered their identification unreliable. Instead, the individual coordinate frames were directly registered with each other based on relative positions alone (see Subsect. \ref{G347} for details).}
}
\end{table*}

\clearpage
\section{Full results of proper motion and expansion measurements}
\begin{table*}[h!]
\renewcommand{\arraystretch}{1.5}
\caption{Full results of proper motion measurements and astrometric calibration of the individual observations for the six CCOs targeted in this work.}
\label{AllPMTable}
\centering
\begin{tabular}{cccccc}
\hline\hline
SNR & CCO & $\mu_{\alpha}$ & $\mu_{\delta}$ & $\alpha_0 $ & $\delta_0 $\\
 & & $(\si{mas.yr^{-1}})$ & $(\si{mas.yr^{-1}})$ & $(\si{arcsec})$ & $(\si{arcsec})$ \\
\hline
G15.9+0.2 & CXOU J181852.0$-$150213 & $-17.29_{-12.00}^{+11.65}$ & $-4.38_{-9.54}^{+9.51}$ & $0.110_{-0.053}^{+0.052}$ & $0.060_{-0.043}^{+0.042}$ \\
Kes 79 & CXOU J185238.6+004020 & $-2.59_{-10.37}^{+10.49}$ & $-2.72_{-11.65}^{+12.03}$ & $-1.089_{-0.120}^{+0.123}$ & $-0.244_{-0.137}^{+0.147}$ \\ 
Cas A & CXOU J232327.9+584842 & $17.58_{-12.62}^{+12.78}$ & $-35.23_{-17.37}^{+17.33}$ & $0.060_{-0.104}^{+0.103}$ & $-0.346_{-0.134}^{+0.134}$ \\
G330.2+1.0 & CXOU J160103.1$-$513353 & $-2.70_{-5.44}^{+5.30}$ & $-6.39_{-5.43}^{+5.47}$ & $-0.444_{-0.036}^{+0.037}$ & $-0.821_{-0.037}^{+0.037}$ \\ 
RX J1713.7$-$3946 & 1WGA J1713.4$-$3949 & $-3.90_{-23.92}^{+24.43}$ & $-19.63_{-28.93}^{+28.67}$ & ...\tablefootmark{a} & ...\tablefootmark{a} \\
G350.1$-$0.3 & XMMU J172054.5$-$372652 & $-3.06_{-7.98}^{+7.89}$ & $17.42_{-9.42}^{+9.41}$ & $-1.007_{-0.037}^{+0.038}$ & $-0.847_{-0.032}^{+0.032}$ \\ 
\hline\hline
SNR & Observation ID & $\Delta x_{i}$ & $\Delta y_{i}$  & $r_{i}-1$ & $\theta_{i}$\\
 & & $(\si{arcsec})$ & $(\si{arcsec})$ & $(10^{-4})$ & $(10^{-4} \, \si{rad})$\\
 \hline 
G15.9+0.2 & 5530 & $-0.010_{-0.102}^{+0.105}$ & $0.153_{-0.087}^{+0.086}$ & $5.7_{-5.0}^{+5.1}$ & $6.6_{-4.9}^{+5.2}$ \\ 
 & 6288 & $-0.066_{-0.142}^{+0.143}$ & $0.129_{-0.117}^{+0.112}$ & $-9.7_{-7.2}^{+7.4}$ & $6.1_{-7.4}^{+7.3}$ \\ 
 & 6289 & $0.171_{-0.097}^{+0.096}$ & $0.176_{-0.081}^{+0.081}$ & $2.3_{-4.8}^{+4.7}$ & $5.4_{-5.1}^{+5.0}$ \\ 
 & 16766 & $0.020_{-0.047}^{+0.047}$ & $-0.260_{-0.033}^{+0.033}$ & $5.8_{-2.5}^{+2.5}$ & $-2.5_{-1.8}^{+1.7}$ \\ 
\hline 
Kes 79 & 17453 & $0.128_{-0.117}^{+0.115}$ & $-0.323_{-0.137}^{+0.131}$ & $1.2_{-7.1}^{+6.4}$ & $-4.5_{-5.7}^{+6.1}$ \\ 
 & 17659 & $0.548_{-0.124}^{+0.121}$ & $-0.156_{-0.151}^{+0.136}$ & $-2.5_{-7.2}^{+6.5}$ & $-6.0_{-11.0}^{+8.2}$ \\ 
 & 1982 & $-0.164_{-0.091}^{+0.085}$ & $0.026_{-0.097}^{+0.093}$ & $2.5_{-5.0}^{+4.1}$ & $-8.2_{-3.5}^{+3.5}$ \\ 
\hline 
Cas A & 1505 & $-0.243_{-0.072}^{+0.071}$ & $0.056_{-0.110}^{+0.108}$ & $0.6_{-4.2}^{+4.1}$ & $-10.7_{-4.9}^{+5.0}$ \\ 
 & Late\tablefootmark{b} & $-0.526_{-0.102}^{+0.104}$ & $0.267_{-0.134}^{+0.134}$ & $10.9_{-4.8}^{+4.7}$ & $-15.9_{-5.9}^{+5.9}$ \\ 
\hline 
G330.2+1.0 & 19163 & $-0.164_{-0.036}^{+0.035}$ & $0.606_{-0.036}^{+0.036}$ & $2.4_{-2.2}^{+2.2}$ & $-1.5_{-2.1}^{+2.1}$ \\ 
 & 20068 & $-0.104_{-0.037}^{+0.036}$ & $0.216_{-0.036}^{+0.036}$ & $2.0_{-1.8}^{+1.8}$ & $-4.1_{-2.0}^{+2.0}$ \\ 
 & 6687 & $0.027_{-0.043}^{+0.044}$ & $0.221_{-0.043}^{+0.043}$ & $1.4_{-1.9}^{+1.9}$ & $-2.0_{-1.6}^{+1.6}$ \\ 
\hline 
RX J1713.7$-$3946 & $5559-13284$\tablefootmark{a} & $-0.292_{-0.192}^{+0.189}$ & $-0.255_{-0.228}^{+0.225}$ & $0.8_{-8.1}^{+8.0}$ & $-4.9_{-9.2}^{+8.9}$ \\ 
\hline 
G350.1$-$0.3 & 10102 & $-0.072_{-0.062}^{+0.061}$ & $0.568_{-0.079}^{+0.077}$ & $0.9_{-3.7}^{+3.7}$ & $4.4_{-3.3}^{+3.3}$ \\ 
 & 20312 & $0.423_{-0.040}^{+0.040}$ & $0.094_{-0.034}^{+0.035}$ & $-1.4_{-4.7}^{+4.7}$ & $-15.3_{-5.4}^{+5.3}$ \\ 
 & 20313 & $0.438_{-0.042}^{+0.042}$ & $0.087_{-0.036}^{+0.036}$ & $2.0_{-6.5}^{+6.2}$ & $-21.5_{-10.1}^{+9.8}$ \\ 
 & 21118 & $0.343_{-0.036}^{+0.036}$ & $0.098_{-0.032}^{+0.031}$ & $4.0_{-3.1}^{+3.0}$ & $-4.4_{-4.7}^{+4.6}$ \\ 
 & 21119 & $0.531_{-0.039}^{+0.039}$ & $0.269_{-0.034}^{+0.033}$ & $9.1_{-3.7}^{+3.6}$ & $1.3_{-4.4}^{+4.5}$ \\ 
 & 21120 & $0.449_{-0.040}^{+0.040}$ & $0.158_{-0.034}^{+0.034}$ & $7.9_{-3.7}^{+3.7}$ & $-0.7_{-4.7}^{+4.6}$ \\ 
\hline 
\end{tabular}
\tablefoot{The upper part of the table displays the best-fit astrometric solution for each CCO -- proper motion and position at a reference time -- as resulting from our fit. 
In the lower part, we show for all systems the optimal determined astrometric calibration parameters -- describing translation, stretch and rotation of the coordinate system -- for each individual observation.  
\tablefoottext{a}{For RX J1713.7$-$3946, only relative astrometric analysis was performed, via direct subtraction of positions from the two epochs. Therefore, our fit provides no absolute source positions.}
\tablefoottext{b}{For Cas A, the closely spaced individual observations (IDs 11240, 12057, 12058, 12059) were merged to create a single late-time data set.}

The tangent points, which we used for conversion of celestial coordinates to a (right-handed) local Cartesian coordinate system and as center of scaling and rotation, were 
$(18^{h} 18^{m} 52.\!\!^{s}080,\,-15^{\circ} 02^{\prime} 14.\!\!^{\prime\prime}11)$ for G15.9+0.2, 
$(18^{h} 52^{m} 38.\!\!^{s}4888,\, +00^{\circ} 40^{\prime} 19.\!\!^{\prime\prime}848)$ for Kes 79,  
$(23^{h} 23^{m} 27.\!\!^{s}940,\,58^{\circ} 48^{\prime} 42.\!\!^{\prime\prime}40)$ for Cas A, 
$(16^{h} 01^{m} 03.\!\!^{s}100,\,-51^{\circ} 33^{\prime} 53.\!\!^{\prime\prime}00)$ for G330.2+1.0, 
$(17^{h} 13^{m} 28.\!\!^{s}320,\,-39^{\circ} 49^{\prime} 53.\!\!^{\prime\prime}34)$ for RX J1713.7$-$3946, 
and $(17^{h} 20^{m} 54.\!\!^{s}500,\,-37^{\circ} 26^{\prime} 52.\!\!^{\prime\prime}00)$ for G350.1$-$0.3. 
}
\end{table*}

\clearpage

\begin{table*}[h!]
\renewcommand{\arraystretch}{1.5}
\caption{Results of expansion measurements in individual regions for G15.9+0.2, Kes 79,  and G350.1$-$0.3}
\label{ExpTable}
\centering
\begin{tabular}{cccccc}
\hline\hline
SNR & Region & $\vartheta$ & $\mu_{\rm exp} $ & $\tau^{-1} $ & $\tau$ \\
 & & $(\si{arcsec})$ & $(\si{mas.yr^{-1}})$ & $(10^{-4}\,\si{yr^{-1}})$ & $(\si{yr})$ \\
\hline
G15.9+0.2 & A & $ 152^{+15}_{-15}$ & $ 57^{+61}_{-70}$ & $ 3.66^{+4.38}_{-4.83}$ & ...\\ 
 & B & $ 134^{+15}_{-15}$ & $ 80^{+39}_{-37}$ & $ 5.99^{+3.30}_{-3.15}$ & ... \\ 
 & C & $ 134^{+15}_{-15}$ & $ -16^{+59}_{-63}$ & $ -1.23^{+4.68}_{-5.13}$ & ... \\ 
 & D & $ 144^{+15}_{-15}$ & $ 106^{+63}_{-59}$ & $ 7.44^{+4.71}_{-4.29}$ & ... \\ 
 & E & $ 156^{+15}_{-15}$ & $ 3^{+84}_{-62}$ & $ 0.24^{+5.60}_{-4.21}$ & ... \\ 
 & F & $ 139^{+15}_{-15}$ & $ 138^{+75}_{-74}$ & $ 9.97^{+5.68}_{-5.55}$ & ... \\ 
 & G & $ 140^{+15}_{-15}$ & $ -25^{+230}_{-183}$ & $ -1.58^{+16.11}_{-13.55}$ & ... \\ 
\hline
Kes 79 & A & $ 248^{+53}_{-52}$ & $ 105^{+24}_{-23}$ & $ 4.24^{+1.74}_{-1.34}$ & ... \\ 
 & B & $ 263^{+52}_{-51}$ & $ -21^{+26}_{-28}$ & $ -0.80^{+1.20}_{-1.30}$ & ... \\ 
 & C & $ 166^{+52}_{-50}$ & $ 73^{+81}_{-72}$ & $ 4.41^{+5.73}_{-4.50}$ & ... \\ 
 & D & $ 178^{+53}_{-52}$ & $ -1^{+54}_{-55}$ & $ -0.08^{+3.38}_{-3.38}$ & ... \\ 
 & E & $ 126^{+57}_{-52}$ & $ -31^{+71}_{-66}$ & $ -2.36^{+5.80}_{-6.36}$ & ... \\ 
 & F & $ 205^{+54}_{-54}$ & $ 17^{+168}_{-255}$ & $ 0.80^{+8.61}_{-12.55}$ & ... \\
 & G & $ 153^{+51}_{-50}$ & $ 208^{+143}_{-157}$ & $ 13.05^{+10.06}_{-10.13}$ & ... \\
 & H & $ 161^{+56}_{-55}$ & $ 46^{+38}_{-38}$ & $ 2.85^{+3.36}_{-2.60}$ & ... \\ 
\hline
G350.1$-$0.3 & A & $ 190^{+6}_{-6}$ & $ 253^{+58}_{-65}$ & $ 13.39^{+3.07}_{-3.59}$ & $745^{+263}_{-138}$ \\ 
 & B & $ 182^{+6}_{-6}$ & $ 257^{+9}_{-9}$ & $ 14.09^{+0.89}_{-0.87}$ & $710^{+47}_{-42}$ \\ 
 & C & $ 177^{+6}_{-6}$ & $ 226^{+13}_{-12}$ & $ 12.78^{+1.03}_{-0.99}$ & $782^{+66}_{-58}$ \\ 
 & D & $ 166^{+6}_{-6}$ & $ 174^{+9}_{-11}$ & $ 10.51^{+0.94}_{-0.94}$ & $952^{+94}_{-78}$ \\ 
 & E & $ 155^{+6}_{-6}$ & $ 268^{+17}_{-17}$ & $ 17.28^{+1.46}_{-1.44}$ & $579^{+52}_{-45}$ \\ 
 & F & $ 137^{+6}_{-6}$ & $ 193^{+7}_{-8}$ & $ 14.04^{+1.14}_{-1.11}$ & $712^{+61}_{-54}$ \\ 
 & G & $ 134^{+5}_{-6}$ & $ 140^{+9}_{-9}$ & $ 10.47^{+1.14}_{-1.11}$ & $955^{+113}_{-94}$ \\ 
 & H & $ 120^{+6}_{-6}$ & $ 128^{+14}_{-15}$ & $ 10.75^{+1.64}_{-1.59}$ & $930^{+161}_{-123}$ \\ 
 & I & $ 150^{+5}_{-6}$ & $ 146^{+7}_{-7}$ & $ 9.72^{+0.96}_{-0.93}$ & $1029^{+109}_{-92}$ \\ 
 & J & $ 153^{+5}_{-6}$ & $ 137^{+4}_{-4}$ & $ 9.01^{+0.85}_{-0.83}$ & $1110^{+112}_{-95}$ \\ 
 & K & $ 157^{+5}_{-6}$ & $ 180^{+6}_{-7}$ & $ 11.47^{+0.92}_{-0.89}$ & $872^{+74}_{-65}$ \\ 
 & L & $ 159^{+5}_{-6}$ & $ 180^{+8}_{-7}$ & $ 11.30^{+0.93}_{-0.90}$ & $885^{+77}_{-67}$ \\ 
 & M & $ 87^{+6}_{-6}$ & $ 123^{+37}_{-38}$ & $ 14.13^{+4.77}_{-4.64}$ & $706^{+336}_{-177}$ \\ 
 & N & $ 118^{+6}_{-6}$ & $ 100^{+44}_{-45}$ & $ 8.40^{+3.82}_{-3.92}$ & $1154^{+772}_{-345}$ \\ 
 & O & $ 88^{+6}_{-6}$ & $ 71^{+36}_{-36}$ & $ 8.11^{+4.29}_{-4.31}$ & $1172^{+860}_{-381}$ \\ 
 & P & $ 64^{+6}_{-6}$ & $ 84^{+27}_{-26}$ & $ 13.13^{+4.76}_{-4.46}$ & $760^{+382}_{-201}$ \\ 
 & Q & $ 77^{+6}_{-6}$ & $ 59^{+31}_{-30}$ & $ 7.73^{+4.24}_{-4.37}$ & $1211^{+943}_{-395}$ \\ 
 & R & $ 61^{+6}_{-6}$ & $ 51^{+23}_{-24}$ & $ 8.26^{+4.35}_{-4.32}$ & $1157^{+854}_{-378}$ \\ 
\hline

\end{tabular}
\tablefoot{The table shows the quantities underlying Figs.~\ref{G15Exp}, \ref{KesExp} and \ref{G350Exp}. To ease comparison between the systems, we provide the resulting constraints on the expansion rate $\tau^{-1}$ for all three SNRs. In addition, we provide its inverse, the free expansion age $\tau$, only for G350.1$-$0.3, since the expansion rate has to be significantly different from zero to obtain a sensible value for individual regions.}
\end{table*}

\section{Illustration of SNR expansion measurements in individual regions}

\begin{figure*}[h!]
\centering
\includegraphics[width=15.2cm]{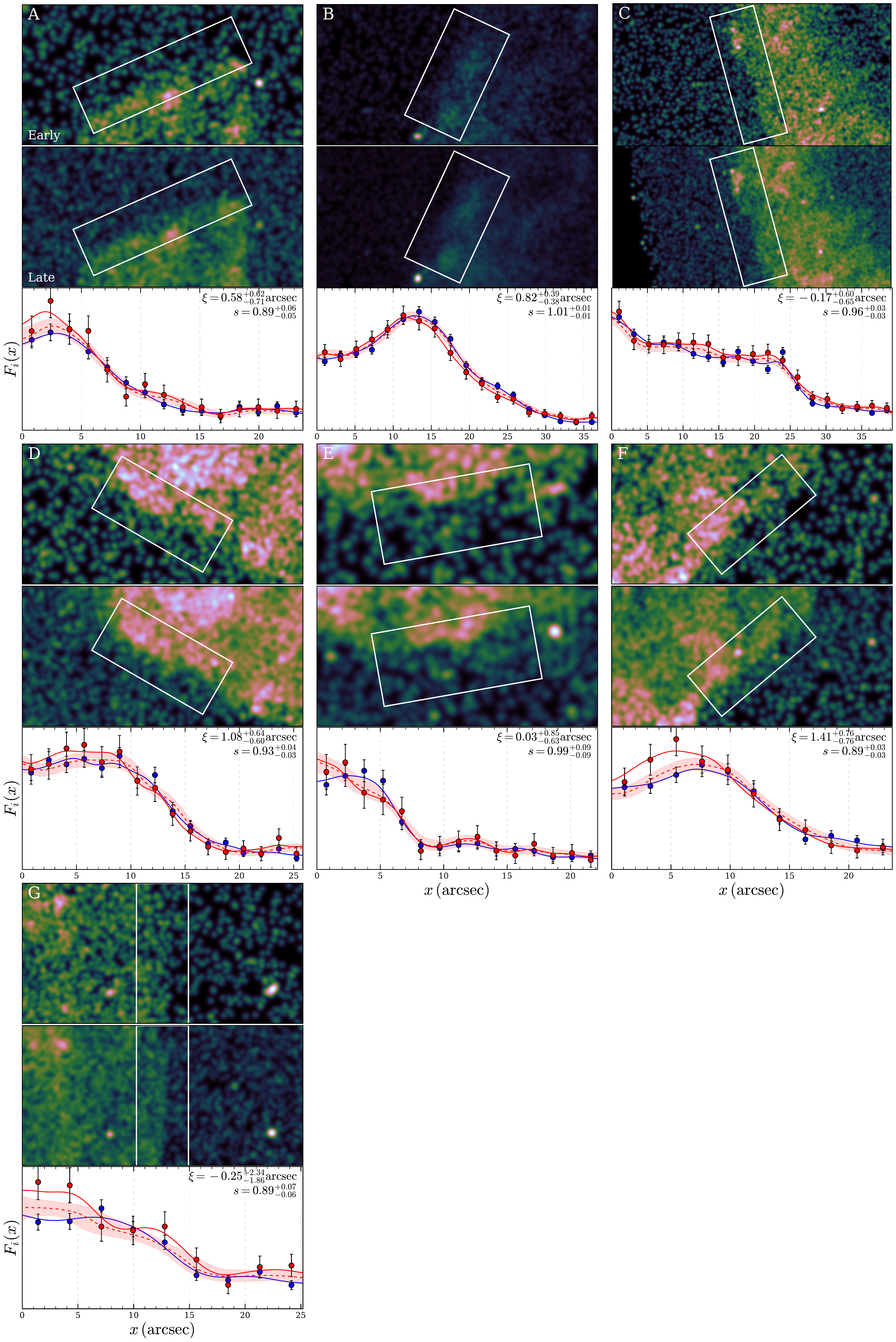}
\caption{Direct comparison of flux profiles in different regions of G15.9+0.2. Each panel is labelled with the letter corresponding to the respective region. Top and middle sub-panels show the exposure corrected images at the early and late epoch, and the white rectangle outlines the measurement region. 
The bottom sub-panel depicts the one-dimensional flux profiles $F_{i}(x)$ at the early (red) and late (blue) epoch. 
The $x$ coordinate increases outwards along the radial direction of the SNR.
The markers display the binned distribution, and the solid line reflects the flux profile smoothed with a Gaussian kernel. 
The dashed red line corresponds to the profile of the early epoch, shifted and rescaled by the parameters indicated in the upper right corner, with the shaded area displaying the associated uncertainty. 
}
\label{G15FullExp}
\end{figure*}

\begin{figure*}[h!]
\centering
\includegraphics[width=15.2cm]{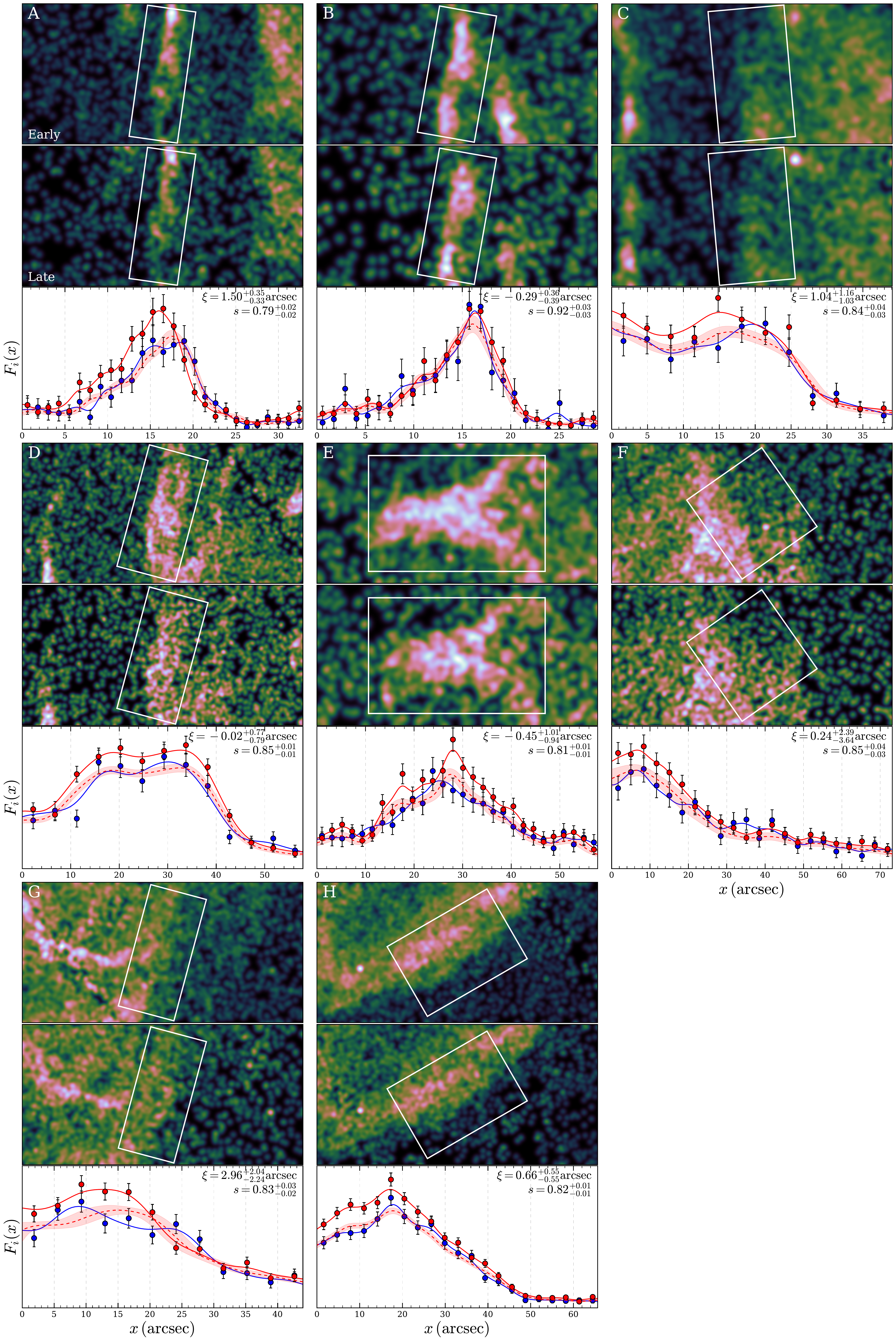}
\caption{Direct comparison of flux profiles in different regions of Kes 79 (as in Fig.~\ref{G15FullExp}).}
\label{KesFullExp}
\end{figure*}

\begin{figure*}[h!]
\centering
\includegraphics[width=15.2cm]{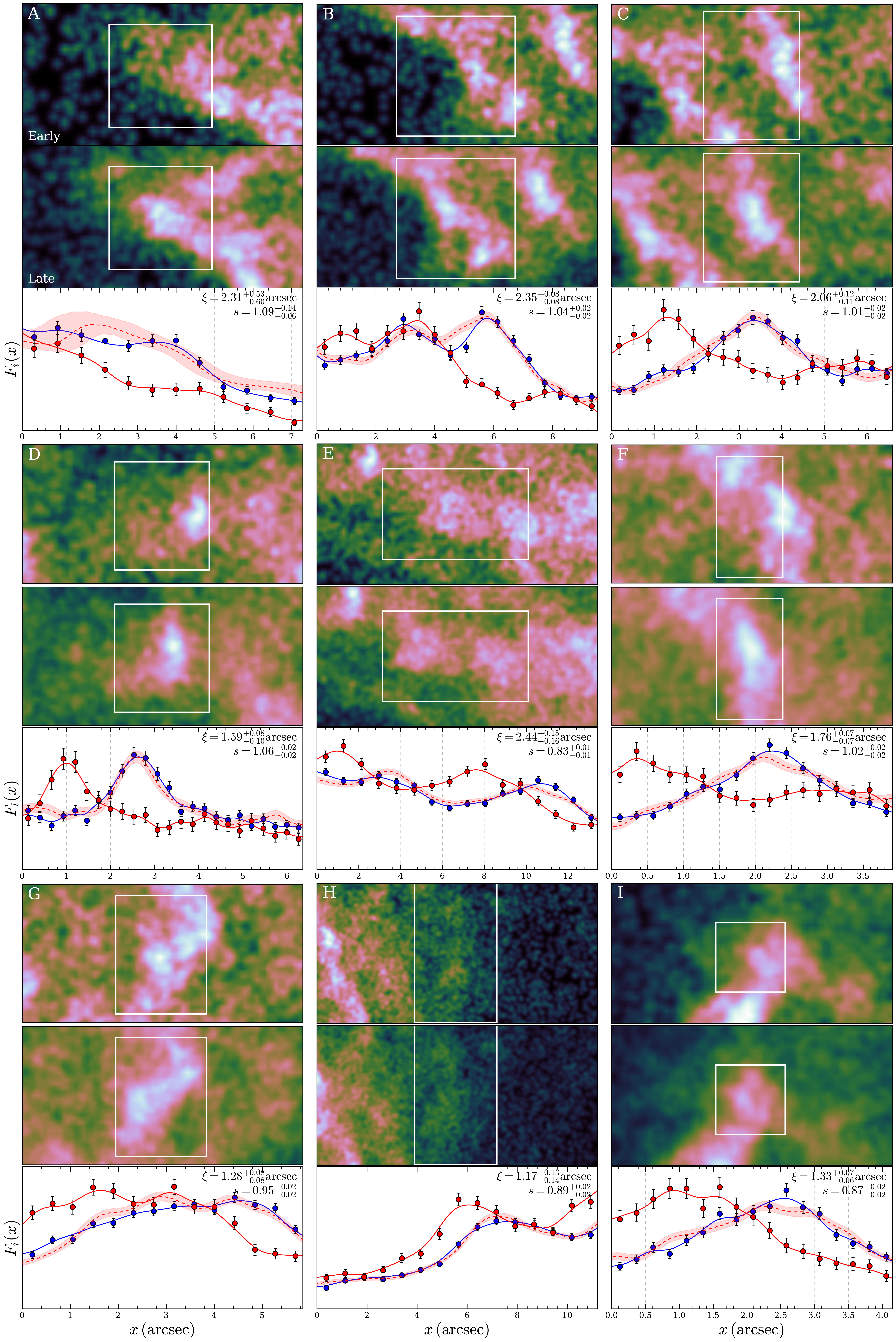} 
\caption{Direct comparison of flux profiles of G350.1$-$0.3 (as in Fig.~\ref{G15FullExp}).}
\label{G350FullExp_1}
\end{figure*}

\addtocounter{figure}{-1} 

\begin{figure*}[h!]
\centering
\includegraphics[width=15.2cm]{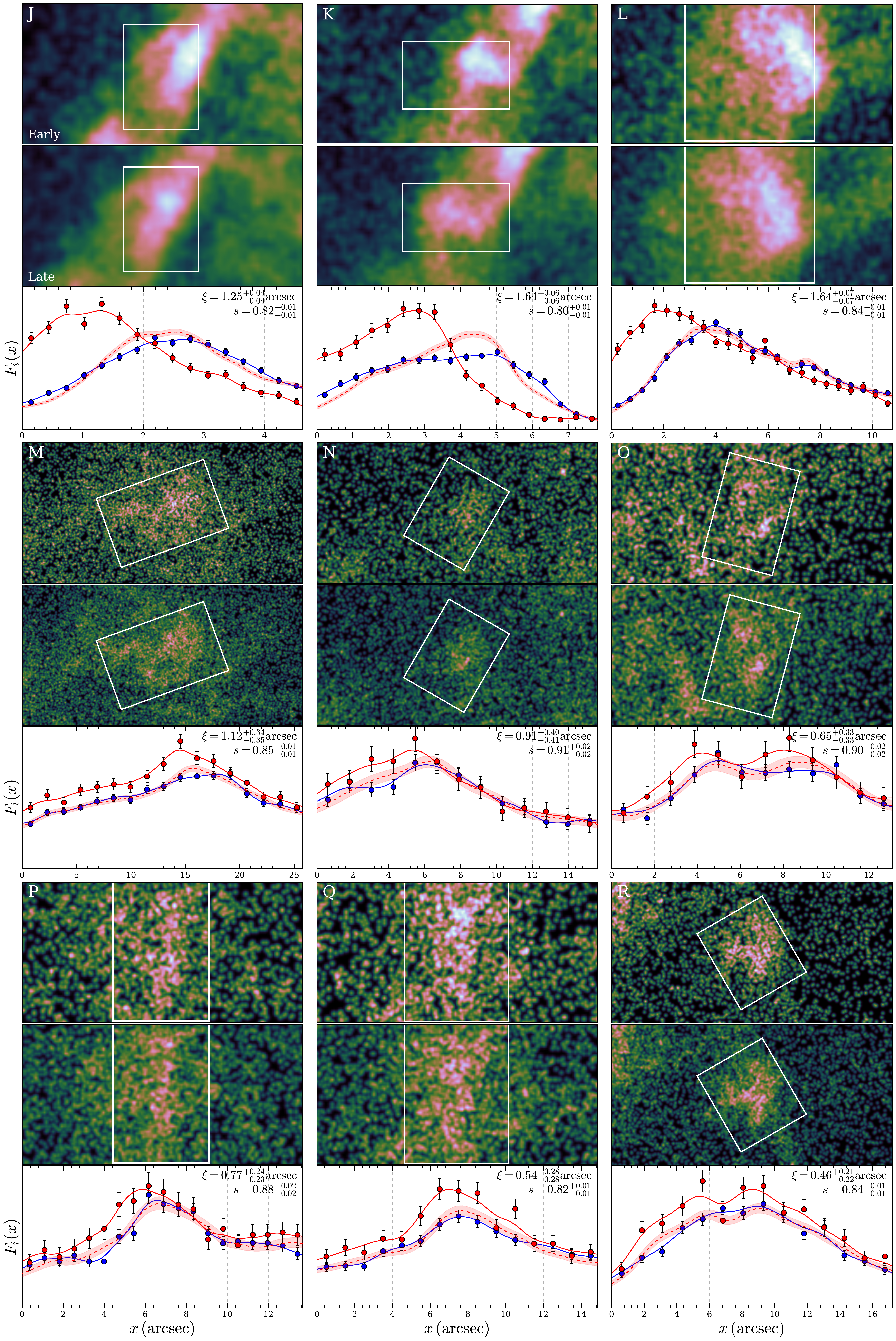}
\caption{- Continued}
\label{G350FullExp_2}
\end{figure*}

\end{appendix}
\end{document}